\documentclass[twocolumn]{aastex631}

\newcommand{\sqcm}{cm$^{-2}$}  
\newcommand{\kms}{$\rm km~s^{-1}$} 
\newcommand{\lya}{Ly$\alpha$}
\newcommand{\lyb}{Ly$\beta$}

\newcommand{\HI}{\mbox{H\,{\sc i}}}

\newcommand{\OI}{\mbox{O\,{\sc i}}}
\newcommand{\OII}{\mbox{O\,{\sc ii}}}

\newcommand{\OVI}{\mbox{O\,{\sc vi}}}

\newcommand{\logm}{${\rm log}_{10}(M_{\star}/\rm M_{\odot})$}
\newcommand{\Msun}{$\rm M_{\odot}$} 

\newcommand{\lognovi}{${\rm log}_{10}(N_{\rm OVI}/{\rm cm}^{-2})$}

\newcommand{\novie}{N_{\rm O\mbox{\sc vi}}}

\usepackage[flushleft]{threeparttable}
\usepackage{ragged2e}
\usepackage{graphicx}	
\usepackage{amsmath}	
\usepackage{amssymb}	
\usepackage{tabularx}
\usepackage{newtxtext,newtxmath,ulem}
\usepackage{multirow}
\usepackage{cases}
\usepackage{booktabs}
\usepackage[T1]{fontenc}
\usepackage{xcolor}
\usepackage{soul}
\usepackage{adjustbox}

\usepackage{dcolumn}
    \newcolumntype{d}[1]{D{.}{.}{#1}}

\begin{document}

\title{MUSEQuBES: The column density, covering fraction, and mass of \OVI-bearing gas in and around low-redshift galaxies} 

\author[0009-0000-0797-7365]{Sayak Dutta}
\affiliation{Inter-University Centre for Astronomy \& Astrophysics, Post Bag 04, Pune, India 411007}

\author[0000-0003-3938-8762]{Sowgat Muzahid}
\affiliation{Inter-University Centre for Astronomy \& Astrophysics, Post Bag 04, Pune, India 411007}

\author{Joop Schaye}
\affiliation{Leiden Observatory, Leiden University, PO Box 9513, NL-2300 RA Leiden, the Netherlands}

\author{Sean Johnson}
\affiliation{Department of Astronomy, University of Michigan, 1085 S. University Ave, Ann Arbor, MI 48109, USA}

\author{Nicolas F. Bouch\'e}
\affiliation{Centre de Recherche Astrophysique de Lyon (CRAL) UMR5574, Univ Lyon1, Ens de Lyon, CNRS, 69230 Saint-Genis-Laval,
France}

\author{Hsiao-Wen Chen} 
\affiliation{Department of Astronomy \& Astrophysics, The University of Chicago, 5640 S. Ellis Avenue, Chicago, IL 60637, USA} 

\author{Sebastiano Cantalupo}
\affiliation{Department of Physics, University of Milan Bicocca, Piazza della Scienza 3, 20126, Milano, Italy}

\begin{abstract}

\noindent 
We present a study of \OVI-bearing gas around 247 low-mass (median \logm~$=8.7$) galaxies at low redshifts $(0.1 < z < 0.7)$ using background quasars as part of the MUSE Quasar-fields Blind Emitters Survey (MUSEQuBES). 
We find that the average \OVI\ column density, ${\rm log}_{10}\left<N(\OVI)/{\rm cm}^{-2}\right> = 14.14^{+0.09}_{-0.10}$, measured within the virial radius for our sample, is significantly lower than for $L_*$ galaxies. 
Combining 253 star-forming galaxies (mostly more massive) from the literature with 176 star-forming galaxies from MUSEQuBES, we find that both $\left<N(\OVI)\right>$ and the average covering fraction peak at \logm $\approx9.5$. The virial temperature corresponding to this stellar mass is ideal for \OVI\ production via collisional ionization.  
However, we argue that photoionization and/or non-equilibrium processes 
are necessary to produce the \OVI\ associated with low-mass, dwarf galaxies (\logm~$<9$). The average \OVI\ mass within the virial radius of dwarf galaxies is measured to be $ 10^{5.2_{-0.1}^{+0.1}}$~\Msun.   
 The characteristic normalized impact parameter at which the \OVI\ covering fraction drops to half of its peak value is largest ($\approx 1.1$) for galaxies with stellar mass \logm~$\approx9.5$.   
We report the presence of a highly ionized metal floor with $\log_{10} N(\OVI)/{\rm cm^{-2}} = 13.2$ outside the virial radius of dwarf galaxies inferred from median spectral stacking.  

\end{abstract}

\keywords{galaxies: formation – galaxies: evolution – galaxies: haloes – (galaxies:) quasars: absorption lines}

\section{Introduction} \label{sec:intro}

The gaseous halos surrounding galaxies, known as the circumgalactic medium (CGM), are thought to retain evidence of both gas accretion and galactic outflows—the two poorly understood aspects of galaxy evolution models. The CGM consists of metal-rich, kinematically complex, multiphase gas that extends at least out to the virial radius of galaxies \citep[see][for a review]{Tumlinson_2017}. The total gas and metal masses across the different phases of the CGM can potentially account for the ``missing baryons'' and ``missing metals'' in galaxy halos \citep[e.g.,][]{McGaugh_2010, Peeples_2014}. Therefore, probing and characterizing different gas phases in the CGM is crucial. Owing to its tenuous nature, the CGM is challenging to study in emission. With the advent of state-of-the-art integral field spectrographs, however, the situation is changing dramatically \citep[see e.g.,][]{Dutta_2024, Guo_2023, Zabl_2021}.

The atomic and ionic transitions serving as sensitive diagnostics for the physical and chemical conditions of the CGM predominantly fall within the rest-frame ultraviolet (UV) region.
Absorption line spectroscopy of bright background quasars has paved the way for CGM studies \citep[e.eg.,][]{Bergeron_1986, Petitjean_1990}.  
With the availability of the far-UV (FUV) sensitive Cosmic Origins Spectrograph (COS) onboard the Hubble Space Telescope ($HST$), multiple surveys have been conducted to investigate the connection between low-$z$ galaxies and the gas surrounding them. The rest-frame UV transitions of the fifth ionization state of oxygen (i.e., the \OVI~$\lambda\lambda1031,1037$ doublet) are among the favourite line diagnostics for studying the metal enrichment in diffuse gas across different cosmic times \citep[see e.g.,][]{Aguirre_2008,Tripp_2008,Muzahid_2012}. This is owing to the suitable rest-frame wavelengths, the large oscillator strengths of the transitions, and the large cosmic abundance of oxygen. However, \OVI\ is not the dominant ionization state of oxygen. The ionization fraction of \OVI, $f_{\OVI}$, reaches its peak ($\approx0.2$) at ionization parameter of $10^{-1}$ under photoionization equilibrium  \citep[PIE;][]{Oppenheimer_2013} or at a gas temperature of $T\approx10^{5.5}$~K under collisional ionization equilibrium \citep[CIE;][]{Gnat_2007,Oppenheimer_2013}. Since the cooling function peaks at around similar temperatures, the cooling time for collisionally ionized, \OVI-bearing gas is rather short, particularly when the metallicity is relatively high. In case of non-equilibrium, when the gas can cool faster than it recombines, or if photoionization dominates, \OVI\ can trace lower temperatures \citep[]{Oppenheimer_2013}. Therefore, a wide range of physical processes give rise to \OVI\  \citep[see ][]{Dopita_1996, Edgar_1986, Heckman_2002, Begelman_1990, Borkowski_1990}.

The connection between bright galaxies and \OVI\ absorption has been investigated extensively using statistically significant samples of galaxy-absorber pairs. Studies have reported a near unity covering fraction of \OVI\ absorption extending out to $\approx100$~kpc around bright galaxies \citep[e.g.,][] {Prochaska_2011,Tumlinson_2011,Johnson_15, Kacprzak_2015}. Moreover, the late-type, star-forming, blue galaxies in these studies exhibited a significantly higher covering fraction and higher $N(\OVI)$ compared to the early-type, passive, red galaxies. \citet[]{Chen_2009} observed a stronger clustering of \OVI\ absorbers around star-forming galaxies compared to passive, absorption-line-dominated galaxies upto a projected distance of 3$h^{-1}$ Mpc \citep[but see][]{Finn_2016}. The reasons for this dichotomy is debated in the literature. Some theoretical studies have attributed the cause to the difference in halo mass between the observational samples of star-forming and passive galaxies \citep[e.g.,][]{Oppenheimer_2016,Sanchez_2019}. This conclusion is supported by the fact that halos with virial temperatures of  $\approx10^{5.5}$~K are conducive to \OVI\ formation under collisional ionization. The observational study of \citet{Tchernyshyov_2023}, however, showed that such a dichotomy exists even when the halo mass is controlled for the star-forming and passive samples. They advocated that a transformation of the CGM occurs concurrently with galaxy quenching, as predicted by simulations \citep[see][]{Oppenheimer_2016}.  \citet{Suresh_2017} also argued that the dichotomy stems from the effect of AGN feedback in their simulation. Theoretical studies of Milky-Way type galaxies further suggested that \OVI\ could arise in supernova-driven outflows \citep[e.g.,][]{Liang_2016, Li_2020} and photoionized inflows \citep[e.g.,][]{Stern_2018}. It is therefore crucial to study \OVI-bearing gas around galaxies with a large dynamic range in properties such as stellar mass ($M_{\star}$) and star formation rate (SFR) to better understand its origin.

The bulk of the circumgalactic \OVI\ observations in the literature are for high-mass galaxies ($M_{\star} \gtrsim 10^{9.5}~M_{\odot} $). Consequently, almost all of the theoretical studies aiming to reproduce the \OVI\ observables focus on Milky-Way type galaxies. A wide range of physical models can reproduce such observables for Milky-Way type galaxy, and a consensus on the origin of \OVI\ absorbers in galaxy halos is yet to emerge. With the advent of state-of-the-art integral field spectrographs (IFS) such as MUSE \citep[]{Bacon_2010} on VLT/UT4 and KCWI \citep[]{Morrissey_2018} on Keck, it is now efficient to find and identify low-mass galaxies around bright background quasars. As a consequence, the availability of \OVI\ measurements around low-mass galaxies is gradually increasing \citep[see e.g.,][]{Qu_2024}.

In this study, we revisit the connection between low-$z$ galaxies and \OVI-bearing gas surrounding them using a unique, low-mass sample of galaxies drawn from the low-$z$ part of the MUSEQuBES survey \citep[]{Dutta_2024}. Exploiting a total of $\approx65$~h of GTO observations, MUSEQuBES conducted a galaxy survey around 16 UV-bright background quasars that have high-quality HST/COS spectra. Each MUSE field yields $\approx90,000$ spectra from each $0.2'' \times 0.2''$ spaxel, enabling the identification of low-mass galaxies without the need for any preselection. Our deep MUSE observations, with on-source exposure times ranging from 2 to 10 hours, yielded a homogeneous sample primarily composed of low-mass (median $M_{\star} \approx 10^{8.7}~\rm M_{\odot}$),  sub-$L_*$ galaxies. A significant fraction of these galaxies exhibited impact parameters from the background quasars, $D < R_{\rm vir}$, where $R_{\rm vir}$ is the inferred virial radius of the halo hosting the intervening galaxy.

This paper is organized as follows: In Section~\ref{sec:data} we describe the galaxy sample used in this paper. Section~\ref{sec:absorption_analysis} provides the details of the absorption line data and galaxy-absorber pairs. The results are presented in Section~\ref{sec:results} followed by a discussion in Section~\ref{sec:discussion}. Our key findings are summarized in Section~\ref{sec:summary}. Throughout the paper, we adopt a $\Lambda$-CDM cosmology with $\Omega_{\rm m} = 0.3$, $\Omega_{\Lambda} = 0.7$, and a Hubble constant of $H_0 = 70~ \rm km~ s^{-1}~ Mpc^{-1}$. All distances are in physical units unless specified otherwise.

\section{Data} 
\label{sec:data}

A galaxy survey centered on UV-bright quasars is essential for our study of circumgalactic \OVI\ absorption. In this work, we primarily used the galaxies from the MUSEQuBES survey at low $z$ \citep[]{Dutta_2024}. This provides a set of 247 new quasar-galaxy pairs for which \OVI\ absorption can be studied. An overview of the survey design and analysis procedure for MUSEQuBES is given in \citet[]{Dutta_2024}. A brief summary is given below.

\begin{figure}
    \centering
    \includegraphics[width=1.03\linewidth]{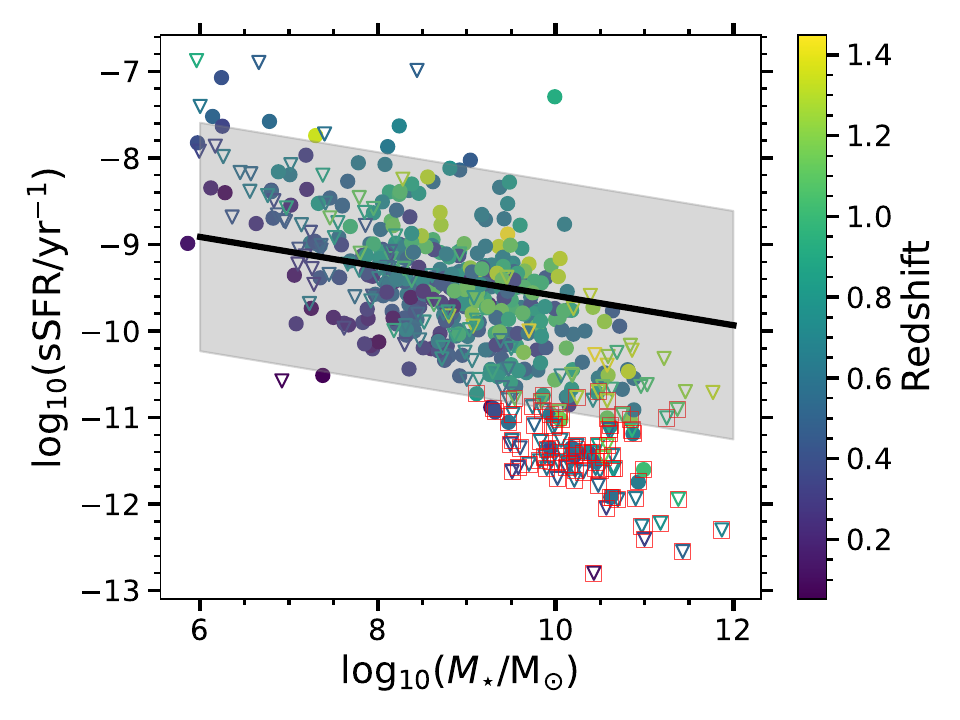}
    \caption{The sSFR of all MUSEQuBES galaxies plotted against the stellar mass. The filled circles and open triangles represent the detection of and 3$\sigma$ upper limits on the sSFR, respectively. The points are color-coded by the redshift of the galaxies. The points with red squares around them represent quenched galaxies, defined as galaxies with sSFR measurements (detection or 3$\sigma$ upper limits) below 3$\sigma$ scatter of redshift dependant star-forming main sequence \citet[]{Boogard_18}. The SFMS relation at $z=0.5$ and its 3$\sigma$ scatter are shown with the black solid line and the gray shaded region.}  
    \label{fig:prop_sfms}
\end{figure}

\subsection{The MUSEQuBES galaxy survey}
\label{sec:MUSEQuBES_galaxy_survey}

We carried out a blind spectroscopic redshift survey in the fields centered on 16 quasars using the MUSE instrument, which is mounted on the VLT/UT4, and has a 1x1 arcmin field-of-view (FoV). The selected 16 quasars (with emission redshift, $z_{\rm qso}$ ranging from $0.42-1.44$) have the best signal-to-noise ($S/N$; 15 - 85 per resolution element) FUV spectra available in the $HST$ Spectroscopic Legacy Archive \citep[HSLA]{Peeples_2017}. With on-source exposure time ranging from $2-10$ hours per MUSE field, we detected a total of 413 galaxies foreground to the quasars (at least 2000~\kms\ bluewards of the quasar redshifts). These galaxies, with secured spectroscopic redshifts, are detected in the MUSE white-light images \citep[see][for details]{Dutta_2024}. The 1D spectra of these continuum-detected galaxies are extracted from MUSE cubes and examined through MARZ \citep[]{Hinton_2016} to determine the redshift based on the spectral features. The line fluxes of available emission and absorption lines are measured using PLATEFIT \citep[]{Tremonti_2004}.

The star-formation rates (SFR) are determined by the H$\alpha$ or [\OII] (when H$\alpha$ is not covered or not detected at  $>3\sigma$) line fluxes returned by PLATEFIT using the \citet[]{Kennicutt_1998} relations (for H$\alpha$)  or \citet[]{Kewley_2004} relations (for \OII), corrected to the \citet[]{Chabrier_2003} initial mass function. Further, a dust correction is applied to SFRs measured from H$\alpha$ whenever the H$\beta$ line is detected with $>3\sigma$ significance. For galaxies without $S/N>3$ detection of H$\alpha$, H$\beta$  (in combination with H$\gamma$), or [\OII], we could not estimate the SFR. For these galaxies, we use the PLATEFIT error on the H$\alpha$ (for $z < 0.42$) or [\OII] (for $z > 0.42$) line flux to derive a 3$\sigma$ upper limit on the SFR.

Stellar masses ($M_{\star}$) for the galaxy sample are calculated using the spectral energy distribution fitting code FAST \citep[]{Kriek_2009}. We created a series of 11 boxcar filters from the full available wavelength range of MUSE to perform the fit. Once the stellar mass is estimated, we use the abundance matching relation from \citet[]{Moster_2013} to obtain the halo mass ($M_{\rm halo}$) and virial radius ($R_{\rm vir}$) which we define as the radius at which the mean enclosed density is 200 times the critical density of the universe.

The specific star-formation rate (sSFR~$\equiv {\rm SFR}/M_{*}$) is plotted against the \logm\ for the complete MUSEQuBES sample in Fig.~\ref{fig:prop_sfms}. The 3$\sigma$ upper limits and detections are shown with hollow downward triangles and filled circles, color-coded by the redshift of the galaxies. The passive galaxies, defined as galaxies located 3$\sigma$ below the redshift-dependent star-forming main sequence \citep[SFMS; see][]{Boogard_18}, are marked with red squares. These passive/quenched galaxies are flagged as `E'. Among the remaining galaxies, those with measured SFR and 3$\sigma$ upper limits on the SFR are classified as star-forming (`SF') and unclassified (`U') galaxies, respectively. Note that the galaxies in our sample show a large dynamic range in stellar mass and SFR, and the vast majority of them lie along the main-sequence relation. The color gradient for the points at a given stellar mass indicates the cosmic evolution of SFR \citep[see e.g.,][]{Lehnert_2015}.

The top, middle, and bottom panel of Fig.~\ref{fig:prop2} shows the SFR, sSFR, and stellar mass plotted against galaxy redshift with grey circles. The stellar masses of these galaxies range from $10^{7.9}-10^{9.4}~\rm M_{\odot}$ (68 percentile), with a median stellar mass of $10^{8.7}~\rm M_{\odot}$. Note that the galaxies for which the circumgalactic \OVI\ absorption can be studied with the COS FUV spectra are indicated by the blue color. In the next section, we discuss the subsample of MUSEQuBES galaxies used in this study.

\subsection{Selection of MUSEQuBES galaxies for \OVI\ study}  

In order to study the CGM of the MUSEQuBES galaxies using the \OVI~$\lambda\lambda~1031,1037$ doublet, we used the following selection criteria: 

\begin{itemize}

    \item A foreground galaxy (at least 2000 \kms\ bluewards of the quasar redshift) is selected if the redshifted wavelength of the \OVI\ doublet falls within the coverage of the COS spectra of the corresponding background quasar.
 
    \item Galaxies for which the redshifted \OVI\ absorption falls within the geocoronal \lya\ emission or Galactic absorption (1205--1220~\AA) or geocoronal \OI\ emission (1301--1305~\AA) are excluded from our analysis.

    \item There are Lyman-limit systems (LLS) in two of the sightlines used in this work (TEX0206-048 and Q1354+048). Galaxies in these fields for which the  redshifted \OVI\ lines fall below the respective Lyman limits are excluded from this work. 

    \item Additionally, the redshifted \OVI\ lines for 8 foreground galaxies towards quasar PKS0552--640 fall within the COS spectral gap. All these galaxies are excluded from this work.
    
\end{itemize}

\begin{figure}
    \centering
    \includegraphics[width=1.0\linewidth]{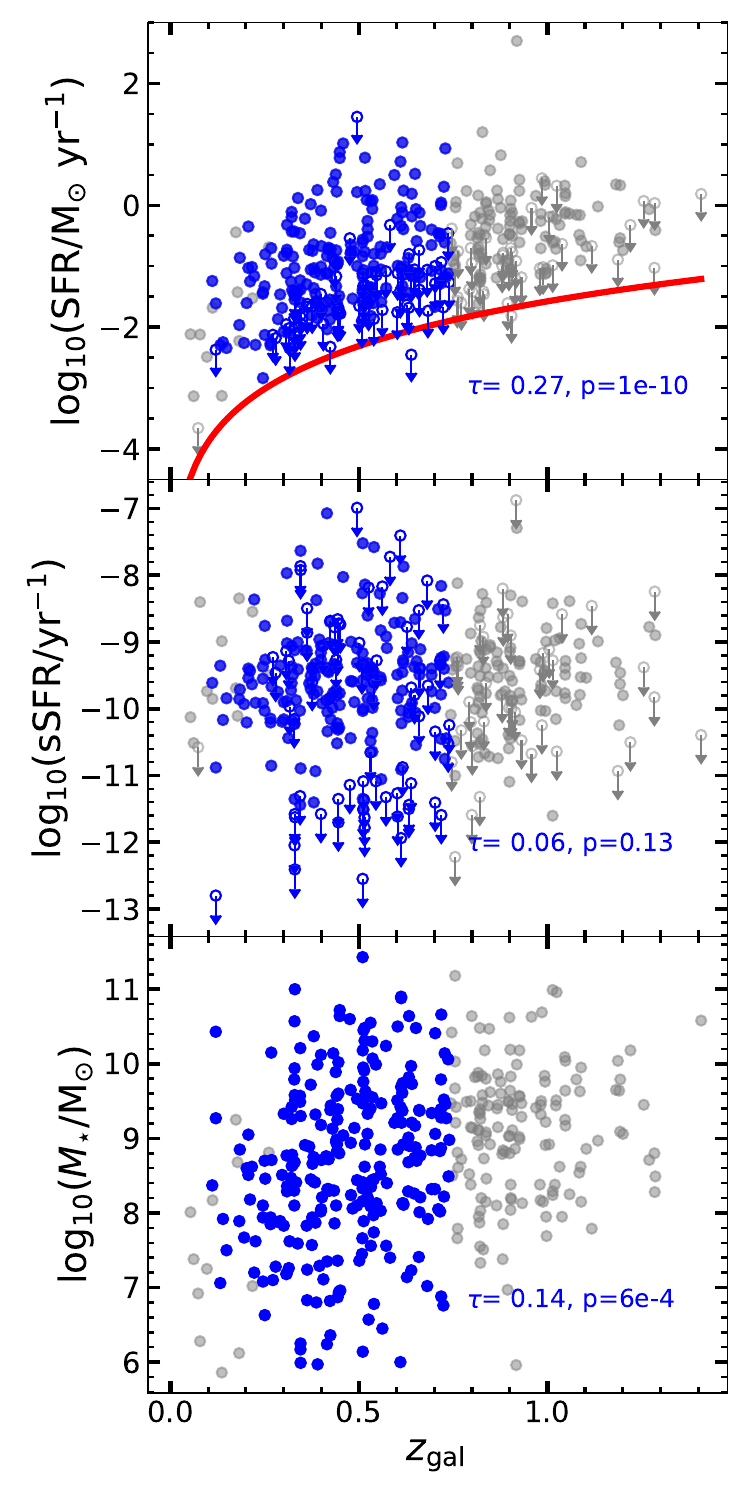}
    \caption{The SFR (top), sSFR (middle), and stellar mass (bottom) of all MUSEQuBES galaxies plotted against redshift. The subsample of galaxies used in this work is shown in blue. The upper limits are indicated by the downward arrows in the top and middle panels. The red solid line in the top panel represents the limiting SFR for a fiducial H$\alpha$ line flux of $10^{-18}~{\rm erg~cm^{-2}s^{-1}}$. The results of generalized Kendall's-$\tau$ rank correlation tests for the subsample used in this work (including the upper limits) are indicated in each panel.} 
    \label{fig:prop2}
\end{figure}

A sample of 247 galaxies from our MUSEQuBES survey satisfies the aforementioned conditions. Among these, 176, 29, and 42 galaxies are classified as `SF', `E', and `U', respectively. The SFR, sSFR, and $M_{\star}$ of these galaxies are plotted against the redshift in Fig.~\ref{fig:prop2} with blue circles. The gray circles indicate the galaxies in the MUSEQuBES catalog, that are not included in this work.

In the top panel of Fig.~\ref{fig:prop2}, a mild correlation is seen between SFR and redshift for the galaxies included in this work (generalized Kendall-$\tau = 0.27, p \ll 0.1$). The presence of a lower envelope suggests that this correlation is likely driven by the flux limit of our MUSE data. The red solid line, which adequately explains the lower envelope, shows the limiting SFR for a fixed H$\alpha$ line flux of $10^{-18}~\rm erg~cm^{-2}~s^{-1}$. The relatively weak correlation seen between \logm\ and $z$ ($\tau = 0.14, p\ll 0.1$) in the bottom panel could be a manifestation of the underlying correlation between SFR and $z$ propagated through the SFMS. The middle panel shows that no significant trend is seen between sSFR (SFR/$M_{\star}$) and $z$ ($\tau=0.06, p = 0.13$). This is expected as the stellar mass dependence on redshift cancels the sSFR dependence on redshift. 

\begin{figure*}
\centering
\begin{tabular}{cc}
\adjustbox{valign=b}{\begin{tabular}{@{}c@{}}
      \includegraphics[width=.4\linewidth]{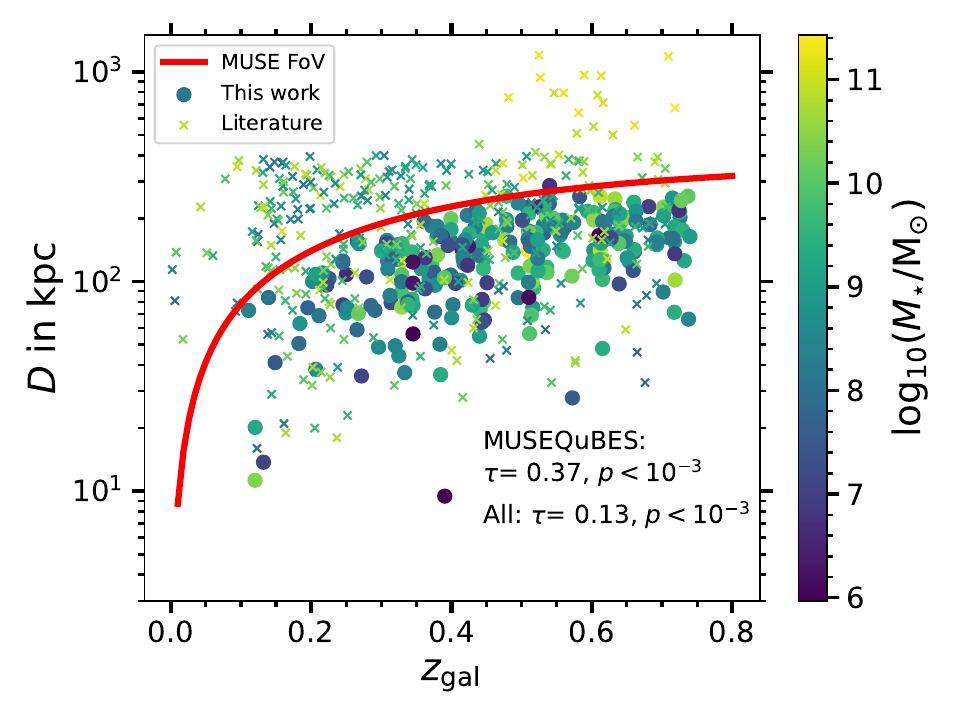} \\
      \includegraphics[width=.4\linewidth]{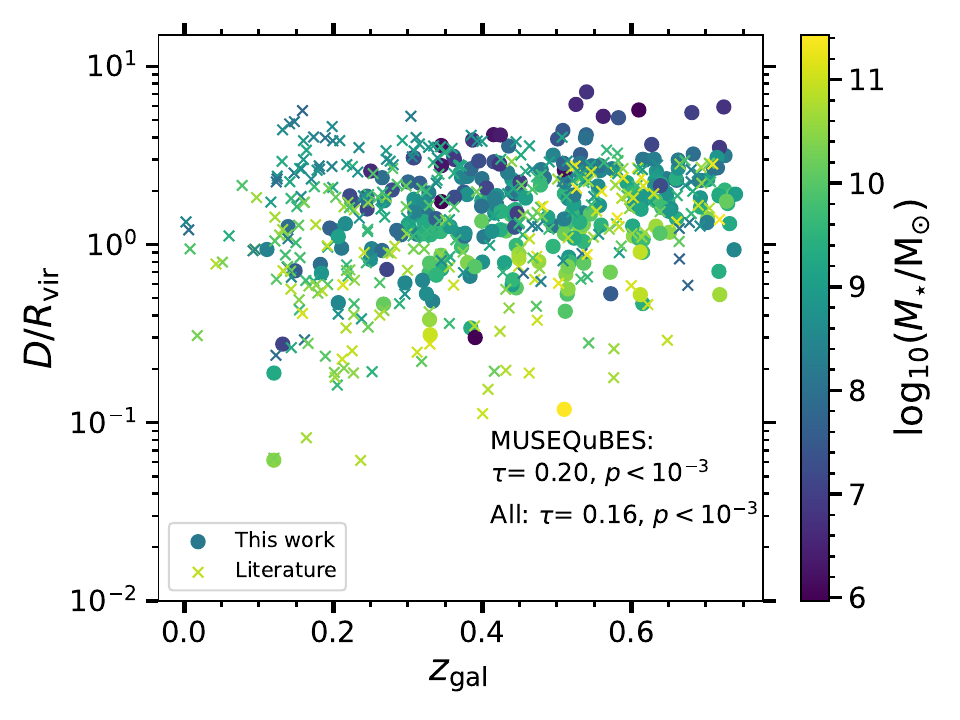}
\end{tabular}}
\hskip -12pt
\adjustbox{valign=b}{
      \includegraphics[width=.6\linewidth,height=11.1cm]{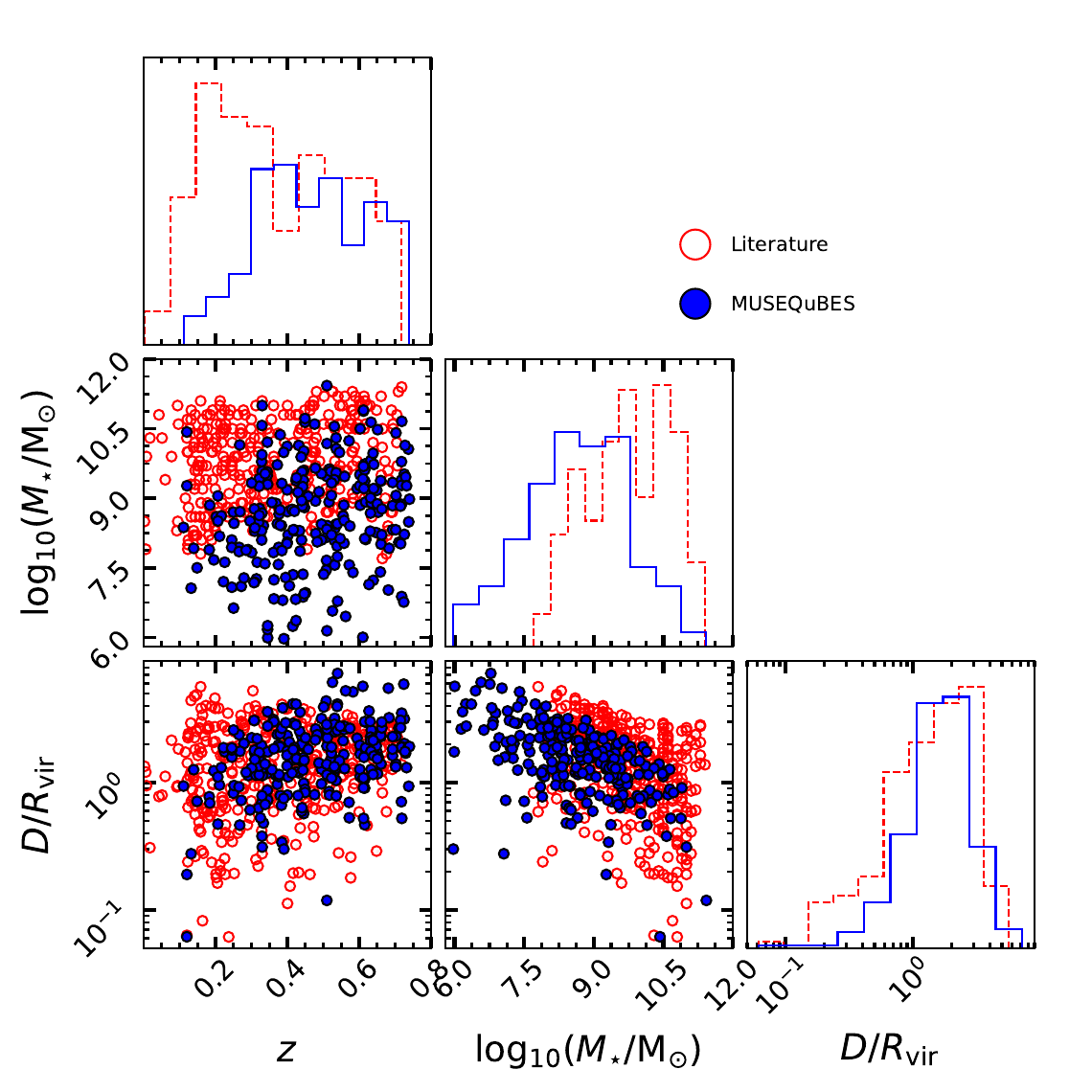}}   
\end{tabular}
\caption{ {\tt Left:} The impact parameter (top left) and normalized impact parameter (bottom left) are plotted against the redshift of galaxies used in this work. The points are color-coded by the stellar mass of the galaxies. The filled circles and crosses represent galaxies from MUSEQuBES and literature, respectively. The red solid line in the top left panel shows the maximum physical distance corresponding to the MUSE FoV as a function of redshift. The impact parameters (both physical and normalized) of the MUSEQuBES galaxies used in this work show a moderate correlation with redshift, as indicated by Kendall's $\tau$ coefficients. The correlation is relatively weak in the combined sample of galaxies from MUSEQuBES and literature. 
{\tt Right:} The $z$, \logm, and $D/R_{\rm vir}$ distribution of the combined sample of literature and MUSEQuBES plotted against each other, with histograms plotted along the diagonal. The MUSEQuBES sample is shown with filled blue circles and solid blue lines while the literature sample is shown with hollow red circles and dashed red lines. }
\label{fig:d_z}
\end{figure*}

The MUSE FoV of $1'\times1'$ restricts the impact parameter to $\approx300$ kpc at $z\approx0.6$ ($\approx$100 kpc at $z\approx0.1$). In the top left and bottom left panels of Fig.~\ref{fig:d_z}, we show the impact parameter ($D$) and normalized impact parameter ($D/R_{\rm vir}$), respectively, plotted against the redshift of MUSEQuBES galaxies ($z_{\rm gal}$) included in this work with solid circles. The data points are color-coded by the stellar mass. The red line in the top left panel represents the maximum impact parameter probed by the MUSE FoV. The moderate correlation between $D$ and $z_{\rm gal}$ ($\tau = 0.37,~p \ll 0.1$) most likely arises due to this envelope restricting high-impact parameter galaxies from being detected at lower redshifts. The crosses in the two panels of Fig.~\ref{fig:d_z} are the galaxies selected from literature for this work (see Sec.~\ref{sec:gal_lit} for more details). Many of these galaxies are observed in non-IFU surveys with a larger FoV. Consequently, their impact parameters go beyond our MUSEQuBES galaxies. The right panel of Fig.~\ref{fig:d_z} shows the $z$, \logm, and $D/R_{\rm vir}$ distribution of the combined sample of literature and MUSEQuBES plotted against each other. The corresponding histograms are plotted along the diagonal. The MUSEQuBES sample is shown with filled blue circles and solid blue lines while the literature sample is shown with hollow red circles and dashed red lines.

\begin{figure*}
    \centering
    \includegraphics[width=0.5\linewidth]{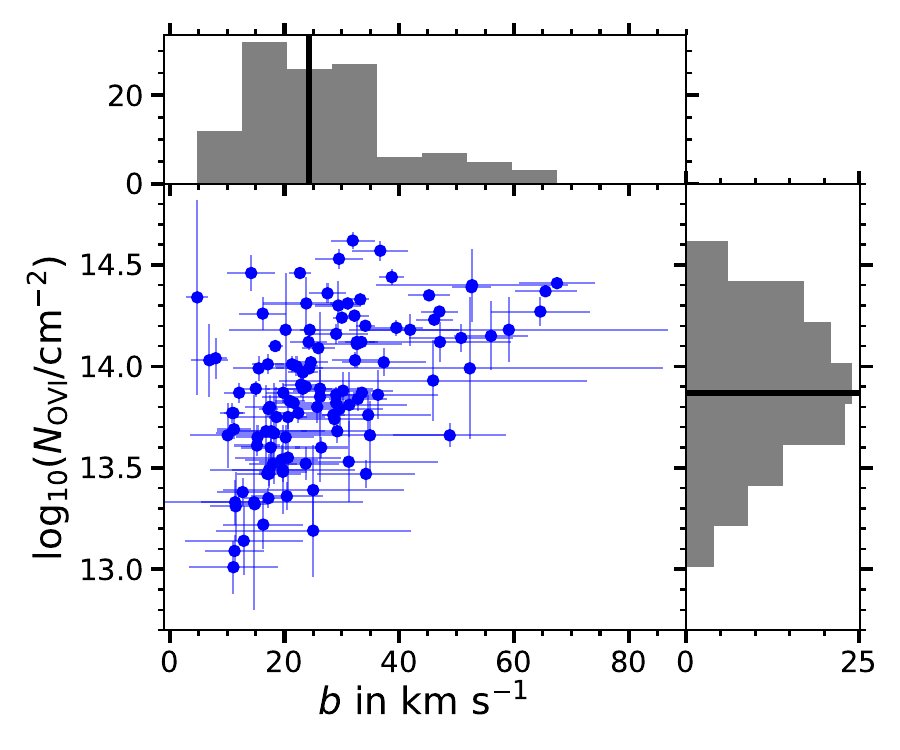}%
    \includegraphics[width=0.5\linewidth]{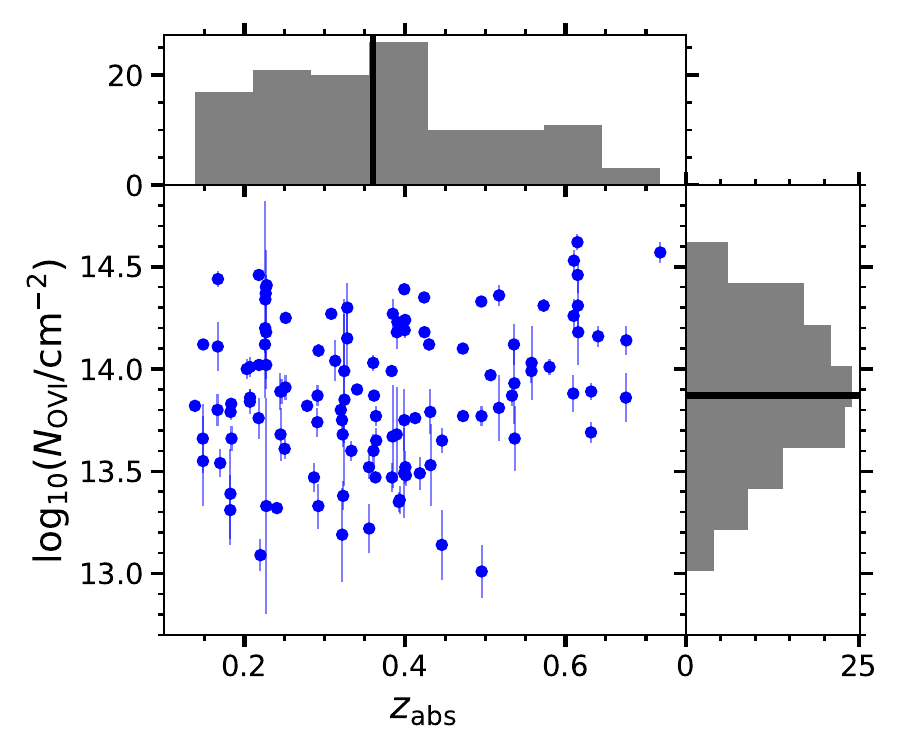}
    \caption{The column densities of the \OVI\ components obtained in the blind search across 16 background quasar spectra used in the MUSEQuBES survey plotted against $b-$parameter ({\tt left}) and redshift ({\tt right}). The histograms on the top represent $b$ and redshift distribution in the {\tt left} and {\tt right} panels, respectively. The histograms on the right represent the column density distribution in both panels.} 
    \label{fig:all_ind_comp}
\end{figure*}

\subsection{Galaxies from the literature for \OVI\ study}
\label{sec:gal_lit}

Although our analysis primarily uses galaxies from MUSEQuBES, the relatively smaller number of galaxies with high masses prompted us to combine our MUSEQuBES sample with galaxy samples from the literature. Recently, \citet[]{Tchernyshyov_2022} compiled $N(\OVI)$ measurements for a sample of 249 galaxies from several surveys in the literature alongside their own galaxy survey using Gemini-GMOS \citep[][]{Wilde_2021}.

\citet[]{Tchernyshyov_2022} used either public photometry, when available, to measure the galaxies' $M_{\star}$ and sSFR (e.g., the galaxies in \citet[][COS-Halos]{Tumlinson_2011}, \citet[][eCGM]{Johnson_15}, \citet{Keeney_2018}, \citet[COS-LRG]{Chen_2018, Zahedy_2019}, \citet[][RDR]{Berg_2019}) or the measurements reported in the original papers (e.g., \citet[][]{Johnson_17}, \citet[][]{Keeney_2017}, \citet[][QSAGE]{Bielby_2019}). \citet[]{Tchernyshyov_2022} provided the $M_{\star}$, $R_{\rm vir}$, and star formation flags (`SF' for star-forming and `E' for quenched)\footnote{They used a fixed sSFR cut of $10^{-10}~ \rm yr^{-1}$ for defining the `SF' and `E' galaxy populations} for all the 249 galaxies. Among these, 197 galaxies are classified as `SF' and 52 as `E'.

Additionally, we included $N(\OVI)$ measurements (detection and upper limits) for 98 out of 103 unique galaxies from the CUBS survey \citep[][]{Chen_20, Qu_2024} for which SFR measurements are available.\footnote{ We note here that the CUBS catalog associates the galaxy and $N(\OVI)$ absorption based on the smallest $D/R_{\rm vir}$ when multiple galaxies reside in a group environment.} Using the SFR and $M_{\star}$  from their catalog, we classified the galaxies with `SF', `E', or `U' based on their position relative to the SFMS, following the conditions adopted for the MUSEQuBES galaxies in Sec.~\ref{sec:MUSEQuBES_galaxy_survey}. In the CUBS dataset, there are 56 galaxies classified as `SF', 39 galaxies as `E', and 3 galaxies as `U'.

These 347 (249+98) galaxies from the literature are combined with the 247 MUSEQuBES galaxies to investigate the stellar mass dependence of the column density ($N(\OVI)$) and covering fraction ($\kappa$) profiles. In total, we compiled a sample of 594 galaxies for which circumgalactic \OVI\ measurements are available. Among these, 429 galaxies are categorized as `SF', 120 as `E', and 45 are classified as `U'.

\section{Absorption data analysis}
\label{sec:absorption_analysis}

We study the circumgalactic \OVI\ absorption for the galaxies observed with MUSE using the spectra of UV-bright background quasars with $z_{\rm qso} = 0.42-1.44$. The G130M+G160M grating spectra of COS covering 1150-1800~\AA\ provide \OVI\ coverage in the redshift range of $\approx 0.13-0.73$. All of the 16 quasars in this work have G130M (1150--1450~\AA) and G160M (1450--1800~\AA) grating spectra available in the HSLA archive \citep[]{Peeples_2017}. They are spliced together at a wavelength where the $S/N$ becomes approximately equal for the two grating spectra. The median $S/N$ and other information regarding the 16 quasar spectra are given in Table~1 of \citet[]{Dutta_2024}.

\subsection{Blind \OVI\ absorber catalog}

A blind catalog of \OVI\ absorbers in the 16 QSO spectra is created through visual inspection of the spectra using the doublet matching technique. The candidate absorbers are then decomposed into individual Voigt profile components using {\sc Vpfit} \citep[][]{vpfit} using the wavelength-dependent line spread function (LSF) of COS.  Each identified \OVI\ component is assigned a quality flag ($Q$), where $Q=3$ indicates the high-confidence \OVI\ components with both the doublet components detected without contamination and the corresponding \HI\ absorption also being detected. The components for which one of the doublet members is mildly blended or weak/noisy are classified as $Q=2$. \HI\ or other metal lines may be present but are not required for classification. Finally, $Q=1$ are tentative components for which both the doublet members are blended, or very weak, or noisy. Only \OVI\ absorption components with $Q=2$ and $3$ are used in this work.

There are a total of 118 \OVI\ components detected in the spectra of 16 quasars in the redshift range of $0.13 < z_{\rm abs} < 0.71$ ({74 are flagged as $Q=3$ and 44 as $Q=2$}). The best-fit component column densities and $b$-parameters returned by {\sc vpfit} are shown on the left panel of Fig.~\ref{fig:all_ind_comp}. The median $N(\OVI)$ and $b$ values are $\approx10^{13.9}~{\rm cm}^{-2}$ and $\approx25$ \kms, respectively. A moderate correlation (Kendall-$\tau=0.38,~p<0.001$) between $N(\OVI)$ and $b$ is observed.
However, this correlation does not persist when the components with $N(\OVI)\gtrsim10^{14}~{\rm cm}^{-2}$ are considered, below which our survey suffers from incompleteness.

The right panel of Fig.~\ref{fig:all_ind_comp} shows no significant correlation between $z_{\rm abs}$ and $N(\OVI)$ (Kendall-$\tau=0.1$). However, we noticed a dearth of data points with $N(\OVI) < 10^{13.7}$~\sqcm\ and $z>0.4$. This can be attributed to the lower $S/N$ of the G160M grating spectra used in this work.

\begin{figure*}
    \centering
    \includegraphics[width=0.5\linewidth]{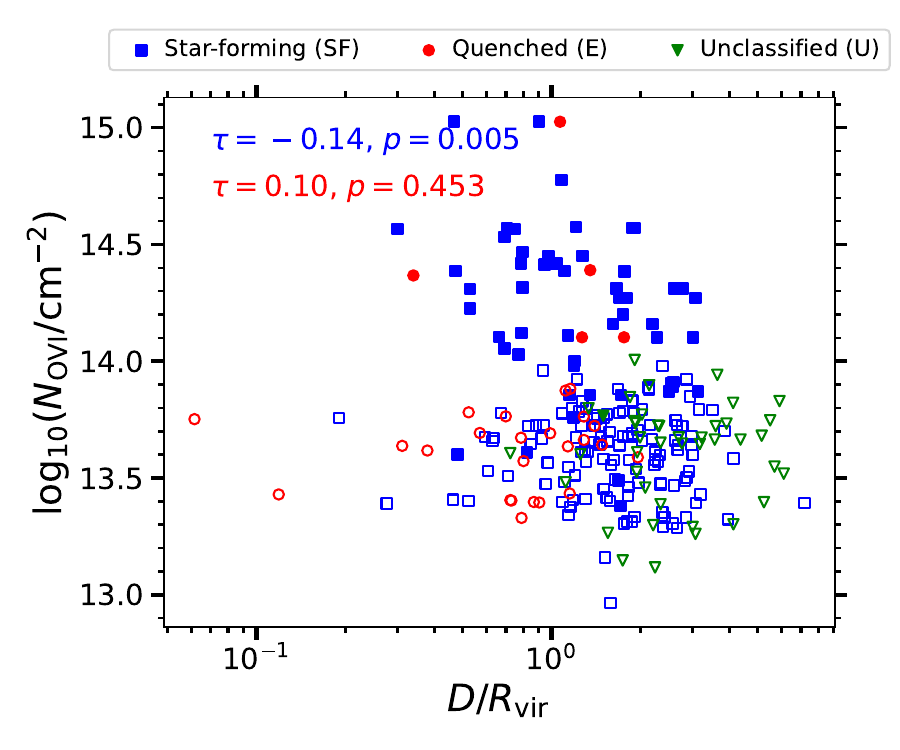}%
\includegraphics[width=0.5\linewidth]{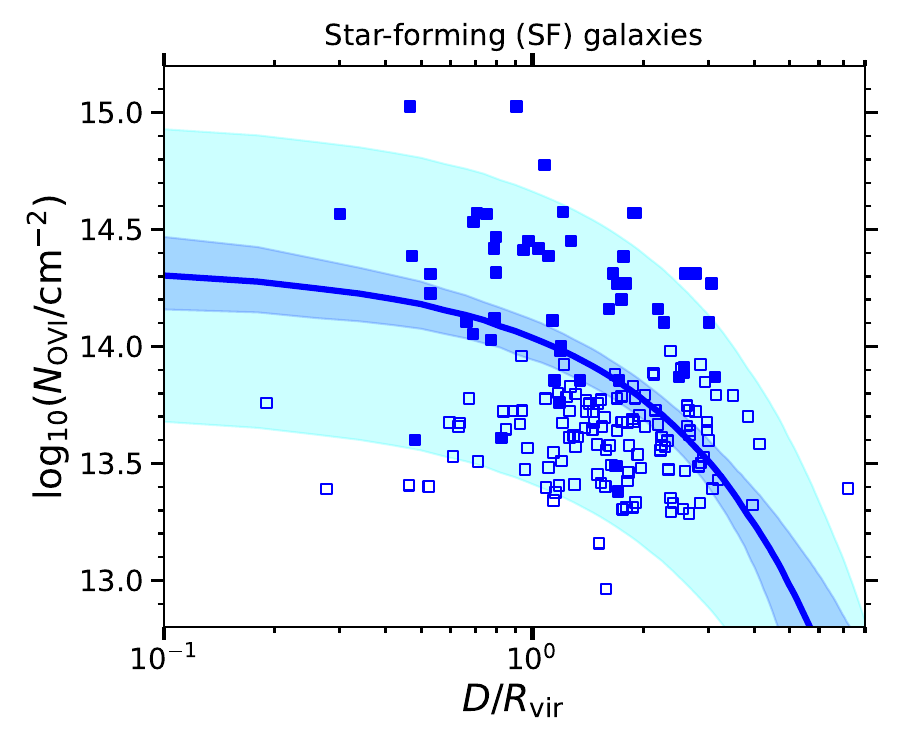}
    \caption{{\tt Left}: $N(\OVI)$ plotted against $D/R_{\rm vir}$ for the galaxies in MUSEQuBES. The hollow points represent the 3$\sigma$ upper limits on $N(\OVI)$ for non-detections. The blue squares, red circles, and green triangles indicate star-forming (SF), quenched (E), and unclassified (U) galaxies, respectively. The results of the Kendall-$\tau$ correlation test, including the upper limits, between $N(\OVI)$ and $D/R_{\rm vir}$ for the `SF' and `E' galaxies are indicated inside the panel. A weak anti-correlation between $N(\OVI)$ and $D/R_{\rm vir}$ is observed for the `SF' galaxies. {\tt Right}: $N(\OVI)$ plotted against $D/R_{\rm vir}$ only for the `SF' galaxies in MUSEQuBES. The solid line represents the best-fit model median (see text). The blue-shaded region represents the 68\% confidence intervals on the best-fit model median. The cyan-shaded region indicates the intrinsic scatter.} 
    \label{fig:Novi-prof}
\end{figure*}

\subsection{Associating galaxies with absorbers}
\label{sec:gal-q pair}

For galaxy-absorber association, we cross-match our galaxy catalog with the blind \OVI\ absorber catalog with a linking velocity of $\pm300$~\kms\ centered on each galaxy redshift in a given field. The column densities of all absorption components within the velocity windows are summed to obtain the total column density associated with a galaxy.

Out of the 247 galaxies, 60 exhibit associated \OVI\ absorption with a median of 2 components per system\footnote{Note that a single system can be shared by multiple galaxies}. For the remaining 187 galaxies, we estimated the $3\sigma$ upper limits on \OVI\ column density over a velocity window of $\pm60$~\kms\ using the standard deviation of flux from regions devoid of any absorption. The velocity window used here is twice the typical $b$-parameter of \OVI\ absorbers \citep[see e.g.,][]{Werk_2013,Johnson_15}. The kinematics of the \OVI-bearing gas phase around the 60 galaxies with detected \OVI\ absorption is presented in a separate paper (Dutta et. al, submitted).

Imposing a Friends-of-Friend algorithm, we found that 124 out of the 247 galaxies do not have a neighboring galaxy within a line-of-sight (LOS) velocity of $\pm$500 \kms\ and 500 kpc transverse distance. These are referred to as `isolated galaxies'. The remaining 123 galaxies have one or more neighboring galaxies within the aforementioned cuts and are categorized as `group galaxies'. Among the 60 galaxies with detected \OVI\ absorption, 25 and 35 galaxies are classified as `isolated' and `group' galaxies, respectively. Note that, in our study, multiple `group' galaxies can share the absorption system.

\begin{table*}
\centering
\caption{Best-fit model parameters of $N(\OVI)$-profiles}   
\label{tab:best-fit-post-Novi}
\begin{tabular*}{\textwidth}{l@{\extracolsep{\fill}}ccccccr} 
\hline
 Parameter & \multicolumn{6}{c}{Best-fit value of the parameter for bins with \logm~ } &     \\ \cline{2-7} 
  &  $6.0-8.0$ & $8.0-8.6$ & $8.6-9.2$ & $9.2-9.8$ & $9.8-10.4$ & $10.4-11.2$ & MUSEQuBES$^b$  \\ \hline 
  \hline
  
${\rm log}_{10}N_0$ & $14.0_{-0.2}^{+0.2}$ & $13.9_{-0.1}^{+0.2}$ & $13.9_{-0.1}^{+0.1}$ &  $14.6_{-0.2}^{+0.3}$ &  $14.6_{-0.2}^{+0.3}$ & $14.3_{-0.2}^{+0.2}$ & $14.3_{-0.2}^{+0.2}$  \\ 

$\gamma$  & $1.0_{-0.5}^{+1.2}$ & $1.4_{-0.9}^{+1.6}$ & $3.1_{-0.8}^{+0.6}$ & $1.2_{-0.4}^{+0.5}$ & $1.1_{-0.4}^{+0.8}$ & $0.8_{-0.3}^{+0.8}$ & $1.0_{-0.3}^{+0.5}$  \\

$L_{\rm s}$ & $3.0_{-1.3}^{+0.7}$ & $3.2_{-1.3}^{+0.6}$ & $2.6_{-0.3}^{+0.2}$ & $1.2_{-0.6}^{+0.4}$ & $0.9_{-0.5}^{+0.5}$ & $1.6_{-0.8}^{+1.2}$ & $1.5_{-0.8}^{+0.7}$  \\ 

$\sigma_{\rm intr} $ & $1.3_{-0.3}^{+0.4}$ &  $1.3_{-0.3}^{+0.4}$ & $1.2_{-0.2}^{+0.3}$ & $1.2_{-0.2}^{+0.3}$ & $0.9_{-0.2}^{+0.3}$ & $1.4_{-0.4}^{+0.4}$ & $1.4_{-0.2}^{+0.3}$  \\ \hline 
${\rm N}_{\rm gal}^{a}$ & 40 & 76 & 104 & 111 & 58 & 40 & 176 \\ \hline
\end{tabular*}
\justify
\item Notes-- For each parameter, the best-fit estimate represents the median value of the corresponding posterior distributions. The quoted uncertainties are 68\% confidence intervals of the posterior probability distributions. $^a$The number of galaxies in different mass bins. $^b$ The best-fit parameters are for the full star-forming galaxy sample from MUSEQuBES only.  
\end{table*}

\section{Results} 
\label{sec:results}

In this section, we investigate the spatial distribution of \OVI\ bearing gas around galaxies through the column density profile and covering fraction profiles.

\subsection{\OVI\ column density profile}
\label{N_prof}

Probing the variation of total \OVI\ column density with impact parameter is the first step to mapping the \OVI\ distribution in and around galaxies. In the left panel of Fig.~\ref{fig:Novi-prof}, the total column density of \OVI\ is plotted against the normalized impact parameter ($D/R_{\rm vir}$) of the MUSEQuBES galaxies. Since our galaxies show a large dynamic range in stellar mass, it is essential to scale the impact parameter by $R_{\rm vir}$ to minimize scatter due to varying halo sizes. The different symbols with different colors indicate the different galaxy types, as indicated in the legend (see Sec.~\ref{sec:MUSEQuBES_galaxy_survey} for details).  The solid and open symbols denote the detection and $3\sigma$ upper limits on column densities, respectively.

To explore potential correlations between $N(\OVI)$ and $D/R_{\rm vir}$, we conduct a Kendall-$\tau$ survival analysis using the {\sc Nada} package in {\sc R}, including the upper limits. The Kendall-$\tau$ test reveals a weak anti-correlation between $N(\OVI)$ and $D/R_{\rm vir}$ for the star-forming (`SF') galaxies ($\tau=-0.14$, $p=0.005$). We note that several galaxies closely separated in redshift (residing in a `group') sharing a common \OVI\ absorption system increases the scatter in this relation. Consistent with \citet[]{Qu_2024}, we find a tighter anti-correlation for the star-forming galaxies when the smallest $D/R_{\rm vir}$ galaxy of a group is associated with the measured column density ($\tau=-0.24$, $p<0.001$). No significant correlation is seen for the quenched (`E') galaxies ($p=0.453$).

Prompted by the weak anti-correlation seen for the `SF' galaxies, we proceed to model the corresponding $N(\OVI)$-profile, adopting an empirical $N(\OVI)$ profile given by a modified exponential function: 
\begin{equation}
    \label{eq:novi_prof}
    N_{\rm O~VI} (D/R_{\rm vir}) = N_0~{\rm exp}\left[- \left(\frac{D/R_{\rm vir}}{L_{\rm s}}\right)^{\gamma}\right]~, 
\end{equation}
where $N_0$, $L_{\rm s}$, and $\gamma$ are the free parameters of the model, signifying the maximum $N(\OVI)$, the length scale of exponential decline, and the slope of the profile, respectively.

Here we used all the `SF' galaxies, but we verified that our results do not change appreciably when only the smallest $D/R_{\rm vir}$ galaxies are considered for groups.

The likelihood function for producing the observed dataset with $m$ detections, and $n$ non-detections with the model parameters $N_0,~L_s,~\gamma$ is given by, 

\begin{equation}
\begin{split}
    \mathcal{L} \propto \prod_{i=1}^{m} \frac{1}{\sqrt{2\pi \sigma^2}} {\rm exp}\left(-\frac{\left(N_i - N_{\rm O~VI}\right)^2} { 2\sigma^2 }
    \right) \\ 
    \times    
  \prod_{i=1}^{n} \left[ \left( \frac{1}{\sqrt{2\pi\sigma_p^2} } \int_{0}^{N_i^{{\rm uplim}}} {\rm exp} \left(- \frac{ (y-N_{\rm O~VI})^2}{2\sigma_p^2} \right) dy \right) \right]~, 
\end{split}
\end{equation}
where $\sigma^2=\sigma_i^2 + \sigma_p^2$, with $\sigma_p = N_{\rm O~VI}~\sigma_{\rm intr}$ denoting the intrinsic scatter, assuming a constant intrinsic scatter $\sigma_{\rm intr}$ in the logarithmic space and $\sigma_i$ denoting the uncertainty in individual measurements.

We use uniform priors to construct the posteriors of
individual model parameters using MCMC method, which is implemented using the {\sc emcee} package. The model median, alongside the 68\% confidence interval, is shown with the blue solid line and shaded region in the right panel of Fig.~\ref{fig:Novi-prof}.

We have also adopted a couple of other empirical models used in the literature to quantify the $N(\OVI)-$profile. A comparison between the different models is discussed in Appendix~\ref{sec:model_comp}. However, our main conclusions do not change appreciably when different models are adopted to fit the $N(\OVI)-$profile.

\begin{figure*}
    \centering
    \includegraphics[width=0.9\textwidth]{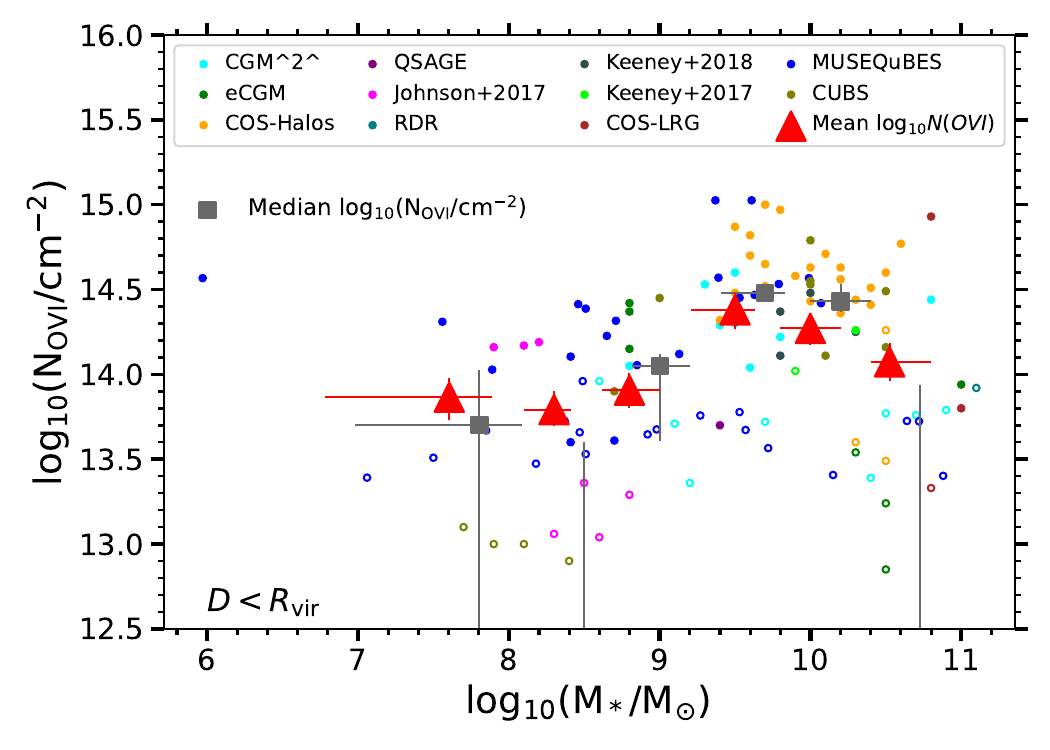}    
    \vskip-0.2cm 
    \caption{$N(\OVI)$ measured within the virial radius of the star-forming galaxies plotted against the stellar mass. The circles represent the combined galaxy sample (filled and open represent measurements and $3\sigma$ upper limits, respectively), and the color-coding indicates the survey. The average $N(\OVI)$ values (using Eq.~\ref{eq:avN}) for six stellar mass bins are shown with the red triangles. The gray squares indicate the median $N(\OVI)$ for these stellar mass bins using survival analysis. The x-error bars represent the 68\% confidence intervals of \logm\ in each bin. The y-error bars represent the 68\% confidence intervals of average ${\rm log}_{10}(N_{\rm OVI}/{\rm cm}^{-2})$ propagated from the 68\% confidence intervals of the best-fit parameters for the red triangles, and 68\% confidence intervals for the gray squares. Only upper limits on the median can be obtained for the second and last stellar mass bins since in these bins the number of non-detections is greater than the number of detections.}
    \label{fig:avgOVI_N}
\end{figure*}

This empirical fit to the $N(\OVI)$ measurements allows us to determine the average \OVI\ column density ($\left<N(\OVI)\right>$) within the virial radius;

 \begin{equation}
 \label{eq:avN}
     \left<\novie\right>= \int_0^1 \novie (x') 2 x' dx'~. 
 \end{equation}
Here the variable $x'$ is $D/R_{\rm vir}$. We use the $N(\OVI)$-profile given by Eq.~\ref{eq:novi_prof} with the best-fit parameters obtained with {\sc mcmc} sampling to calculate $\left<N(\OVI)\right>$.

The ${\rm log}_{10}\left<N(\OVI)/{\rm cm}^{-2}\right>$ for the full MUSEQuBES measurement is $14.14_{-0.10}^{+0.09}$. This is significantly lower than the reported average $N(\OVI)$ of $10^{14.6}~{\rm cm}^{-2}$ for $\approx L_*$ galaxies in \citet[]{Tumlinson_2011}. However, their measurements are limited to sightlines within $\approx0.6R_{\rm vir}$. The best-fit $N(\OVI)-$profile for the MUSEQuBES sample yields  ${\rm log}_{10}\left<N(\OVI)/{\rm cm}^{-2}\right>=14.2$ within $0.6R_{\rm vir}$, which is three times lower than for the $\approx L_*$ galaxies.

It is worth mentioning that although the median \logm\ is 8.7 for our sample, galaxies with a large dynamic range of stellar mass contribute (68\% range of \logm\ is 7.9--9.4), which can introduce additional scatter in the $N(\OVI)$-profile. Disentangling the role of stellar mass on the $N(\OVI)$-profile is thus necessary.

\subsubsection{The effect of stellar mass on the $N(\OVI)$-profile for the star forming galaxies} 
\label{sec:4.1.1}

Expressing the \OVI\ column density profile in terms of $D/R_{\rm vir}$ does not entirely eliminate the effects of varying stellar masses (halo sizes). The $N(\OVI)$ at a given $D/R_{\rm vir}$ may depend on the stellar mass due to other physical factors. Since the 176 `SF' galaxies in MUSEQuBES are predominantly low-mass (median [68\% range]  \logm~$=8.7~ [7.9 - 9.4]$), we combine MUSEQuBES with additional 253 `SF' galaxies from the literature (median [68\% range] \logm~$=9.5~ [8.6 - 10.3]$); see Section.~\ref{sec:gal_lit}) to robustly probe the $M_{\star}$ dependence of the $N(\OVI)$-profile.

 We generate the model $N(\OVI)$ profile for different mass bins to quantify the role of stellar mass on the $N(\OVI)-$profiles. 
We divide the 429 (253+176) galaxies into six mass bins. The lowest stellar mass bin comprises galaxies with \logm~$<8$. The galaxies with \logm~$=8-10.4$ are divided into 4 bins of 0.6 dex. The highest stellar mass bin consists of galaxies with \logm~$>10.4$. The median \logm\ of the 6 stellar mass bins are $7.6, 8.3, 8.8, 9.5, 10.0$ and $10.5$. The numbers of galaxies in each bin are $40, 76, 104, 111, 58$, and $40$, respectively. The best-fit posterior values of the model parameters along with their confidence intervals for different mass bins are summarized in Table~\ref{tab:best-fit-post-Novi}.

For each bin, we calculated the $\left<N(\OVI)\right>$ within the virial radius as discussed earlier\footnote{\citet[]{Tchernyshyov_2022} did not report the error in $N(\OVI)$ (i.e., $\sigma_i$) for the 197 `SF' galaxies. We used a uniform $\sigma_i$ of 0.1 dex for these galaxies. We verified that changing the $\sigma_i$ from 0.1 to 0.3 dex does not alter our conclusions}. The $\left<N(\OVI)\right>$ are plotted against the stellar mass in Fig.~\ref{fig:avgOVI_N} with red triangles. The x-error bar represents the 68 percentile range of $M_{\star}$ in each bin. The y-error bar represents the 68\% confidence interval of $\left<N(\OVI)\right>$, which is derived by propagating the 68\% confidence intervals of the posterior distributions. The $\left<N(\OVI)\right>$ shows a peak at \logm~$\approx 9.5$ and declines at both higher and lower mass ends (see also Table~\ref{tab:novi_table}). Nonetheless, the halos of low-mass (\logm~$<9$) galaxies exhibit an appreciable amount of \OVI\ with $N(\OVI)>10^{13.5}$~\sqcm. Note that this is the first time the  $\left<N(\OVI)\right>$ is measured for galaxies with \logm~$\lesssim 8$ with a statistically meaningful sample. A similar trend of $\left<N(\OVI)\right>$ with stellar mass is also shown in \citet[]{Zahedy_2019}. However, while for their sample passive luminous red galaxies (LRGs) dominate the high-mass bin, we observe the suppression of $\left<N(\OVI)\right>$ even for the high-mass, star-forming galaxies.

\begin{table*}
\caption{Summary of $N(\OVI)$ measurements and Kendall-$\tau$ test results.} 
\centering 
\label{tab:novi_table}
\begin{tabular*}{\textwidth}{l@{\extracolsep{\fill}}cccr} 
\hline 
\logm &  ${\rm log}_{10}~\left<N(\OVI)\right>/{\rm cm^{-2}}$  & ${\rm log}_{10}~\overline{N(\OVI)/}{\rm cm^{-2}}$   &  $\tau$ & $p$   \\
(1)   &    (2)        & (3)         & (4)       & (5)     \\ \hline      
$6.0-8.0$   & $13.86^{+0.11}_{-0.14}$  &  $13.7^{+0.3}_{-13.7}$    & $-0.18$ & $0.084$   \\ 
$8.0-8.6$   & $13.79^{+0.11}_{-0.10}$  &  $<13.6$                  & $-0.14$ & $0.069$   \\ 
$8.6-9.2$   & $13.91^{+0.10}_{-0.10}$  &  $14.0^{+0.1}_{-0.4}$     & $-0.20$ & $0.002$   \\ 
$9.2-9.8$   & $14.38^{+0.10}_{-0.10}$  &  $14.5^{+0.1}_{-0.1}$     & $-0.32$ & $<0.001$   \\ 
$9.8-10.4$  & $14.27^{+0.09}_{-0.10}$  &  $14.4^{+0.1}_{-0.1}$     & $-0.50$ & $<0.001$      \\ 
$10.4-11.2$ & $14.07^{+0.11}_{-0.11}$  &  $<13.9$                  & $-0.23$ & $0.027$  \\  \hline
\end{tabular*}
\justify  
Notes-- (1) The stellar mass range of the bin (2) The profile-averaged $N(\OVI)$ calculated within  $R_{\rm vir}$ (3) The median $N(\OVI)$ directly calculated using {\tt Lifelines} (4) The Kentall-$\tau$ parameter for the trend of $N(\OVI)$ with $D/R_{\rm vir}$ (5) The corresponding probability. 
\end{table*}

The filled and hollow circles in Fig.~\ref{fig:avgOVI_N} represent individual \OVI\ detection and $3\sigma$ upper limits, respectively, within the $R_{\rm vir}$ for all the galaxies we compiled here. The median values of these measurements are shown by the gray squares. The median and 68\% confidence intervals are obtained with the {\tt Lifelines} package of {\sc Python}, including the upper limits\footnote{For two bins, the package failed to return a non-zero median value due to the excessive upper limits. We only showed the 68 percentile range for the two bins.}. These directly measured median values are consistent with our model-averaged  $N(\OVI)$. The model-averaged and median $N(\OVI)$ along with the results of the generalized Kendall-$\tau$ correlation test between $N(\OVI)$ and $D/R_{\rm vir}$ in each stellar mass bin are summarized in Table~\ref{tab:novi_table}. The strongest anti-correlation is observed for the bins with \logm~$=9.2-9.8$ and \logm~$=9.8-10.4$, where the $\left<N(\OVI)\right>$ peaks.

Finally, we modeled the $N(\OVI)$ measurements for 120 passive galaxies from the literature and MUSEQuBES using the procedure described above, despite the lack of anticorrelation of $N(\OVI)$ with $D/R_{\rm vir}$ for these galaxies (see Appendix~\ref{sec:novi_sf_e}). The model-averaged and median $N(\OVI)$ within $R_{\rm vir}$ for the passive galaxy sample is significantly lower than that of the star-forming galaxies when the stellar mass is controlled (Fig.~\ref{fig:sf-q_comp}).

\subsection{\OVI\ covering fraction ($\kappa$) profile } 
\label{fc_prof}

The covering fraction ($\kappa$) is the fraction of the sightlines passing through the CGM with measured column density above a certain threshold. This is a measure of the patchiness of the absorbing medium in the projected 2D plane perpendicular to the LOS. In our analysis of the covering fraction, we do not consider the galaxy-absorption pairs for which the $3\sigma$ limiting column density is higher than the threshold column density. As a result,  $0\%$ and $13\%$ of the data points are excluded for the threshold $N(\OVI)=10^{14}~{\rm cm}^{-2}$ and $10^{13.75}~{\rm cm}^{-2}$, respectively.

\begin{figure*}
    \centering
    \includegraphics[width=0.5\linewidth]{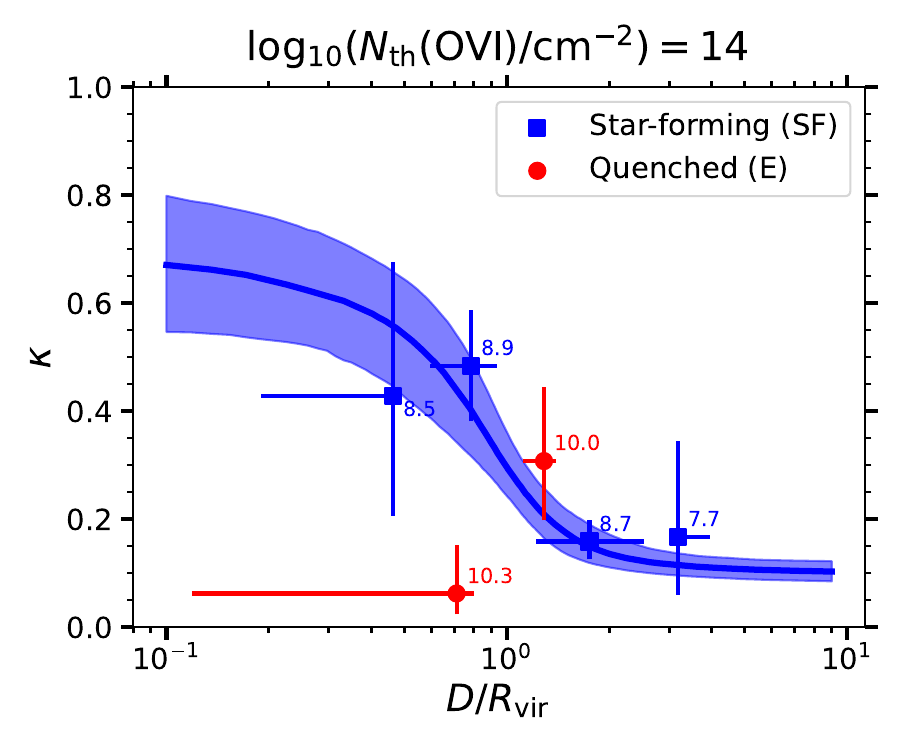}%
    \includegraphics[width=0.5\linewidth]{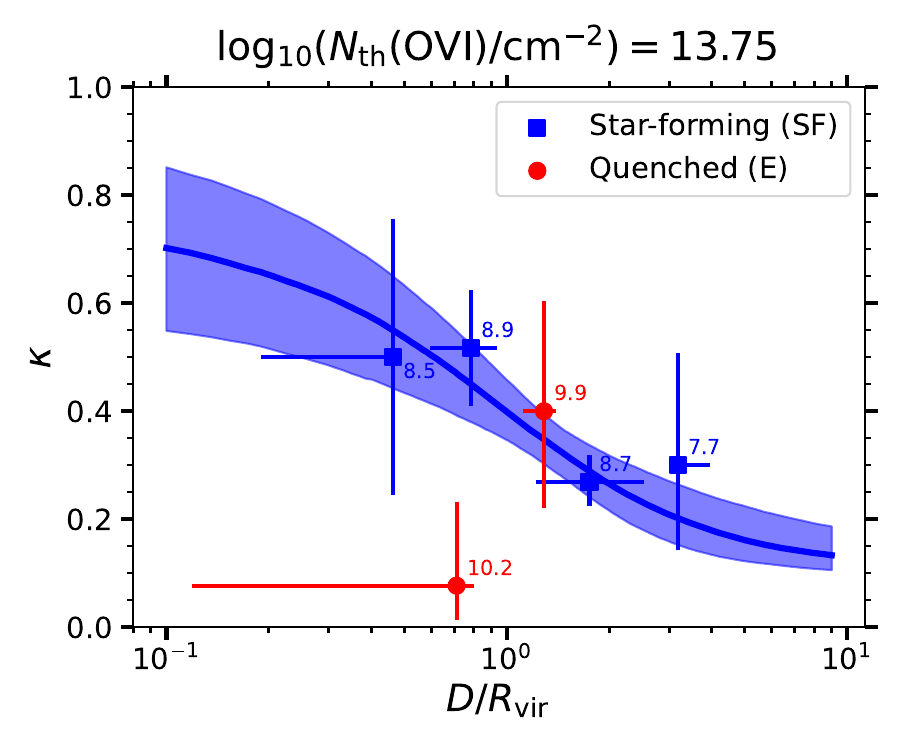} 
        \caption{ {\tt Left}: \OVI\ covering fraction ($\kappa$) plotted against $D/R_{\rm vir}$ for the MUSEQuBES galaxies for a threshold \lognovi~$ =14.0$. The blue squares and red circles indicate $\kappa$ for star-forming and quenched galaxies, respectively. The $D/R_{\rm vir}$ is split into 4 bins with 0.5, 1.0, and 3.0 as the separating value to compute the covering fraction for the star-forming galaxies. The quenched galaxies are split into two bins with $D/R_{\rm vir}=1$ as the bin boundary. The y-error bar represents the 68\% Wilson-score confidence interval. The solid blue line indicates the best-fit model to the unbinned data for star-forming galaxies (see text). The blue shaded regions represent the 68\% confidence interval on the best-fit model. The numbers next to the data points represent the median \logm\ of the galaxies in each bin. {\tt Right:} Same as {\tt left} but for a threshold \lognovi~$=13.75$.}
    \label{fig:fc-prof}
\end{figure*}

The $\kappa$ of `SF' and `E' galaxy populations of the MUSEQuBES sample are plotted against $D/R_{\rm vir}$ in Fig.~\ref{fig:fc-prof} with blue squares and red circles, respectively. The left and right panels show the $\kappa$-profile for a threshold ${\rm log}_{10}(N(\OVI)/{\rm cm}^{-2})$ of 14.0 and 13.75, respectively. The `SF' galaxies are split into four $D/R_{\rm vir}$ bins, with the bin boundaries being 0.5, 1.0, and 3.0. There are 7, 31, 126, and 12 star-forming galaxies in the four $D/R_{\rm vir}$ bins, respectively. The quenched galaxies are split into two $D/R_{\rm vir}$ bins separated by $D/R_{\rm vir}=1$, with 16 (13) galaxies with $D < R_{\rm vir}$ ($D>R_{\rm vir}$). The x and y error bars represent the 68\% interval of the $D/R_{\rm vir}$ distribution and the 68\% Wilson-score interval on $\kappa$ in each bin, respectively.

As an alternate method to obtain the $\kappa$-profile, we used the novel approach introduced by \citet[]{Schroetter_2021} that does not require binning of the data points in the impact parameter. Briefly, we assume that the detection probability of a circumgalactic \OVI\ absorber above a threshold column density follows a logistic function. The conventional definition of the logistic function requires a smooth transition from 0 to 1. We slightly modified the logistic function such that it produces a smooth transition between $\kappa_1$ and  $\kappa_0 + \kappa_1$, where $\kappa_1$ and $\kappa_0$ are free parameters. We used  
\begin{equation}
\label{eq:fc}
    p[Y=1]=\frac{\kappa_0}{1+e^{-t}} + \kappa_1~, 
\end{equation} 
where $p[Y=1]$ represents the probability of a detection (implied by $Y=1$) above the threshold $N(\OVI)$, essentially the $\kappa$. Following \citet[]{Schroetter_2021}, the parameter $t$ is taken to be a function of the independent the variable $D/R_{\rm vir}$. We  adopted 
\begin{equation}
\label{eq:fc_tparam}
    t=\alpha({\rm log}_{10}(D/R_{\rm vir})-\beta)~, 
\end{equation}
where $\alpha$ and $\beta$ are two free parameters. $\alpha$ describes the slope of the covering fraction profiles, and $\beta$ represents ${\rm log}_{10}(D/R_{\rm vir})$ where $\kappa$ (or $p[Y=1]$) $=\kappa_0/2 + \kappa_1 (\equiv \kappa_{50})$, analogous to the zero point at 50 percent covering fraction.

The parameter space is sampled to generate the dichotomous observable (1 if detected and 0 otherwise) based on the Bernoulli distribution. We use the {\sc pymc3} for the {\sc mcmc} sampling in order to obtain the best-fit parameters. The best-fit modeled $\kappa$ on the unbinned data for the star-forming sample is shown with the blue solid line in the left and right panel of Fig.~\ref{fig:fc-prof} for ${\rm log}_{10}(N(\OVI)/{\rm cm}^{-2})$ threshold of 14 and 13.75, respectively. The best-fit parameters are given in Table~\ref{tab:best-fit-post_k}. 
The blue-shaded region represents the  68\% confidence interval on the best-fit model.
It is evident that the logistic model reproduces the binned $\kappa$ profile reliably. A similar analysis could not be done for the `E' galaxies owing to the small numbers.

The best-fit model in the left panel of Fig.~\ref{fig:fc-prof} reveals a smooth transition of $\kappa$ from $\approx70$\% at $\approx 0.1D/R_{\rm vir}$ for star-forming galaxies to a modest value of $\approx30$\% at the virial radius. Similar to the $N(\OVI)$-profile, the modeled $\kappa$-profile enables us to measure the average covering fraction ($\left<\kappa\right>$) within the virial radius as: 
\begin{equation}
    \left<\kappa\right> = \int_0^{1} \kappa(x') 2x' dx'~. 
\end{equation}
We obtain $\left<\kappa\right>=0.46_{-0.08}^{+0.08}$ for a threshold $N(\OVI)=10^{14}~{\rm cm}^{-2}$ and $0.50_{-0.08}^{+0.07}$ for a threshold $N(\OVI)=10^{13.75}~{\rm cm}^{-2}$. Similar to $\left< N(\OVI) \right>$, these numbers are substantially lower compared to $\kappa=0.8-1$ for \OVI\ absorption around $\approx0.6R_{\rm vir}$ of $\approx L_*$ galaxies observed in the COS-Halos survey.  

 \begin{figure*}
     \centering
     \includegraphics[width=1.0\linewidth]{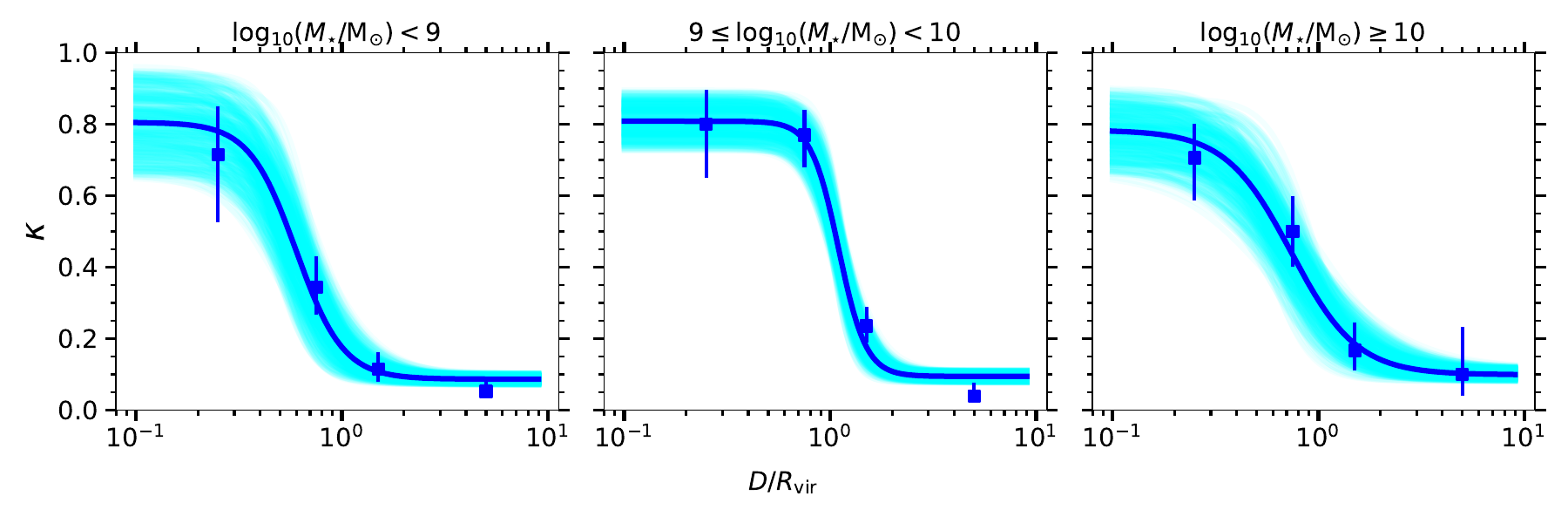} 
     \caption{The covering fraction profile for a threshold \lognovi~$=14$  plotted against the $D/R_{\rm vir}$ for three mass bins with \logm~$<9$ ({\tt left}), $\leq 9$ \logm~$<10$ ({\tt middle}), and \logm~$\geq10$ ({\tt right}) for the combined `SF' galaxy sample (MUSEQuBES+literature). The median \logm\ for the bins are 8.4, 9.5, and 10.4. The blue squares and y-error bars represent the binned $\kappa$ and 68\% confidence intervals.  The blue solid line and the shaded region in cyan represent the model $\kappa$ along with the 68\% confidence intervals.}
     \label{fig:fc_m_prof}
 \end{figure*}

Fig.~\ref{fig:fc-prof} shows that the star-forming galaxies exhibit a significantly enhanced $\kappa$ for both the column density thresholds within $R_{\rm vir}$. However, no significant difference in $\kappa$ between the `SF' and `E' galaxies is seen beyond the virial radius. Similar to our findings, a bimodality in the \OVI\ covering fraction was also observed in the COS-Halos survey between star-forming and passive galaxies. However, we point out that the median stellar mass of the `SF' galaxies is more than an order of magnitude lower than that of the `E' galaxies. The median $M_{\star}$ for each $D/R_{\rm vir}$ bin is indicated next to the data points in Fig.~\ref{fig:fc-prof}. It is therefore important to control the $M_{\star}$ for the `SF' and `E' to disentangle the role of star formation in the \OVI\ covering fraction.

Recently, \citet[]{Tchernyshyov_2023} showed that the bimodality of \OVI\ absorption persists inside the virial radius even when controlled for stellar mass and halo mass. The lack of high-mass, star-forming galaxies in MUSEQuBES prevents us from carrying out a similar analysis robustly. The majority of the passive galaxies in our sample have \logm~$>10$. None of the passive galaxies with \logm~$>10$ show any detectable \OVI\ absorption within $R_{\rm vir}$. However, out of 5 star-forming galaxies with \logm$>10$, only 1 (20\%) show \OVI\ absorption within $R_{\rm vir}$ and it has ${\rm log}_{10}(N(\OVI)/{\rm cm}^{-2})=14.4$.

\begin{table}
\centering
\caption{Best-fit model parameters for the $\kappa$-profiles}   
\label{tab:best-fit-post_k}
\begin{tabular}{lcccr} 
\hline

 Threshold  & \multicolumn{4}{c}{Best-fit value of the parameter}    \\ \cline{2-5} 
 \lognovi &  $\alpha$ & $\beta$ & $\kappa_0$ & $\kappa_1$  \\ \hline 
  \hline
 13.75 & $-3.1_{-2.3}^{+1.4}$ & $-0.08_{-0.3}^{+0.2}$ & $0.66_{-0.2}^{+0.2}$ & $0.1_{-0.02}^{+0.02}$ \\ 
 14.0 & $-7.9_{-4.0}^{+2.3}$ & $-0.1_{-0.1}^{+0.1}$ & $0.58_{-0.1}^{+0.1}$ & $0.1_{-0.02}^{+0.02}$ \\ \hline 
\end{tabular}
\justify  
\item Notes-- For each parameter, the best-fit estimate represents the median value of the corresponding posterior distributions. The quoted uncertainties are 68\% confidence intervals of the posterior probability distributions. 
\end{table}

\begin{table}
\centering
\caption{Best-fit model parameters for the $\kappa$-profiles for a threshold \lognovi~$=14$ for different $M_{\star}$ bins}   
\label{tab:best-fit-post_k_mbin}
\begin{tabular}{lcccr} 
\hline

 \logm  & \multicolumn{4}{c}{Best-fit value of the parameter}    \\ \cline{2-5} 
  &  $\alpha$ & $\beta$ & $\kappa_0$ & $\kappa_1$  \\ \hline 
  \hline
 $<9$ & $-8.6_{-2.8}^{+1.9}$ & $-0.23_{-0.06}^{+0.05}$ & $0.71_{-0.09}^{+0.10}$ & $0.09_{-0.01}^{+0.01}$ \\ 
 $9-10$ & $-15.0_{-3.3}^{+3.0}$ & $0.04_{-0.02}^{+0.02}$ & $0.72_{-0.04}^{+0.04}$ & $0.09_{-0.01}^{+0.01}$ \\  
 $\geq 10$ & $-6.4_{-2.1}^{+1.8}$ & $-0.13_{-0.07}^{+0.09}$ & $0.68_{-0.07}^{+0.08}$ & $0.10_{-0.01}^{+0.01}$ \\   
 \hline 
\end{tabular}
\justify  
\item Notes-- For each parameter, the best-fit estimate represents the median value of the corresponding posterior distributions. The quoted uncertainties are 68\% confidence intervals of the posterior probability distributions. 
\end{table}

\subsubsection{The effect of stellar mass on the $\kappa$-profile for the star-forming galaxies}

In order to investigate the role of stellar mass on the covering fraction of star-forming galaxies, we used the logistic regression method introduced earlier to obtain the $\kappa$-profile for multiple stellar mass bins. For this exercise, we included galaxies from the literature as well (see Section~\ref{sec:gal_lit}). The $M_{\star}$ range of the star-forming galaxies from the combined dataset is split into 3 bins with \logm $<9$, $9\leq$ \logm $<10$, and \logm $\geq$ 10. There are 197, 155, and 77 galaxies in the three bins.  We use the logistic model described in Section~\ref{fc_prof}, allowing the parameters $\kappa_0$, $\kappa_1$, $\alpha$, and $\beta$ to vary across the $M_{\star}$ bins. The best-fit parameters for the three mass bins are tabulated in Table \ref{tab:best-fit-post_k_mbin}. 

\begin{figure}
    \centering
    \includegraphics[width=1.0\linewidth]{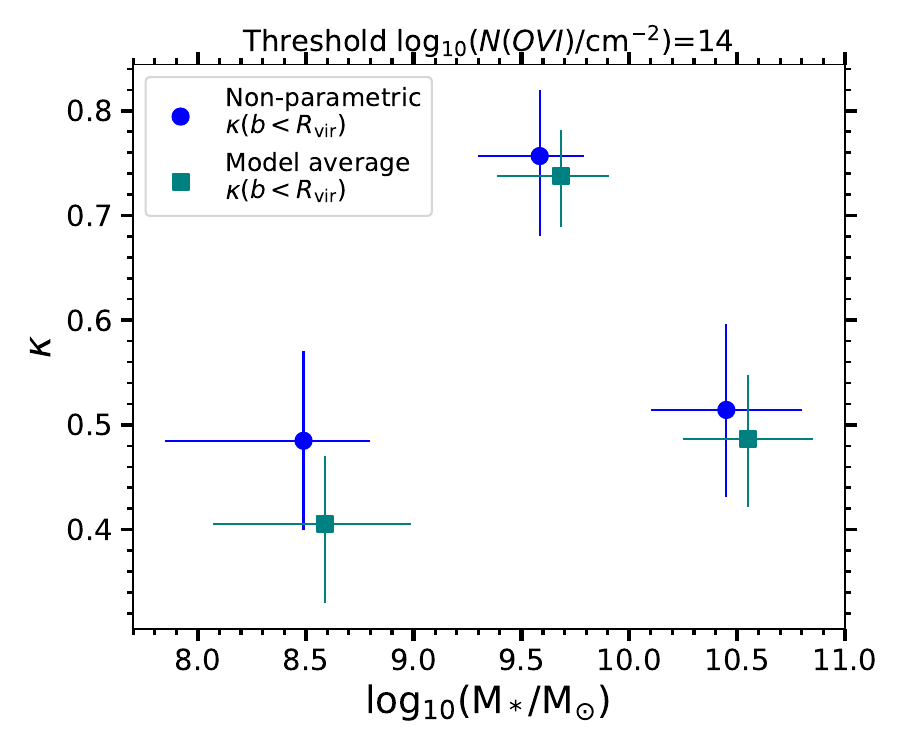}
    \caption{The covering fraction ($\kappa$) measured within $R_{\rm vir}$ for a threshold \lognovi~$=14$ plotted against the median $M_{\star}$ for the combined `SF' galaxy sample. The teal squares represent the mean $\kappa$ obtained using the best-fit logistic models for each stellar mass bin (Eq.~\ref{eq:fc_tparam}). The blue circles represent directly measured $\kappa$ values for galaxies with $D/R_{\rm vir}< 1$. The x- and y-error bars are similar to Fig.~\ref{fig:avgOVI_N}. The slight offset between the green and blue points along the x-axis is for clarity.} 
    \label{fig:avFc_M}
\end{figure}

\begin{figure*}
    \centering
    \includegraphics[width=0.5\linewidth]{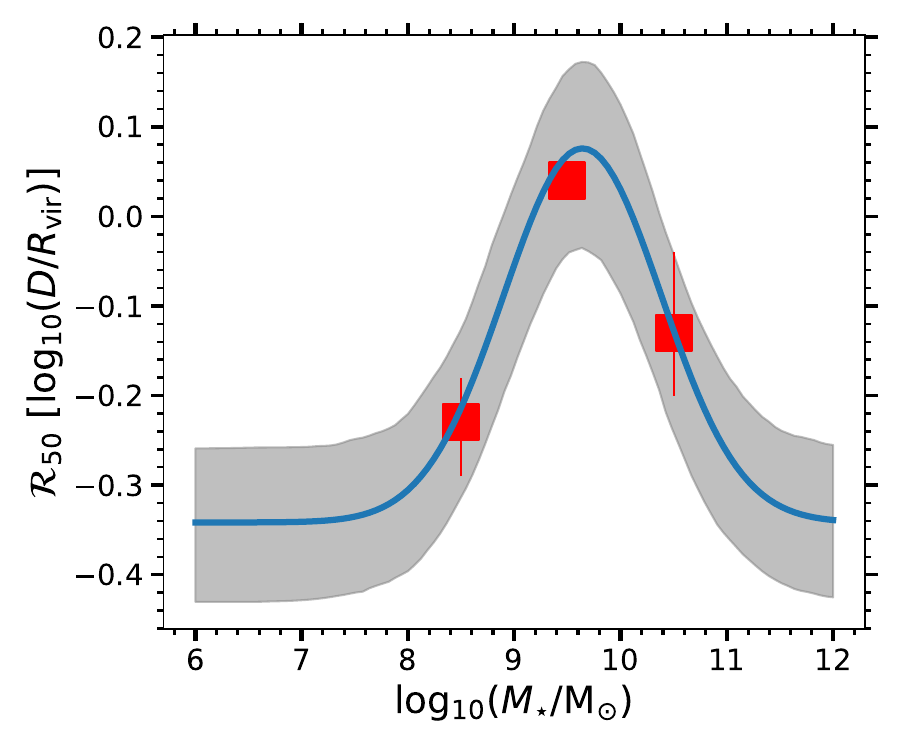}%
    \includegraphics[width=0.5\linewidth]{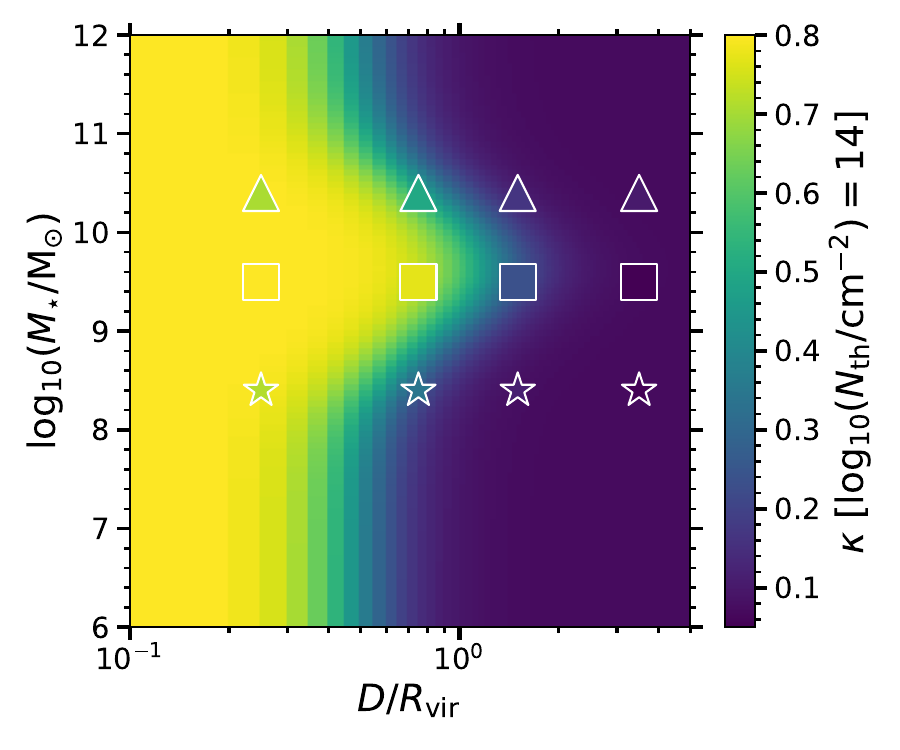}
    \caption{{\tt Left:} The best-fit $\mathcal{R}_{50}$ (see Eq.~\ref{eq:R50}) plotted against the stellar mass with blue solid line. The grey-shaded region represents the 68\% confidence interval on the best-fit model. The red squares represent the $\mathcal{R}_{50}$ (identical to $\beta$) measured for three individual \logm\ bins.
    {\tt Right}: The best-fit $\kappa$ color-map as a function of $D/R_{\rm vir}$ along the x-axis and \logm\ along the y-axis. The star symbols, squares, and triangles are binned $\kappa$ in 4 $D/R_{\rm vir}$ and 3 \logm\ bins. The model reproduces the binned measurements except for the lowest $D/R_{\rm vir}$, where the difference is within 1$\sigma$.}  
    \label{fig:kappa50}
\end{figure*}

In Fig.~\ref{fig:fc_m_prof}, we show the median modeled $\kappa$-profile for the unbinned data (for a threshold $N(\OVI)=10^{14}~{\rm cm}^{-2}$) for the three mass bins with solid blue lines in the three panels. The cyan narrow lines indicate 500 randomly drawn models within 68\% confidence intervals of the posterior distributions. 
We also show the binned covering fractions and 68\% Wilson-score intervals with solid blue squares and error bars.

The $\kappa$-profiles across different $M_{\star}$ bins show the steepest slope for the intermediate mass bin, while the slopes are significantly shallower for both the higher and lower mass bins (see Table \ref{tab:best-fit-post_k_mbin}). Additionally, similar to the $N(\OVI)$, a significant enhancement in the $\kappa$ within the $R_{\rm vir}$ is observed for the intermediate mass bin. The median stellar mass of this bin (\logm$\approx9.5$) is consistent with halos with the virial temperature of $\approx10^{5.5}$ K at which the ion-fraction of \OVI\ peaks in the CIE. 

The $\left<\kappa\right>$ within $R_{\rm vir}$ for galaxies in the three mass bins are obtained using two methods. First, $\left<\kappa\right>$ is obtained by averaging over the best-fit logistic models in each stellar mass bin. These are shown by the teal squares in Fig.~\ref{fig:avFc_M}. The 68\% confidence interval of $\left<\kappa\right>$ is derived from the 68\% confidence intervals of the posterior distributions and shown as the y-error bars. We also show the non-parametric $\kappa$ determined directly using the $N(\OVI)$ measurements for each stellar mass bin for galaxies with $D/R_{\rm vir}<1$. These measurements are shown by the blue circles in Fig.~\ref{fig:avFc_M}. The corresponding y-error bars represent 68\% Wilson score intervals. The mean and median $\kappa$ within $R_{\rm vir}$ are consistent within $1\sigma$ uncertainty.

The parameter $\beta$ is the value of ${\rm log}_{10}(D/R_{\rm vir})$ at $\kappa_{50}$.
Interestingly, the best-fit $\beta\approx0.04$ for the intermediate mass bin suggests a length scale of $\approx1.1R_{\rm vir}$, which is significantly larger compared to the values of $\approx0.6R_{\rm vir}$ and $\approx0.7R_{\rm vir}$ for the lower and higher \logm\ bins, respectively (see Table \ref{tab:best-fit-post_k_mbin}). 
Eq.~\ref{eq:fc_tparam} can be modified to include the variation of $\beta$ with \logm. Guided by the peak of $\beta$ in the intermediate mass bin, we used a simple Gaussian function to modify  Eq.~\ref{eq:fc_tparam} as follows:
\begin{equation}
    \label{eq:fc_m_unbin}
    t=\alpha\left[{\rm log}_{10}(D/R_{\rm vir})-(\Tilde{\beta}~e^{-\frac{\left(m - m_0\right)^2} { 2\sigma_m^2}}- \epsilon)\right]~,
\end{equation}
where $m$ represents \logm, and $m_0,~\Tilde{\beta},$ and $\epsilon$ are three additional free parameters in the fit to account for the variation with \logm. We do not impose an explicit mass dependence on $\kappa_0,~\kappa_1$ or $\alpha$.

 In the left panel of Fig.~\ref{fig:kappa50}, the stellar mass-dependent-$\beta$, denoted by $\mathcal{R}_{50}$, is shown as a function of stellar mass with the blue solid line. $\mathcal{R}_{50}$ is expressed in units of $\log_{10} D/\rm R_{\rm vir}$ and is given by 
\begin{equation}
\label{eq:R50}
 \mathcal{R}_{50} = (\Tilde{\beta}~e^{-\frac{\left(m - m_0\right)^2} { 2\sigma_m^2}}- \epsilon)~. 
\end{equation} 
The grey-shaded region represents the 68\% confidence interval on the best-fit model. The red squares represent the $\mathcal{R}_{50}$ (identical to $\beta$) measured for three individual \logm\ bins (see Table~\ref{tab:best-fit-post_k_mbin}). The best-fit $\mathcal{R}_{50}$ peaks at \logm$\approx9.5$, consistent with the peak observed in  $\left<\kappa\right>$. In the right panel of Fig.~\ref{fig:kappa50}, the modeled $\kappa$ following Eq.~\ref{eq:fc_m_unbin} for a threshold $N(\OVI)$ of $10^{14}~{\rm cm}^{-2}$ is color-coded as a function of $D/R_{\rm vir}$ along the x-axis and \logm\ along the y-axis. The individual star symbols, squares, and triangles represent the binned $\kappa$ with the same threshold for three stellar mass bins and four $D/R_{\rm vir}$ bins, adopting the same color convention. Our model adequately reproduces the binned $\kappa$ within the 1$\sigma$ uncertainty.

We have verified that the $\kappa-$profiles do not change appreciably with redshift, at least within the redshift range of our interest.

\section{Discussion}
\label{sec:discussion}

We have studied the properties of \OVI-bearing gas in the CGM of low-redshift ($0.1-0.7$) galaxies. In particular, we investigated the connection between host galaxy properties and the strength and incidence rate of \OVI\ absorption. In this section, we discuss the main findings from Section~\ref{sec:results}.

\begin{figure*}
    \centering
    \includegraphics[width=0.5\linewidth]{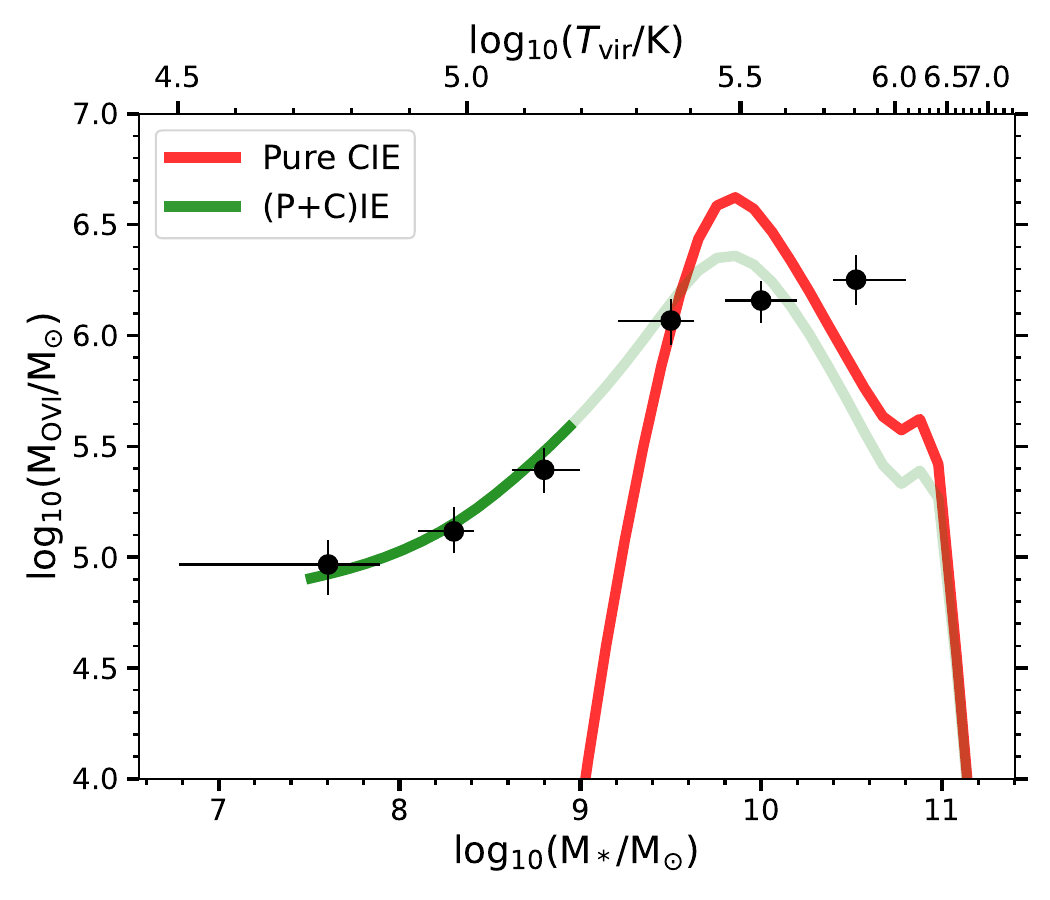}%
    \includegraphics[width=0.5\linewidth]{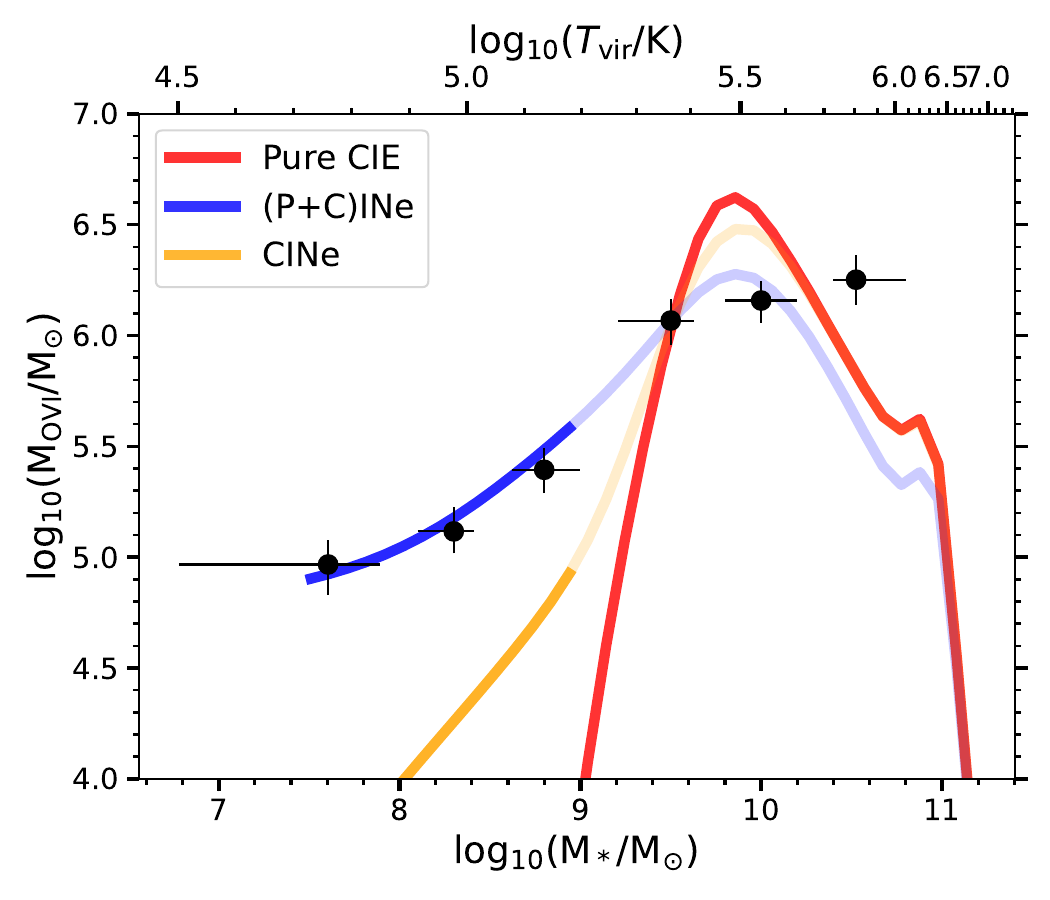}
        
    \caption{ {\tt Left:} Measured average \OVI\ mass within the virial radius of star-forming galaxies, plotted against the stellar mass with black solid circles. The top x-axis indicates the virial temperature of the corresponding halo assuming $z=0.4$. The binning is described in section 4.1. The x-error bars are 68\% confidence intervals in a given mass bin. The y-error bars are 68\% confidence interval on $M(\OVI)$ propagated from the posterior 68\% confidence interval of the parameter space. The red stripe shows the predicted $M(\OVI)$ for collisionally ionized ambient hot halo with $f_{\rm CGM}=0.15$ and metalicity 0.3$Z_{\odot}$ in CIE. The green solid stripe shows $M(\OVI)$ predicted from the same ambient hot halo in the presence of a photoionizing EGB (extragalactic background) assuming $n_{\rm H}=10^{-4.5}~{\rm cm}^{-3}$. 
    {\tt Right}: The data points and the red stripe are the same as the left panel. The blue solid stripe shows the predicted $M(\OVI)$ from the non-equilibrium cooling of hot outflows with 0.3$Z_{\odot}$ and $f_{\rm CGM}=0.15$ in the presence of a photoionizing EGB assuming $n_{\rm H}=10^{-4.5}~{\rm cm}^{-3}$. The orange stripe shows the same but without the presence of the EGB. } 
    \label{fig:movi_mod}
\end{figure*}

\subsection{{\rm O~{\sc vi}} mass in the CGM as a function of stellar mass}  
\label{sec:5.1}

The moderate anti-correlation seen between the $N(\OVI)$ and $D/R_{\rm vir}$ for the  `SF' galaxies in our sample indicates the galactic origin of the \OVI\ absorption. Following \citet[]{Tchernyshyov_2022}, we determined the average \OVI\ column density ($\langle N(\OVI)\rangle$) within the virial radius of `SF' galaxies with a range of stellar masses (Fig.~\ref{fig:avgOVI_N}). The addition of 176 predominantly low-mass galaxies from MUSEQuBES improved the $\langle N(\OVI)\rangle$ measurements significantly, particularly at \logm~$<9$. In fact, $\langle N(\OVI)\rangle$ is for the first time well-measured at \logm~$<8.5$ .

The average \OVI\ mass in the CGM ($<R_{\rm vir}$) of galaxies can be obtained using: 
\begin{equation}
\label{eq:avMovi}
    M(\OVI) = \langle N(\OVI) \rangle \times~ m_{\rm O} \times ~\pi R_{\rm vir}^2~, 
\end{equation}
where $m_{\rm O}$ is the mass of an oxygen atom. In Fig.~\ref{fig:movi_mod}, the derived $M(\OVI)$ is plotted against $M_{\star}$ with black solid circles. Overall, $M(\OVI)$ increases with the stellar mass by an order of magnitude from \logm~$\approx7.5-9.5$. At higher stellar masses probed by our sample, the $M(\OVI)$ remains flat at $\approx10^{6.3}$~\Msun.

As expected, our $M(\OVI)$ measurements are broadly consistent with \citet{Tchernyshyov_2022}, since their data is a subset of our combined sample. However, with the increased sample size, we offer significantly improved measurements, especially for low-mass galaxies. For example, by increasing the number of $N(\OVI)$ measurements within the virial radius by a factor of 2, we are able to constrain the $\log_{10}(M(\OVI)/\rm M_{\odot})$ to  $\approx 5.0^{+0.1}_{-0.1}$ for galaxies with \logm~$<8.0$.

The red stripe in the left panel of Fig.~\ref{fig:movi_mod} shows the expected variation of $M(\OVI)$ with stellar mass, if the \OVI\ originates from the ambient hot halo gas. We predicted $M(\OVI)$ using:

\begin{equation}
\label{eq:movi} 
\begin{split}
M(\OVI) = f_{\OVI}(T) \times {\rm 10^{-3.31+[O/H]}} \times A_{\rm O} \times f_{\rm CGM} \\
\times f_{b} \times M_{\rm halo}~,  
\end{split}
\end{equation}
where $\rm [O/H]$ is the oxygen abundance in solar units, $f_{\rm CGM}$ is the fraction of total galactic baryons in the CGM, $f_b (\equiv \Omega_b/\Omega_{\rm M})$ is the cosmic baryon fraction, and $A_{\rm O}$ is the atomic mass of oxygen. The \OVI\ ion fraction, $f_{\OVI}(T)$, is obtained from the CIE tables of \citet{Oppenheimer_2013}, determined at the virial temperature of the halo\footnote{Using the $T_{\rm vir}-M_{\rm halo}$ relation from \citet[]{Oppenheimer_2016}}. Note that the abundance of the \OVI-bearing gas phase and the $f_{\rm CGM}$ for different stellar masses are largely unknown. If we assume that the $\rm [O/H]$ is similar to that of the cool, photoionized gas phase, i.e., $\rm [X/H] \approx -0.5$ \citep[see][]{Prochaska_2017,Sameer_2024}, then  $f_{\rm CGM}\approx 0.15$ can explain the estimated $M(\OVI)$ for galaxies at \logm~$\approx9.5$.

The pure CIE model under-produces $M(\OVI)$ for the mass bin of \logm$\geq$10.4.  Note that, in the presence of non-thermal pressure support (e.g., turbulence, cosmic rays), the virial temperature for a given halo mass can be overestimated \citep[see e.g.,][]{Lochhaas_2021, Tchernyshyov_2023}. 
Another explanation for this discrepancy could be the projection effect along the line of sight. Some of the detected \OVI\ likely originate from regions outside the virial radius, even if the transverse distance is smaller than $R_{\rm vir}$ \citep[see][]{Ho_2021}. These regions outside the virial radius are conducive to \OVI\ production for such massive halos \citep[e.g.,][]{Oppenheimer_2016}. Alternatively, a higher $f_{\rm CGM}$ and/or $\rm [O/H]$ for the highest mass bin can also account for the data. Additionally, the pure CIE model overproduces $M(\OVI)$ for the mass bin of \logm=9.8-10.4. A lower $f_{\rm CGM}$ and/or $\rm [O/H]$ could resolve the discrepancy.

The pure CIE model fails to reproduce the estimated $M(\OVI)$ for \logm~$<9.0$. A higher $f_{\rm CGM}$ can increase $M(\OVI)$ for galaxies with \logm~$<9$. 
 Even at solar metallicity, the required $f_{\rm CGM}$ exceeds $1$, which is unphysical. Note that the CGM of drawrf galaxies is expected to have subsolar metallicity \citep[e.g.,][]{Suresh_2017}. 
Evidently, the observed $M(\OVI)$ in the CGM of dwarf galaxies requires origin(s) other than the virialized gas in pure CIE since the virial temperatures are too low for collisional ionization.

\OVI\ in the CGM of dwarf galaxies can alternatively be produced via photoionization. Note that the pure CIE model shown in Fig.~\ref{fig:movi_mod} (red stripe) does not consider any external radiation field. The extragalactic UV background (UVB) radiation field can increase the degree of ionization and reduce cooling efficiencies, favoring the production of \OVI\ at lower temperatures. The green stripe in the left panel of Fig.~\ref{fig:movi_mod} shows the inferred $M(\OVI)$ for a `hybrid' model that considers both photoionization and collisional ionization, while keeping all other parameters the same as the pure CIE model.
 The ion-fractions are obtained at $n=10^{-4.5}~{\rm cm}^{-3},~z=0.4$ from an interpolated density-redshift-temperature grid from \citet[]{Oppenheimer_2013}, which are computed in the presence of the UVB HM01.
The green stripe is highlighted at  \logm~$<9$, where the UVB strongly boosts the $M(\OVI)$ estimates as compared to the pure CIE model. Evidently, photoionization equilibrium with the UVB can account for the estimated \OVI\ mass.

Another way to explain the estimated $M(\OVI)$ for dwarf galaxies is through non-equilibrium processes. Non-equilibrium effects become important when gas that was initially heated to very high temperatures ($>10^6$~K) cools faster than it recombines (i.e., the ratio of cooling time to recombination time, $t_{\rm cool}/t_{\rm rec} ~<1$). Consequently, the gas remains overionized for its temperature. In the right panel of Fig.~\ref{fig:movi_mod}, the blue stripe shows the $M(\OVI)$ predicted for such non-equilibrium cooling gas at the virial temperature, with a metallicity of $\rm [X/H] = -0.5$ and a density of  $n_{\rm H}=10^{-4.5}~{\rm cm}^{-3}$ subject to the HM01 UVB at $z\approx0.4$. The ion-fractions are taken from an interpolated density-redshift-temperature grid from \citet[]{Oppenheimer_2013}. The $M(\OVI)$ predicted by this model is in good agreement with the observations for \logm~$<9$. The orange stripe shows the same but without the UVB radiation field.
 It is apparent that non-equilibrium cooling gas with a subsolar metallicity cannot produce the observed $M(\OVI)$ for dwarf galaxies in the absence of the UVB. 
While a higher metallicity (e.g., $\rm [X/H]\gtrsim0$) could account for the observed $M(\OVI)$ without invoking the UVB, such elevated metallicities are not typically observed even in the ISM of dwarf galaxies \citep[]{vanZee_2006}. Thus, the presence of the UVB is crucial in both equilibrium and non-equilibrium scenarios to explain the observed $M(\OVI)$.

In the case of non-equilibrium cooling, it is important to understand the physical processes that can heat the gas to a very high temperature. Such high temperatures can be achieved via shocks, presumably due to star-formation feedback. At higher redshift, there is strong evidence that some circumgalactic \OVI\ traces hot, metal-rich outflows \citep[see][]{Turner_15}. For example, an outflow velocity of $\approx200$ \kms\ can give rise to a temperature of $>10^6$~K in the post-shock medium. Note that the assumed $f_{\rm CGM}$ of $0.15$ would in this case correspond to the fraction of baryons that are shock-heated but not yet cooled. For a galaxy of \logm~$=7$ ($\log_{\rm 10} M_{\rm halo}/\rm M_{\odot}\approx10$), the estimated cold gas in ISM is $\approx10^8~\rm M_{\odot}$ \citep[see][]{Peeples_2014}. Therefore, the baryon fraction for the stellar and ISM components is only 6\%  (i.e., $ 10^{8.04} /(0.16\times 10^{10}$)). It is thus possible that 15\% of the remaining 94\% baryons are in hot outflows. However, we note that only a fraction of the available baryons may be out of equilibrium.

\subsection{Importance of local ionization sources}

In the previous section, we emphasized the importance of photoionization in contributing significantly to the total $M(\OVI)$ for low-mass halos. The HM01 UVB model at $z \approx 0.4$ requires a low-density region with $n_{\rm H} \approx 10^{-4.5}~{\rm cm}^{-3}$, corresponding to an ionization parameter of $\approx 10^{-1}$, to efficiently produce \OVI. Such a density corresponds to an over-density, $\delta$ \footnote{Assuming the relation of mean baryon density $\bar{n}_{\rm H}$ with redshift from \citet[]{Schaye_2001}.} of $\approx70$, which is typical of the bulk of the CGM.


\begin{figure*}
\centering
     \includegraphics[width=0.5\linewidth]{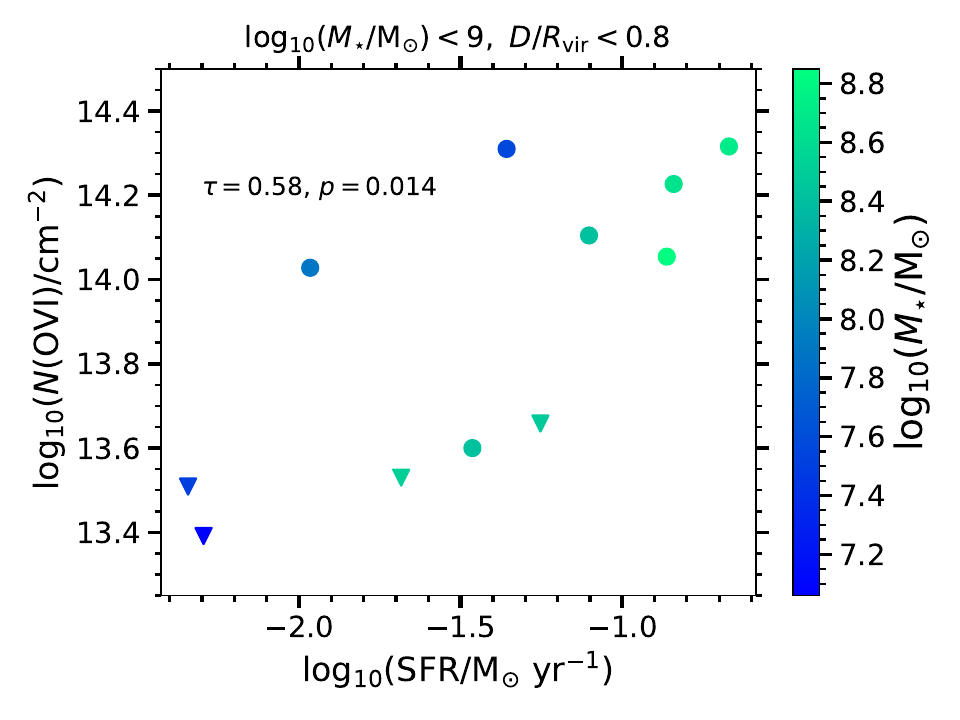}%
     \includegraphics[width=0.5\linewidth]{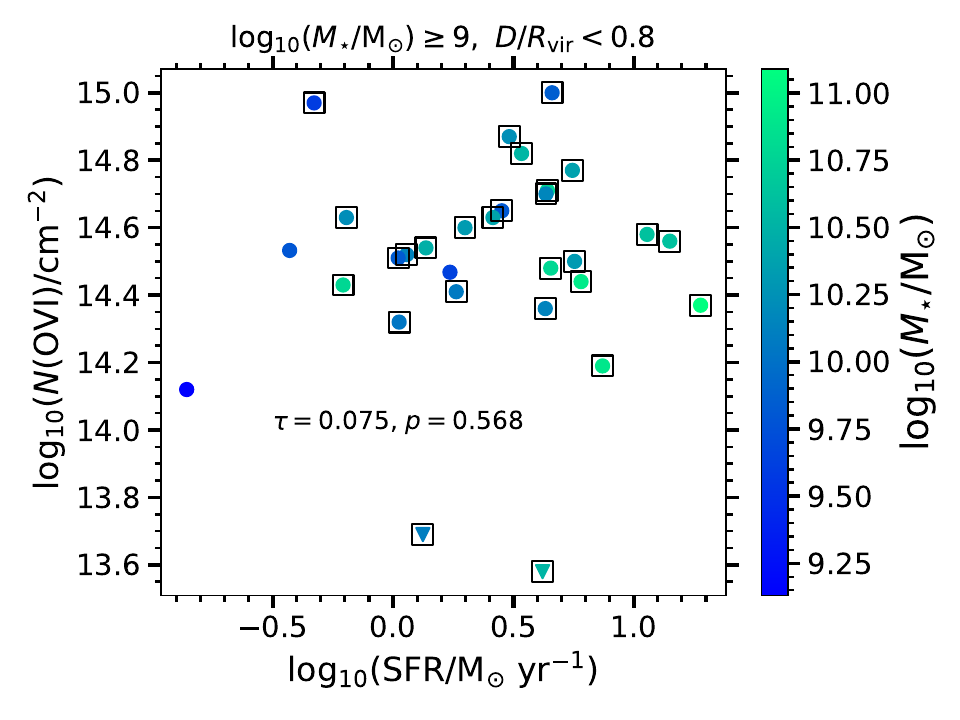}
     \caption{{\tt Left:} \OVI\ column densities and 3$\sigma$ upper limits for low-mass galaxies (\logm~$<9$) with $D/R_{\rm vir}<0.8$  plotted against the SFR by the filled circles and downward triangles, respectively. The data points are color-coded by the stellar mass. Only isolated, star-forming galaxies from MUSEQuBES are shown. The Kendall-$\tau$ correlation reveals a strong correlation ($\tau=0.58$) with $>2\sigma$ significance. {\tt Right:} Same as left, but for the high-mass (\logm~$\geq9$) galaxies. Here we included measurements from star-forming COS-Halos galaxies, marked with black squares. No significant correlation is observed for the high-mass sample.}

\label{fig:sfr-novi}
\end{figure*}

This density contrast demands a cloud size of $\approx22$ kpc to produce the observed $\left<N(\OVI)\right>\approx10^{13.8}~{\rm cm}^{-2}$ for low-mass galaxies, assuming $[{\rm X/H}]\approx -0.5$ and $f_{\OVI}\approx0.2$. A lower metallicity of $[{\rm X/H}]\approx -1$ results in a larger cloud size of $\approx66$ kpc, which is consistent with the halo size of low-mass galaxies. The densities inside the virialized halo, however, can be higher than this, resulting in a smaller cloud size.

Additional photoionization from the local sources is required to maintain the ionization parameter at higher densities. 
Previously, \citet[]{Werk_2016} showed that soft X-ray emission in the form of the local ionizing source alongside the UVB could result in a match to the observed line-ratios without resorting to very low-density and unusually large cloud sizes. They argued that such high-energy photons can be produced by the super-soft X-ray sources (SSSs) and/or hot ISM resulting from supernovae heating. Such high-energy photons from the local ionizing sources may play a key role in galaxy evolution \citep[]{Cantalupo_2010}. A relatively weak anti-correlation of $N(\OVI)$ and $D/R_{\rm vir}$ for the galaxies with \logm~$<9$ ($\tau \gtrsim -0.2$, see Table~\ref{tab:novi_table}) could be a manifestation of both the local ionizing flux and density falling as $\propto 1/D^{2}$, keeping a roughly flat ionization parameter.

Alternatively, a higher $f_{\rm CGM}$ can produce the observed $M(\OVI)$ in low-mass galaxies for higher densities without requiring local ionization sources. In such a case the required $f_{\rm CGM}$ exceeds 50\% for the dwarf galaxies.


\subsection{The trend between $N(\OVI)$ and SFR for `SF' galaxies} 
\label{sec:5.2}

The COS-Halos survey reported a strong bimodality of $N(\OVI)$ with sSFR, with massive ($L_*$) star-forming galaxies showing significantly higher \OVI\ column densities than passive galaxies which show \OVI\ absorption in only a small fraction of the sample \citep[]{Tumlinson_2011}.
We confirmed this trend in Fig.~\ref{fig:fc-prof}  inside $R_{\rm vir}$, but found that quenched galaxies follow the same relation as star-forming galaxies for $D/R_{\rm vir}=1-2$. Subsequently, we focused on the covering fraction of `SF' galaxies only. No significant trend between $N(\OVI)$ and SFR is reported for the massive ($\approx L_*$), star-forming galaxies in the COS-Halos sample. In this section, we will investigate whether there is any trend between the $N(\OVI)$ and SFR for the low-mass, `SF' galaxies in our sample.

In the left panel of Fig.~\ref{fig:sfr-novi}, we show the individual $N(\OVI)$ measurements for the `SF' galaxies with $D < 0.8R_{\rm vir}$  ($<$70~kpc) and \logm~$<9$ against the SFR. The $3\sigma$ upper limits in cases of non-detection are shown by the downward triangles. The $N(\OVI)$ of low-mass galaxies shows a strong correlation with SFR (Kendall-$\tau = 0.58,~p=0.014$) with $>2\sigma$ significance. Such a correlation is expected if star-formation-driven outflows give rise to the \OVI\ in the CGM, as seems to be the case for at least part of the \OVI\ around $z\approx2$ SF galaxies \citep[see][]{Turner_15}. In fact, for a given mass loading factor ($\eta \equiv \dot M_{\rm out}/\rm SFR$) and a fixed star-formation timescale, a linear relation is expected between the \OVI\ mass and the SFR with a slope of unity when plotted on a log-scale. The observed correlation between $N(\OVI)$ and SFR suggests that star-formation-driven outflows can contribute to the \OVI\ budget in the CGM of low-mass galaxies.

The right panel of Fig.~\ref{fig:sfr-novi} shows the $N(\OVI)$ versus SFR for the `SF' galaxies with \logm~$\geq9$. Here we included measurements from \citet[]{Tumlinson_2011} besides MUSEQuBES to increase the sample size. Unlike for the low-mass galaxies in the left panel, no significant correlation is observed here (Kendall-$\tau = 0.07$, $p= 0.57$). We have seen in the previous section that the estimated $M(\OVI)$ for the high-mass galaxies can be explained if \OVI\ arises from the ambient hot halo at the virial temperature.

\subsection{Do {\rm O~{\sc vi}} absorbers trace transient structures in the CGM of dwarf galaxies?}

In Section~\ref{sec:5.1}, we have observed that collisional ionization at the virial temperature of sub-solar metallicity gas, under both equilibrium and non-equilibrium conditions, cannot produce sufficient $M(\OVI)$ to account for the observations of dwarf galaxies (\logm~$<9$). Indeed, photoionization by the UVB or local sources seems to be necessary since the virial temperatures of the halos are too low for \OVI\ production. In the previous section, we reported a correlation between $N(\OVI)$ and SFR for the dwarf galaxies within $\approx R_{\rm vir}$. If such a correlation is driven by metal-rich outflows from these dwarf galaxies, then it is possible to shock-heat the CGM to super-virial temperatures ($\geq10^{6}$~K) depending on the shock speed. Such shock-heated gas can give rise to \OVI\ when cooling down.

A correlation between the $N(\OVI)$ and line width ($\Delta v_{\rm lw}$) is predicted theoretically for such radiatively cooling gas \citep[e.g.,][]{Heckman_2002, Bordoloi_2017}. Briefly, the \OVI\ column density is  given by:
\begin{equation}
\label{eq:cooling_flow}
     N(\OVI) = \frac{3k_BT}{\Lambda(T,Z)} v_{\rm cool}\left({\rm \frac{O}{H}}\right)_{\odot}Zf_{\OVI}~, 
 \end{equation}
where $v_{\rm cool}$ is the velocity of the cooling flow, $\Lambda(T,Z)$ is the cooling function in the units of ${\rm erg~cm^{3}~s^{-1}}$, $T$ indicates the column-density-weighted mean temperature of the cooling gas, and all other symbols have their usual meanings. The line width $\Delta v_{lw}$ in this context can be written as:
\begin{equation}
    \Delta v_{{\rm lw}}^{2} = 2k_BT/m_{\rm O} + v_{\rm cool}^{2}~. 
\end{equation}
We used the cooling function $\Lambda(T, Z)$ and the $f_{\OVI}$ from \citet[]{Oppenheimer_2013} assuming CIE to obtain the relation between $N(\OVI)$ and $\Delta v_{\rm lw}$. Note that the relation in Eq.~\ref{eq:cooling_flow} is 
insensitive to the metallicity provided [${\rm O/H}$] $\gtrsim-2$.

\begin{figure}
    \centering
    \includegraphics[width=1.0\linewidth]{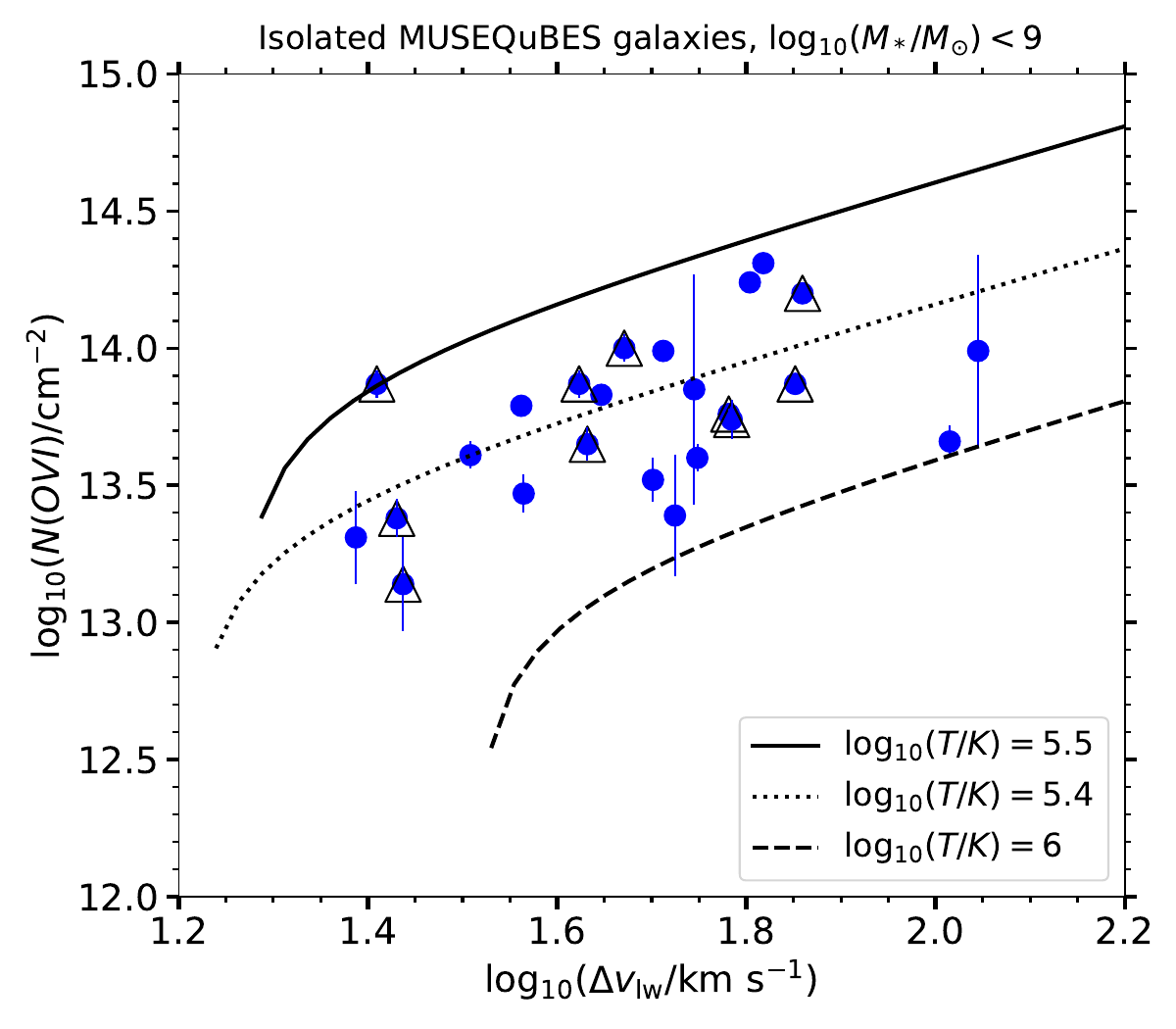}
    \caption{The observed $N(\OVI)$ plotted against the line width ($\Delta v_{\rm lw} \equiv 3b/\sqrt{2}$) for the \OVI\ components associated with isolated, low-mass (\logm~$<9$), MUSEQuBES galaxies. The dashed, dotted, and solid lines represent the predicted $N(\OVI)$ from the cooling flow models of \citet[]{Heckman_2002} for $T=10^{6},~10^{5.4}~{\rm and}~10^{5.5}$~K, respectively. The points with black triangle envelopes indicate \OVI\ components detected at $D>R_{\rm vir}$.}
    \label{fig:N-b_Heck}
\end{figure}

The kinematics of the \OVI\ bearing gas will be presented in Dutta et al. 2024 (submitted). However, in Fig.~\ref{fig:N-b_Heck}, we show the observed $N(\OVI)$ against the $\Delta v_{\rm lw}$, defined as $3b/\sqrt{2}$ following the definition of \citet[]{Bordoloi_2017}, for the \OVI\ components associated with isolated, low-mass (\logm~$<9$), MUSEQuBES galaxies. The dashed, dotted, and solid black lines represent the predicted $N(\OVI)$ from the cooling flow models (Eq.~\ref{eq:cooling_flow}) for $T=10^{6},~10^{5.4}~{\rm and}~10^{5.5}$K, respectively. The observed \OVI\ column densities in the CGM of dwarf galaxies are broadly consistent with the cooling flow model at column-density-weighted mean temperature of $T \approx 10^{5.4}$~K. We found that 12 out of 14 (7 out of 10) \OVI\ components detected at $D<R_{\rm vir}$ ($D \geq R_{\rm vir}$) show $b$-parameters broad enough to be consistent with such a temperature. This suggests that a fraction of these \OVI\ components may be tracing transient structures in the CGM cooling down to $\approx 10^{5.4}$~K from an initial hot phase, presumably triggered by star-formation-driven shocks. 
However, it is worthwhile to note that the observed $N(\OVI)-b$ correlation may partly result from the inability to detect broad, low column density components.

\subsection{ The robustness of $M(\OVI)$ in the CGM of dwarf galaxies }

The $M(\OVI)$ in Section~\ref{sec:5.1} were estimated from the $N(\OVI)$ measurements in the CGM within $R_{\rm vir}$. Therefore, the robustness of $M(\OVI)$ depends on the way we associate an absorber with a galaxy. Recall that we used a velocity window of $\pm300$~\kms\ centered on the galaxy redshift to determine the total $N(\OVI)$.
The $M(\OVI)$ for low-mass (\logm$<9$) MUSEQuBES galaxies within $R_{\rm vir}$ with this fiducial velocity window is  ${\rm log}_{10}(M(\OVI)/{\rm M_{\odot}})=5.2^{+0.1}_{-0.1}$. Using a smaller velocity window of $\pm100$ \kms\ provides consistent results.

Our $M(\OVI)$ estimates are marginally lower than the mass  
measured for a sample of ``isolated'' dwarf (\logm$<9$) galaxies in the CUBS survey \citep[i.e., ${\rm log}_{10}(M(\OVI)/{\rm M_{\odot}})\approx5.6^{+0.1}_{-0.3}$ within $R_{\rm vir}$;][]{Mishra_2024}. Note that they selected only the isolated galaxies for the $M(\OVI)$ measurements. Additionally, they obtained the mean $M(\OVI)$ with a non-parametric measure (Eq.~4 in their paper). Following the same method, we obtain ${\rm log}_{10}(M(\OVI)/{\rm M_{\odot}})=5.45$ and ${\rm log}_{10}(M(\OVI)/{\rm M_{\odot}})=5.37$ for a velocity window of $\pm300$ \kms\ and $\pm100$ \kms, respectively. These values are consistent with their measurement within the $1\sigma$ uncertainty.

With such a fixed LOS velocity window, it is possible that a fraction of the detected absorbers may reside outside the virial radius, as the exact location along the LOS is largely uncertain. This could lead to an overestimation of the measured column density within the CGM. Owing to their relatively smaller virial radii, this effect becomes more serious for dwarf galaxies \citep[see][]{Ho_2021}.

Additionally, the contributions from neighboring massive galaxies outside the MUSE FOV can also lead to an overestimation of the $M(\OVI)$ for the dwarf galaxies. 
In order to test the robustness of the $M(\OVI)$ estimates, we select a subsample of galaxies from our MUSEQuBES sample that has no neighbor detected within $\pm500$~\kms\ LOS separation within the MUSE FoV. The $\langle N(\OVI)\rangle$ and $M(\OVI)$ estimates remain unaffected even with this sample. Next, we used data from a wide-field ($\approx 20''$) galaxy survey with the Magellan telescope (Johnson et al., in prep) for 6/16 quasar fields in our sample. Magellan observations show that 3 of the 9 low-mass galaxies showing detectable \OVI\ absorption in these 6 fields have a massive neighbor outside the MUSE FOV with lower $D/R_{\rm vir}$. Thus, while the effects of massive galaxies outside the MUSE FOV are not severe, it is likely not negligible. To be conservative, the $\left<N(\OVI)\right>$ and  $M(\OVI)$ for galaxies with \logm~$<9$ could be treated as conservative upper limits.

\subsection{{\rm O~\sc vi}-floor in the outskirts of dwarf galaxies}

A lower covering fraction of \OVI\ around low-mass galaxies (\logm~$<9$) for a threshold $N(\OVI)=10^{14}~{\rm cm}^{-2}$ does not necessarily rule out the presence of \OVI\ absorbers below this threshold, which are undetected because of the limited SNR of the COS spectra. In fact, most \OVI\ upper limits are above ${\rm log}_{10}(N(\OVI)/{\rm cm}^{-2})=13.2$.

Spectral stacking is extensively used in the literature to detect weak signals that are otherwise hidden in noise \citep[see e.g.,][]{ Schaye_2000,Steidel_2010,Turner_14, Chen_20}. With the same motivation, we produced an \OVI\ stack using the \logm~$<9$ galaxies in our sample that do not exhibit any detectable \OVI\ absorption in their CGM (124 out of 247 cases). To obtain the median stack, we adopted a method similar to \citet[]{Dutta_2024}. Briefly, (a) For each galaxy, we select a region of $\pm1500$~\kms\ around its redshifted \OVI\ wavelength in the normalized spectrum of the corresponding background quasar. (b) We calculate the median normalized flux in 40~\kms\ wide velocity bins for all such quasar-galaxy pairs. This is referred to as the observed median-stacked flux spectrum. (c) Next, a local pseudo-continuum is estimated and subtracted from the observed median-stacked flux spectrum. We use the median flux far away from the galaxy redshifts ($\pm$~5,000-10,000 \kms) as the pseudo-continuum. Finally, (d) the \OVI\ rest-frame equivalent widths are measured by direct integration of the pseudo-continuum-subtracted median-stacked spectra for LOS velocity windows of $\pm300$~\kms\ ($W_{r,300}$) from the line center.

\begin{figure}
    \centering
    \includegraphics[width=1.0\linewidth]{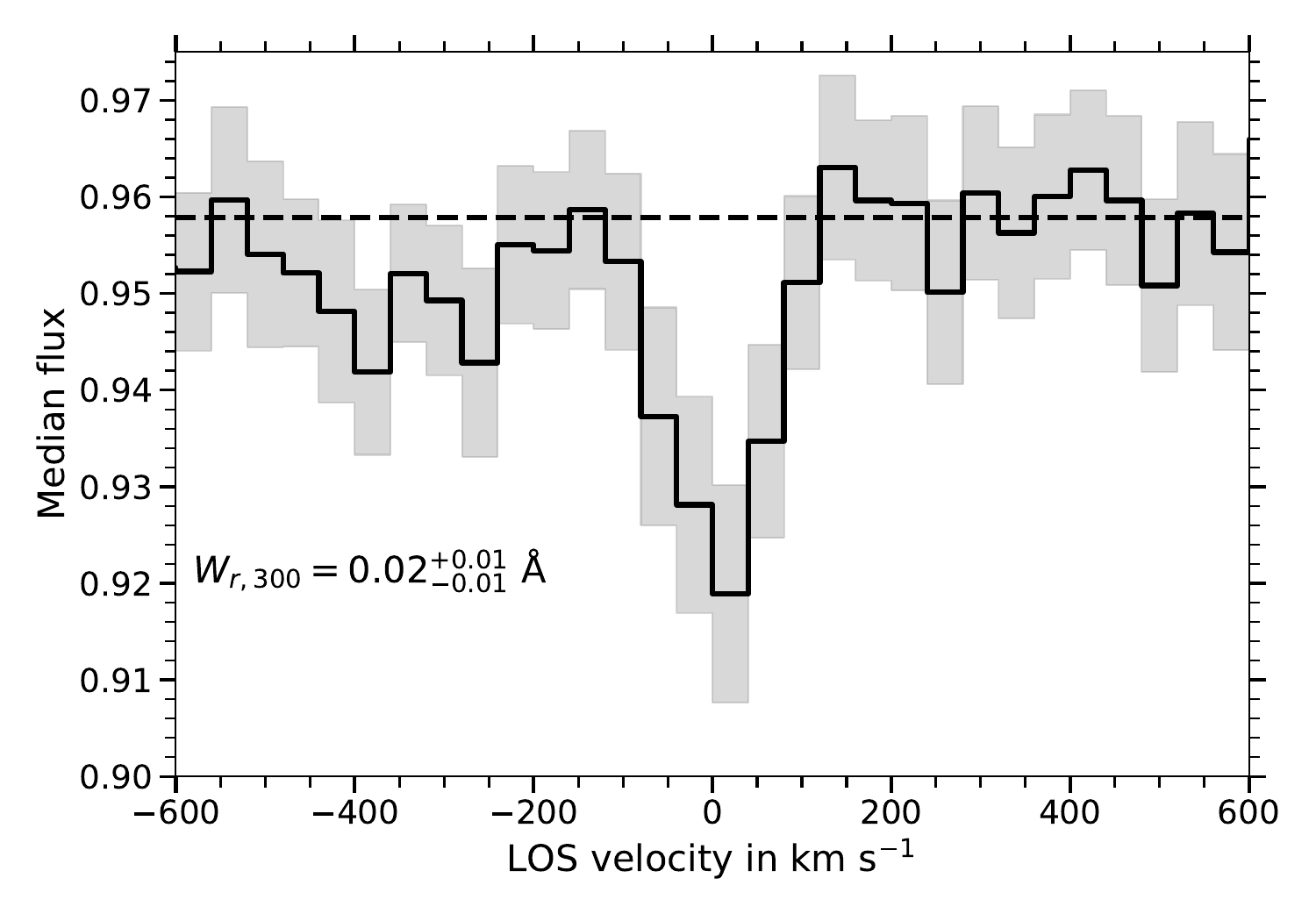}
    \caption{The median-stacked \OVI\ absorption profile for the low-mass (\logm~$<9$) MUSEQuBES galaxies without any individually detected \OVI\ is shown with the black histogram. The black horizontal line represents the pseudo-continuum used to obtain the rest-frame equivalent width. The gray-shaded region represents the standard deviation of flux obtained from 1000 bootstrap realizations.} 
  \label{fig:nondet_stack}
\end{figure}

The median-stacked flux profile, along with the standard deviation, is shown in Fig.~\ref{fig:nondet_stack}. \OVI\ is detected in the median stack with $>90$\% confidence, with a rest-frame equivalent width of $0.02\pm0.01$~\AA. The detection significance and the error on the $W_{r,300}$ measurement (68\% confidence interval) are obtained from the $W_{r,300}$ distribution obtained from 1000 bootstrap resampled median spectra. For such a weak line\footnote{Recall that absorption is not detected in individual spectra.}, it is appropriate to convert the $W_{r,300}$ to column density, assuming the linear part of the curve-of-growth. This yields ${\rm log}_{10}(N(\OVI)/{\rm cm}^{-2})=13.2^{+0.2}_{-0.3}$. We point out here that the majority of the galaxies (113/124) contributing to the stack have $D > R_{\rm vir}$ with a median $D/R_{\rm vir}\approx2$. Thus, the detection of \OVI\ absorption in the stacked spectrum may suggests that there may be
an \OVI-floor far outside the virial radius of low-mass galaxies with an average column density of $\approx10^{13.2}$~\sqcm.

\citet[]{Prochaska_2011} found that the virialized halos of sub-$L_*$ galaxies can account for almost all of the strong \OVI\ absorption ($W_{r,1031}>100~{\rm m}$\AA) seen in the quasar spectra. A similar conclusion was reached by \citet[]{Tchernyshyov_2022} who showed that \OVI\ absorbers with $N(\OVI)>10^{14.5}~{\rm cm}^{-2}$ are mostly contributed by the halos of star-forming galaxies with $M_{\star}$ ranging from $10^{8.8}-10^{10.4}~\rm M_{\odot}$. However, both \citet[]{Prochaska_2011} and \citet[]{Tchernyshyov_2022} argued that the halos of star-forming galaxies cannot account for the low column density \OVI\ absorbers ($N(\OVI)<10^{13.5}~{\rm cm}^{-2}$). \citet[]{Tchernyshyov_2022} speculated the existence of an extended CGM out to 300 kpc around galaxies where these low column density  absorbers might reside. The detected \OVI\ with $N(\OVI)\approx10^{13.2}~{\rm cm}^{-2}$ in the stacked spectrum is in line with their speculations.

To understand the physical and chemical properties of the \OVI-bearing gas detected in the stack, we estimated the median \HI\ column density using two approaches. Firstly, we measured \HI\ column densities (47 cases) and upper limits (77 cases) for all the galaxies. Secondly, we produced a stack of \lyb\ absorption for galaxies without individually detected \lyb\ absorption. In both cases, we obtained a consistent $N(\HI)$ of $10^{13.6}$~\sqcm, indicating a mild overdensity ($\delta \sim 10$) of the absorbing medium, under the assumption that hydrogen is photoionized \citep[e.g.,][]{Schaye_2001}. Under CIE at $T\approx10^{5.5}$~K, the \OVI\ and \HI\ column densities suggest an oxygen abundance of $\rm [O/H] \approx -2.3$ and total hydrogen column density of $\sim 10^{19}$~\sqcm. Note that, the average metallicity of the cool enriched gas phase is observed to be $\rm [X/H] \approx -0.5 ~{\rm to}~-1.0 $ within $\approx 0.5 - 1 R_{\rm vir}$ of massive galaxies \citep[see][]{Prochaska_2017,Pointon_2019,Sameer_2024}. Therefore, it is unlikely that the gas temperature is $<10^{5.3}$~K, since it yields an unreasonably high oxygen abundance (i.e., $\rm [O/H] > -1$) at impact parameters of $\approx 2R_{\rm vir}$ from dwarf galaxies (\logm~$<9$). The expected low metallicity suggest that the detected \OVI\ absorption may be tracing the relatively cooler phase of the canonical warm-hot intergalactic medium \citep[WHIM; e.g.,][]{Dave_2001,Cen_2001}. Indeed, \citet[]{Dave_2001} argued that WHIM gas is heated to $T>10^5$~K primarily by shock heating of gas accreting onto large-scale filamentary structures that are not yet virialized or in dynamical equilibrium. This is consistent with the large impact parameters ($\approx 2 R_{\rm vir}$) of the dwarf galaxies contributing to the stack, which may be embedded in such filaments.

On the contrary, if the gas detected in the stack is in photoionization equilibrium with the extragalactic UV background radiation, the inferred oxygen abundance would be $\rm [O/H] \gtrsim -1.0$ for all plausible densities.  Such a high metallicity is unlikely to be present at $\approx 2R_{\rm vir}$ from dwarf galaxies \citep[]{Suresh_2017}. However, the metallicity of the oxygen-bearing phase can be much higher than the rest of the CGM as the metal can be poorly mixed \citep[see][]{Schaye_2007,Tepper_2011}.

\section{Summary}
\label{sec:summary}

In this paper, we present a study of the \OVI-bearing gas around 247 low-redshift ($0.1\lesssim z\lesssim0.7$) galaxies with a median \logm~$=8.7$  observed in 16 quasar fields in the MUSEQuBES survey. The 60 detected \OVI\ absorption systems, along with the stringent 3$\sigma$ upper-limits for the non-detections, enabled us to investigate the connections between galaxy properties on the \OVI\ column density ($N(\OVI)$) and covering fraction ($\kappa$) profiles. Our key findings are:

\begin{itemize}

    \item The \OVI\ column-density shows a weak anti-correlation with $D/R_{\rm vir}$ ($\tau=-0.15$, $p=0.003$) for the star-forming galaxies, which is not present for the quenched galaxies ($\tau=0.10$, $p=0.453$) in our sample (Fig.~\ref{fig:Novi-prof}).

    \item  The \OVI\ covering fraction profiles for MUSEQuBES galaxies show a significantly higher covering fraction within the virial radius of star-forming galaxies compared to quenched galaxies. No significant difference in covering fraction is observed at $D/R_{\rm vir}=1-2$ for these two galaxy populations (Fig.~\ref{fig:fc-prof}).

    \item Including the \OVI\ measurements around 253 star-forming galaxies from the literature, we provide fitting formulae and best fit parameters for (i) the \OVI\ column density profile for different mass bins (Table~\ref{tab:best-fit-post-Novi}) and (ii) the \OVI\ covering fraction profiles at threshold $N(\OVI)=10^{14}~{\rm cm}^{-2}$ for different mass bins (Table~\ref{tab:best-fit-post_k_mbin}). The impact parameter averaged \OVI\ column density ${\rm log}_{10}\left<N(\OVI)/{\rm cm}^{-2}\right>\approx 14.1$ within $R_{\rm vir}$ for our MUSEQuBES sample is significantly lower than for $\approx L_*$ galaxies.

    \item Using the combined sample of star-forming galaxies, we show that both $\left<N(\OVI)\right>$ and $\left<\kappa(\OVI)\right>$ peak for galaxies with \logm~$\approx9.5$ (Figs.~\ref{fig:avgOVI_N} and \ref{fig:avFc_M}). The virial temperature for these galaxies is optimal for \OVI\ under collisional ionization. Moreover, these galaxies show the strongest anti-correlation of $N(\OVI)$ with $D/R_{\rm vir}$ compared to their lower- and higher-mass counterparts (Table \ref{tab:novi_table}).

    \item The average \OVI\ mass estimated within $R_{\rm vir}$ shows a peak at \logm~$\geq9.5$ and is almost an order of magnitude lower at \logm~$\leq8.5$. CIE at the virial temperature can explain the estimated $M(\OVI)$ for the high-mass galaxies but not for the dwarf galaxies. Photoionization and/or collisionally ionized cooling gas transitioning through $\approx10^{5.4}$~K can account for the $N(\OVI)$ and $M(\OVI)$ measurements for the dwarf galaxies (Fig.~\ref{fig:movi_mod}).  

    \item The $N(\OVI)$ of low-mass (\logm$<9$), star-forming galaxies within 0.8$R_{\rm vir}$ shows a strong correlation with the SFR of the host galaxies ($\tau=0.58,~p=0.014$), which is not present for the high-mass, star-forming counterparts (Fig.~\ref{fig:sfr-novi}).  
 
    \item Using spectral stacking, we report the presence of a highly ionized metal floor outside the virial radius of dwarf galaxies with ${\rm log}_{10}(N(\OVI)/{\rm cm}^{-2}) = 13.2$ (Fig.~\ref{fig:nondet_stack}). We argued that the detected \OVI\ signal may be tracing the relatively cooler phase of the canonical WHIM gas in the cosmic filaments in which the dwarf galaxies are embedded.

\end{itemize}

\begin{acknowledgments}
We thank Marijke Segers, Lorrie Straka, and Monica Turner for their early contributions to the MUSEQuBES project. SC, SD, and SM   acknowledge support from the Indo-Italian Executive Programme of Scientific and Technological Cooperation 2022—2024 (TPN: 63673). SD and SM acknowledge R. Srianand and Aseem Paranjape for insightful discussions. SD acknowledges Alankar Dutta and Prateek Sharma for insightful discussions. 
\end{acknowledgments}

\bibliography{all_ref_ovi_apj}{}

\begin{thebibliography}{}
\expandafter\ifx\csname natexlab\endcsname\relax\def\natexlab#1{#1}\fi
\providecommand{\url}[1]{\href{#1}{#1}}
\providecommand{\dodoi}[1]{doi:~\href{http://doi.org/#1}{\nolinkurl{#1}}}
\providecommand{\doeprint}[1]{\href{http://ascl.net/#1}{\nolinkurl{http://ascl.net/#1}}}
\providecommand{\doarXiv}[1]{\href{https://arxiv.org/abs/#1}{\nolinkurl{https://arxiv.org/abs/#1}}}

\bibitem[{{Aguirre} {et~al.}(2008){Aguirre}, {Dow-Hygelund}, {Schaye}, \& {Theuns}}]{Aguirre_2008}
{Aguirre}, A., {Dow-Hygelund}, C., {Schaye}, J., \& {Theuns}, T. 2008, \apj, 689, 851, \dodoi{10.1086/592554}

\bibitem[{{Bacon} {et~al.}(2010){Bacon}, {Accardo}, {Adjali}, {Anwand}, {Bauer}, {Biswas}, {Blaizot}, {Boudon}, {Brau-Nogue}, {Brinchmann}, {Caillier}, {Capoani}, {Carollo}, {Contini}, {Couderc}, {Daguis{\'e}}, {Deiries}, {Delabre}, {Dreizler}, {Dubois}, {Dupieux}, {Dupuy}, {Emsellem}, {Fechner}, {Fleischmann}, {Fran{\c{c}}ois}, {Gallou}, {Gharsa}, {Glindemann}, {Gojak}, {Guiderdoni}, {Hansali}, {Hahn}, {Jarno}, {Kelz}, {Koehler}, {Kosmalski}, {Laurent}, {Le Floch}, {Lilly}, {Lizon}, {Loupias}, {Manescau}, {Monstein}, {Nicklas}, {Olaya}, {Pares}, {Pasquini}, {P{\'e}contal-Rousset}, {Pell{\'o}}, {Petit}, {Popow}, {Reiss}, {Remillieux}, {Renault}, {Roth}, {Rupprecht}, {Serre}, {Schaye}, {Soucail}, {Steinmetz}, {Streicher}, {Stuik}, {Valentin}, {Vernet}, {Weilbacher}, {Wisotzki}, \& {Yerle}}]{Bacon_2010}
{Bacon}, R., {Accardo}, M., {Adjali}, L., {et~al.} 2010, in Society of Photo-Optical Instrumentation Engineers (SPIE) Conference Series, Vol. 7735, Ground-based and Airborne Instrumentation for Astronomy III, ed. I.~S. {McLean}, S.~K. {Ramsay}, \& H.~{Takami}, 773508, \dodoi{10.1117/12.856027}

\bibitem[{{Begelman} \& {Fabian}(1990)}]{Begelman_1990}
{Begelman}, M.~C., \& {Fabian}, A.~C. 1990, \mnras, 244, 26P

\bibitem[{{Berg} {et~al.}(2019){Berg}, {Howk}, {Lehner}, {Wotta}, {O'Meara}, {Bowen}, {Burchett}, {Peeples}, \& {Tejos}}]{Berg_2019}
{Berg}, M.~A., {Howk}, J.~C., {Lehner}, N., {et~al.} 2019, \apj, 883, 5, \dodoi{10.3847/1538-4357/ab378e}

\bibitem[{{Bergeron} \& {Stasi{\'n}ska}(1986)}]{Bergeron_1986}
{Bergeron}, J., \& {Stasi{\'n}ska}, G. 1986, \aap, 169, 1

\bibitem[{{Bielby} {et~al.}(2019){Bielby}, {Stott}, {Cullen}, {Tripp}, {Burchett}, {Fumagalli}, {Morris}, {Tejos}, {Crain}, {Bower}, \& {Prochaska}}]{Bielby_2019}
{Bielby}, R.~M., {Stott}, J.~P., {Cullen}, F., {et~al.} 2019, \mnras, 486, 21, \dodoi{10.1093/mnras/stz774}

\bibitem[{{Boogaard, Leindert A.} {et~al.}(2018){Boogaard, Leindert A.}, {Brinchmann, Jarle}, {Bouch\'e, Nicolas}, {Paalvast, Mieke}, {Bacon, Roland}, {Bouwens, Rychard J.}, {Contini, Thierry}, {Gunawardhana, Madusha L. P.}, {Inami, Hanae}, {Marino, Raffaella A.}, {Maseda, Michael V.}, {Mitchell, Peter}, {Nanayakkara, Themiya}, {Richard, Johan}, {Schaye, Joop}, {Schreiber, Corentin}, {Tacchella, Sandro}, {Wisotzki, Lutz}, \& {Zabl, Johannes}}]{Boogard_18}
{Boogaard, Leindert A.}, {Brinchmann, Jarle}, {Bouch\'e, Nicolas}, {et~al.} 2018, A\&A, 619, A27, \dodoi{10.1051/0004-6361/201833136}

\bibitem[{{Bordoloi} {et~al.}(2017){Bordoloi}, {Wagner}, {Heckman}, \& {Norman}}]{Bordoloi_2017}
{Bordoloi}, R., {Wagner}, A.~Y., {Heckman}, T.~M., \& {Norman}, C.~A. 2017, \apj, 848, 122, \dodoi{10.3847/1538-4357/aa8e9c}

\bibitem[{{Borkowski} \& {Shull}(1990)}]{Borkowski_1990}
{Borkowski}, K.~J., \& {Shull}, J.~M. 1990, \apj, 348, 169, \dodoi{10.1086/168225}

\bibitem[{{Cantalupo}(2010)}]{Cantalupo_2010}
{Cantalupo}, S. 2010, \mnras, 403, L16, \dodoi{10.1111/j.1745-3933.2010.00806.x}

\bibitem[{{Carswell} \& {Webb}(2014)}]{vpfit}
{Carswell}, R.~F., \& {Webb}, J.~K. 2014, {VPFIT: Voigt profile fitting program}, Astrophysics Source Code Library, record ascl:1408.015

\bibitem[{{Cen} {et~al.}(2001){Cen}, {Tripp}, {Ostriker}, \& {Jenkins}}]{Cen_2001}
{Cen}, R., {Tripp}, T.~M., {Ostriker}, J.~P., \& {Jenkins}, E.~B. 2001, \apjl, 559, L5, \dodoi{10.1086/323721}

\bibitem[{Chabrier(2003)}]{Chabrier_2003}
Chabrier, G. 2003, The Astrophysical Journal, 586, L133, \dodoi{10.1086/374879}

\bibitem[{{Chen} \& {Mulchaey}(2009)}]{Chen_2009}
{Chen}, H.-W., \& {Mulchaey}, J.~S. 2009, \apj, 701, 1219, \dodoi{10.1088/0004-637X/701/2/1219}

\bibitem[{{Chen} {et~al.}(2018){Chen}, {Zahedy}, {Johnson}, {Pierce}, {Huang}, {Weiner}, \& {Gauthier}}]{Chen_2018}
{Chen}, H.-W., {Zahedy}, F.~S., {Johnson}, S.~D., {et~al.} 2018, \mnras, 479, 2547, \dodoi{10.1093/mnras/sty1541}

\bibitem[{Chen {et~al.}(2020)Chen, Steidel, Hummels, Rudie, Dong, Trainor, Bogosavljević, Erb, Pettini, Reddy, Shapley, Strom, Theios, Faucher-Giguère, Hopkins, \& Kereš}]{Chen_20}
Chen, Y., Steidel, C.~C., Hummels, C.~B., {et~al.} 2020, Monthly Notices of the Royal Astronomical Society, 499, 1721, \dodoi{10.1093/mnras/staa2808}

\bibitem[{{Dav{\'e}} {et~al.}(2001){Dav{\'e}}, {Cen}, {Ostriker}, {Bryan}, {Hernquist}, {Katz}, {Weinberg}, {Norman}, \& {O'Shea}}]{Dave_2001}
{Dav{\'e}}, R., {Cen}, R., {Ostriker}, J.~P., {et~al.} 2001, \apj, 552, 473, \dodoi{10.1086/320548}

\bibitem[{{Dopita} \& {Sutherland}(1996)}]{Dopita_1996}
{Dopita}, M.~A., \& {Sutherland}, R.~S. 1996, \apjs, 102, 161, \dodoi{10.1086/192255}

\bibitem[{{Dutta} {et~al.}(2024){Dutta}, {Muzahid}, {Schaye}, {Mishra}, {Chen}, {Johnson}, {Wisotzki}, \& {Cantalupo}}]{Dutta_2024}
{Dutta}, S., {Muzahid}, S., {Schaye}, J., {et~al.} 2024, \mnras, 528, 3745, \dodoi{10.1093/mnras/stae206}

\bibitem[{{Edgar} \& {Chevalier}(1986)}]{Edgar_1986}
{Edgar}, R.~J., \& {Chevalier}, R.~A. 1986, \apjl, 310, L27, \dodoi{10.1086/184775}

\bibitem[{{Finn} {et~al.}(2016){Finn}, {Morris}, {Tejos}, {Crighton}, {Perry}, {Fumagalli}, {Bielby}, {Theuns}, {Schaye}, {Shanks}, {Liske}, {Gunawardhana}, \& {Bartle}}]{Finn_2016}
{Finn}, C.~W., {Morris}, S.~L., {Tejos}, N., {et~al.} 2016, \mnras, 460, 590, \dodoi{10.1093/mnras/stw918}

\bibitem[{Gnat \& Sternberg(2007)}]{Gnat_2007}
Gnat, O., \& Sternberg, A. 2007, The Astrophysical Journal Supplement Series, 168, 213, \dodoi{10.1086/509786}

\bibitem[{{Guo} {et~al.}(2023){Guo}, {Bacon}, {Bouch{\'e}}, {Wisotzki}, {Schaye}, {Blaizot}, {Verhamme}, {Cantalupo}, {Boogaard}, {Brinchmann}, {Cherrey}, {Kusakabe}, {Langan}, {Leclercq}, {Matthee}, {Michel-Dansac}, {Schroetter}, \& {Wendt}}]{Guo_2023}
{Guo}, Y., {Bacon}, R., {Bouch{\'e}}, N.~F., {et~al.} 2023, \nat, 624, 53, \dodoi{10.1038/s41586-023-06718-w}

\bibitem[{{Heckman} {et~al.}(2002){Heckman}, {Norman}, {Strickland}, \& {Sembach}}]{Heckman_2002}
{Heckman}, T.~M., {Norman}, C.~A., {Strickland}, D.~K., \& {Sembach}, K.~R. 2002, \apj, 577, 691, \dodoi{10.1086/342232}

\bibitem[{Hinton {et~al.}(2016)Hinton, Davis, Lidman, Glazebrook, \& Lewis}]{Hinton_2016}
Hinton, S., Davis, T.~M., Lidman, C., Glazebrook, K., \& Lewis, G. 2016, Astronomy and Computing, 15, 61, \dodoi{https://doi.org/10.1016/j.ascom.2016.03.001}

\bibitem[{Ho {et~al.}(2021)Ho, Martin, \& Schaye}]{Ho_2021}
Ho, S.~H., Martin, C.~L., \& Schaye, J. 2021, The Astrophysical Journal, 923, 137, \dodoi{10.3847/1538-4357/ac2c73}

\bibitem[{{Huang} {et~al.}(2021){Huang}, {Chen}, {Shectman}, {Johnson}, {Zahedy}, {Helsby}, {Gauthier}, \& {Thompson}}]{Huang_2021}
{Huang}, Y.-H., {Chen}, H.-W., {Shectman}, S.~A., {et~al.} 2021, \mnras, 502, 4743, \dodoi{10.1093/mnras/stab360}

\bibitem[{Johnson {et~al.}(2015)Johnson, Chen, \& Mulchaey}]{Johnson_15}
Johnson, S.~D., Chen, H.-W., \& Mulchaey, J.~S. 2015, Monthly Notices of the Royal Astronomical Society, 449, 3263, \dodoi{10.1093/mnras/stv553}

\bibitem[{{Johnson} {et~al.}(2017){Johnson}, {Chen}, {Mulchaey}, {Schaye}, \& {Straka}}]{Johnson_17}
{Johnson}, S.~D., {Chen}, H.-W., {Mulchaey}, J.~S., {Schaye}, J., \& {Straka}, L.~A. 2017, \apjl, 850, L10, \dodoi{10.3847/2041-8213/aa9370}

\bibitem[{{Kacprzak} {et~al.}(2015){Kacprzak}, {Muzahid}, {Churchill}, {Nielsen}, \& {Charlton}}]{Kacprzak_2015}
{Kacprzak}, G.~G., {Muzahid}, S., {Churchill}, C.~W., {Nielsen}, N.~M., \& {Charlton}, J.~C. 2015, \apj, 815, 22, \dodoi{10.1088/0004-637X/815/1/22}

\bibitem[{{Keeney} {et~al.}(2017){Keeney}, {Stocke}, {Danforth}, {Shull}, {Pratt}, {Froning}, {Green}, {Penton}, \& {Savage}}]{Keeney_2017}
{Keeney}, B.~A., {Stocke}, J.~T., {Danforth}, C.~W., {et~al.} 2017, \apjs, 230, 6, \dodoi{10.3847/1538-4365/aa6b59}

\bibitem[{Keeney {et~al.}(2018)Keeney, Stocke, Pratt, Davis, Syphers, Danforth, Shull, Froning, Green, Penton, \& Savage}]{Keeney_2018}
Keeney, B.~A., Stocke, J.~T., Pratt, C.~T., {et~al.} 2018, The Astrophysical Journal Supplement Series, 237, 11, \dodoi{10.3847/1538-4365/aac727}

\bibitem[{{Kennicutt}(1998)}]{Kennicutt_1998}
{Kennicutt}, Robert~C., J. 1998, \apj, 498, 541, \dodoi{10.1086/305588}

\bibitem[{Kewley {et~al.}(2004)Kewley, Geller, \& Jansen}]{Kewley_2004}
Kewley, L.~J., Geller, M.~J., \& Jansen, R.~A. 2004, The Astronomical Journal, 127, 2002, \dodoi{10.1086/382723}

\bibitem[{{Kriek} {et~al.}(2009){Kriek}, {van Dokkum}, {Labb{\'e}}, {Franx}, {Illingworth}, {Marchesini}, \& {Quadri}}]{Kriek_2009}
{Kriek}, M., {van Dokkum}, P.~G., {Labb{\'e}}, I., {et~al.} 2009, \apj, 700, 221, \dodoi{10.1088/0004-637X/700/1/221}

\bibitem[{{Lehnert} {et~al.}(2015){Lehnert}, {van Driel}, {Le Tiran}, {Di Matteo}, \& {Haywood}}]{Lehnert_2015}
{Lehnert}, M.~D., {van Driel}, W., {Le Tiran}, L., {Di Matteo}, P., \& {Haywood}, M. 2015, \aap, 577, A112, \dodoi{10.1051/0004-6361/201322630}

\bibitem[{{Li} \& {Tonnesen}(2020)}]{Li_2020}
{Li}, M., \& {Tonnesen}, S. 2020, \apj, 898, 148, \dodoi{10.3847/1538-4357/ab9f9f}

\bibitem[{{Liang} {et~al.}(2016){Liang}, {Kravtsov}, \& {Agertz}}]{Liang_2016}
{Liang}, C.~J., {Kravtsov}, A.~V., \& {Agertz}, O. 2016, \mnras, 458, 1164, \dodoi{10.1093/mnras/stw375}

\bibitem[{{Lochhaas} {et~al.}(2021){Lochhaas}, {Tumlinson}, {O'Shea}, {Peeples}, {Smith}, {Werk}, {Augustin}, \& {Simons}}]{Lochhaas_2021}
{Lochhaas}, C., {Tumlinson}, J., {O'Shea}, B.~W., {et~al.} 2021, \apj, 922, 121, \dodoi{10.3847/1538-4357/ac2496}

\bibitem[{{McGaugh} {et~al.}(2010){McGaugh}, {Schombert}, {de Blok}, \& {Zagursky}}]{McGaugh_2010}
{McGaugh}, S.~S., {Schombert}, J.~M., {de Blok}, W.~J.~G., \& {Zagursky}, M.~J. 2010, \apjl, 708, L14, \dodoi{10.1088/2041-8205/708/1/L14}

\bibitem[{Mishra {et~al.}(2024)Mishra, Johnson, Rudie, Chen, Schaye, Qu, Zahedy, Boettcher, Cantalupo, Chen, Faucher-Giguère, Greene, Li, Zhuoqi, Liu, Lopez, \& Petitjean}]{Mishra_2024}
Mishra, N., Johnson, S.~D., Rudie, G.~C., {et~al.} 2024, The Cosmic Ultraviolet Baryon Survey (CUBS) IX: The enriched circumgalactic and intergalactic medium around star-forming field dwarf galaxies traced by O VI absorption.
\newblock \doarXiv{2408.11151}

\bibitem[{{Morrissey} {et~al.}(2018){Morrissey}, {Matuszewski}, {Martin}, {Neill}, {Epps}, {Fucik}, {Weber}, {Darvish}, {Adkins}, {Allen}, {Bartos}, {Belicki}, {Cabak}, {Callahan}, {Cowley}, {Crabill}, {Deich}, {Delecroix}, {Doppman}, {Hilyard}, {James}, {Kaye}, {Kokorowski}, {Kwok}, {Lanclos}, {Milner}, {Moore}, {O'Sullivan}, {Parihar}, {Park}, {Phillips}, {Rizzi}, {Rockosi}, {Rodriguez}, {Salaun}, {Seaman}, {Sheikh}, {Weiss}, \& {Zarzaca}}]{Morrissey_2018}
{Morrissey}, P., {Matuszewski}, M., {Martin}, D.~C., {et~al.} 2018, \apj, 864, 93, \dodoi{10.3847/1538-4357/aad597}

\bibitem[{{Moster} {et~al.}(2013){Moster}, {Naab}, \& {White}}]{Moster_2013}
{Moster}, B.~P., {Naab}, T., \& {White}, S. D.~M. 2013, \mnras, 428, 3121, \dodoi{10.1093/mnras/sts261}

\bibitem[{{Muzahid} {et~al.}(2012){Muzahid}, {Srianand}, {Bergeron}, \& {Petitjean}}]{Muzahid_2012}
{Muzahid}, S., {Srianand}, R., {Bergeron}, J., \& {Petitjean}, P. 2012, \mnras, 421, 446, \dodoi{10.1111/j.1365-2966.2011.20324.x}

\bibitem[{{Oppenheimer} \& {Schaye}(2013)}]{Oppenheimer_2013}
{Oppenheimer}, B.~D., \& {Schaye}, J. 2013, \mnras, 434, 1043, \dodoi{10.1093/mnras/stt1043}

\bibitem[{{Oppenheimer} {et~al.}(2016){Oppenheimer}, {Crain}, {Schaye}, {Rahmati}, {Richings}, {Trayford}, {Tumlinson}, {Bower}, {Schaller}, \& {Theuns}}]{Oppenheimer_2016}
{Oppenheimer}, B.~D., {Crain}, R.~A., {Schaye}, J., {et~al.} 2016, \mnras, 460, 2157, \dodoi{10.1093/mnras/stw1066}

\bibitem[{{Peeples} {et~al.}(2017){Peeples}, {Tumlinson}, {Fox}, {Aloisi}, {Fleming}, {Jedrzejewski}, {Oliveira}, {Ayres}, {Danforth}, {Keeney}, \& {Jenkins}}]{Peeples_2017}
{Peeples}, M., {Tumlinson}, J., {Fox}, A., {et~al.} 2017, {The Hubble Spectroscopic Legacy Archive}, Instrument Science Report COS 2017-4, 8 pages

\bibitem[{{Peeples} {et~al.}(2014){Peeples}, {Werk}, {Tumlinson}, {Oppenheimer}, {Prochaska}, {Katz}, \& {Weinberg}}]{Peeples_2014}
{Peeples}, M.~S., {Werk}, J.~K., {Tumlinson}, J., {et~al.} 2014, \apj, 786, 54, \dodoi{10.1088/0004-637X/786/1/54}

\bibitem[{{Petitjean} \& {Bergeron}(1990)}]{Petitjean_1990}
{Petitjean}, P., \& {Bergeron}, J. 1990, \aap, 231, 309

\bibitem[{{Pointon} {et~al.}(2019){Pointon}, {Kacprzak}, {Nielsen}, {Muzahid}, {Murphy}, {Churchill}, \& {Charlton}}]{Pointon_2019}
{Pointon}, S.~K., {Kacprzak}, G.~G., {Nielsen}, N.~M., {et~al.} 2019, \apj, 883, 78, \dodoi{10.3847/1538-4357/ab3b0e}

\bibitem[{Prochaska {et~al.}(2011)Prochaska, Weiner, Chen, Mulchaey, \& Cooksey}]{Prochaska_2011}
Prochaska, J.~X., Weiner, B., Chen, H.-W., Mulchaey, J., \& Cooksey, K. 2011, The Astrophysical Journal, 740, 91, \dodoi{10.1088/0004-637x/740/2/91}

\bibitem[{{Prochaska} {et~al.}(2017){Prochaska}, {Werk}, {Worseck}, {Tripp}, {Tumlinson}, {Burchett}, {Fox}, {Fumagalli}, {Lehner}, {Peeples}, \& {Tejos}}]{Prochaska_2017}
{Prochaska}, J.~X., {Werk}, J.~K., {Worseck}, G., {et~al.} 2017, \apj, 837, 169, \dodoi{10.3847/1538-4357/aa6007}

\bibitem[{Qu {et~al.}(2024)Qu, Chen, Johnson, Rudie, Zahedy, DePalma, Schaye, Boettcher, Cantalupo, Chen, Faucher-Giguère, Li, Mulchaey, Petitjean, \& Rafelski}]{Qu_2024}
Qu, Z., Chen, H.-W., Johnson, S.~D., {et~al.} 2024.
\newblock \doarXiv{2402.08016}

\bibitem[{{Sameer} {et~al.}(2024){Sameer}, {Charlton}, {Wakker}, {Kacprzak}, {Nielsen}, {Churchill}, {Richter}, {Muzahid}, {Ho}, {Nateghi}, {Rosenwasser}, {Narayanan}, \& {Ganguly}}]{Sameer_2024}
{Sameer}, {Charlton}, J.~C., {Wakker}, B.~P., {et~al.} 2024, \mnras, 530, 3827, \dodoi{10.1093/mnras/stae962}

\bibitem[{{Sanchez} {et~al.}(2019){Sanchez}, {Werk}, {Tremmel}, {Pontzen}, {Christensen}, {Quinn}, \& {Cruz}}]{Sanchez_2019}
{Sanchez}, N.~N., {Werk}, J.~K., {Tremmel}, M., {et~al.} 2019, \apj, 882, 8, \dodoi{10.3847/1538-4357/ab3045}

\bibitem[{{Schaye}(2001)}]{Schaye_2001}
{Schaye}, J. 2001, \apj, 559, 507, \dodoi{10.1086/322421}

\bibitem[{{Schaye} {et~al.}(2007){Schaye}, {Carswell}, \& {Kim}}]{Schaye_2007}
{Schaye}, J., {Carswell}, R.~F., \& {Kim}, T.-S. 2007, \mnras, 379, 1169, \dodoi{10.1111/j.1365-2966.2007.12005.x}

\bibitem[{{Schaye} {et~al.}(2000){Schaye}, {Rauch}, {Sargent}, \& {Kim}}]{Schaye_2000}
{Schaye}, J., {Rauch}, M., {Sargent}, W. L.~W., \& {Kim}, T.-S. 2000, \apjl, 541, L1, \dodoi{10.1086/312892}

\bibitem[{{Schroetter} {et~al.}(2021){Schroetter}, {Bouch{\'e}}, {Zabl}, {Rahmani}, {Wendt}, {Muzahid}, {Contini}, {Schaye}, {Schmidt}, \& {Wisotzki}}]{Schroetter_2021}
{Schroetter}, I., {Bouch{\'e}}, N.~F., {Zabl}, J., {et~al.} 2021, \mnras, 506, 1355, \dodoi{10.1093/mnras/stab1447}

\bibitem[{Steidel {et~al.}(2010)Steidel, Erb, Shapley, Pettini, Reddy, Bogosavljevi{\'{c}}, Rudie, \& Rakic}]{Steidel_2010}
Steidel, C.~C., Erb, D.~K., Shapley, A.~E., {et~al.} 2010, The Astrophysical Journal, 717, 289, \dodoi{10.1088/0004-637x/717/1/289}

\bibitem[{{Stern} {et~al.}(2018){Stern}, {Faucher-Gigu{\`e}re}, {Hennawi}, {Hafen}, {Johnson}, \& {Fielding}}]{Stern_2018}
{Stern}, J., {Faucher-Gigu{\`e}re}, C.-A., {Hennawi}, J.~F., {et~al.} 2018, \apj, 865, 91, \dodoi{10.3847/1538-4357/aac884}

\bibitem[{{Suresh} {et~al.}(2017){Suresh}, {Rubin}, {Kannan}, {Werk}, {Hernquist}, \& {Vogelsberger}}]{Suresh_2017}
{Suresh}, J., {Rubin}, K. H.~R., {Kannan}, R., {et~al.} 2017, \mnras, 465, 2966, \dodoi{10.1093/mnras/stw2499}

\bibitem[{{Tchernyshyov} {et~al.}(2022){Tchernyshyov}, {Werk}, {Wilde}, {Prochaska}, {Tripp}, {Burchett}, {Bordoloi}, {Howk}, {Lehner}, {O'Meara}, {Tejos}, \& {Tumlinson}}]{Tchernyshyov_2022}
{Tchernyshyov}, K., {Werk}, J.~K., {Wilde}, M.~C., {et~al.} 2022, \apj, 927, 147, \dodoi{10.3847/1538-4357/ac450c}

\bibitem[{Tchernyshyov {et~al.}(2023)Tchernyshyov, Werk, Wilde, Prochaska, Tripp, Burchett, Bordoloi, Howk, Lehner, O’Meara, Tejos, \& Tumlinson}]{Tchernyshyov_2023}
Tchernyshyov, K., Werk, J.~K., Wilde, M.~C., {et~al.} 2023, The Astrophysical Journal, 949, 41, \dodoi{10.3847/1538-4357/acc86a}

\bibitem[{{Tepper-Garc{\'\i}a} {et~al.}(2011){Tepper-Garc{\'\i}a}, {Richter}, {Schaye}, {Booth}, {Dalla Vecchia}, {Theuns}, \& {Wiersma}}]{Tepper_2011}
{Tepper-Garc{\'\i}a}, T., {Richter}, P., {Schaye}, J., {et~al.} 2011, \mnras, 413, 190, \dodoi{10.1111/j.1365-2966.2010.18123.x}

\bibitem[{{Tremonti} {et~al.}(2004){Tremonti}, {Heckman}, {Kauffmann}, {Brinchmann}, {Charlot}, {White}, {Seibert}, {Peng}, {Schlegel}, {Uomoto}, {Fukugita}, \& {Brinkmann}}]{Tremonti_2004}
{Tremonti}, C.~A., {Heckman}, T.~M., {Kauffmann}, G., {et~al.} 2004, \apj, 613, 898, \dodoi{10.1086/423264}

\bibitem[{{Tripp} {et~al.}(2008){Tripp}, {Sembach}, {Bowen}, {Savage}, {Jenkins}, {Lehner}, \& {Richter}}]{Tripp_2008}
{Tripp}, T.~M., {Sembach}, K.~R., {Bowen}, D.~V., {et~al.} 2008, \apjs, 177, 39, \dodoi{10.1086/587486}

\bibitem[{Tumlinson {et~al.}(2017)Tumlinson, Peeples, \& Werk}]{Tumlinson_2017}
Tumlinson, J., Peeples, M.~S., \& Werk, J.~K. 2017, Annual Review of Astronomy and Astrophysics, 55, 389, \dodoi{10.1146/annurev-astro-091916-055240}

\bibitem[{{Tumlinson} {et~al.}(2011){Tumlinson}, {Thom}, {Werk}, {Prochaska}, {Tripp}, {Weinberg}, {Peeples}, {O'Meara}, {Oppenheimer}, {Meiring}, {Katz}, {Dav{\'e}}, {Ford}, \& {Sembach}}]{Tumlinson_2011}
{Tumlinson}, J., {Thom}, C., {Werk}, J.~K., {et~al.} 2011, Science, 334, 948, \dodoi{10.1126/science.1209840}

\bibitem[{Turner {et~al.}(2014)Turner, Schaye, Steidel, Rudie, \& Strom}]{Turner_14}
Turner, M.~L., Schaye, J., Steidel, C.~C., Rudie, G.~C., \& Strom, A.~L. 2014, Monthly Notices of the Royal Astronomical Society, 445, 794, \dodoi{10.1093/mnras/stu1801}

\bibitem[{Turner {et~al.}(2015)Turner, Schaye, Steidel, Rudie, \& Strom}]{Turner_15}
---. 2015, Monthly Notices of the Royal Astronomical Society, 450, 2067, \dodoi{10.1093/mnras/stv750}

\bibitem[{{van Zee} \& {Haynes}(2006)}]{vanZee_2006}
{van Zee}, L., \& {Haynes}, M.~P. 2006, \apj, 636, 214, \dodoi{10.1086/498017}

\bibitem[{Werk {et~al.}(2013)Werk, Prochaska, Thom, Tumlinson, Tripp, O{\textquotesingle}Meara, \& Peeples}]{Werk_2013}
Werk, J.~K., Prochaska, J.~X., Thom, C., {et~al.} 2013, The Astrophysical Journal Supplement Series, 204, 17, \dodoi{10.1088/0067-0049/204/2/17}

\bibitem[{{Werk} {et~al.}(2016){Werk}, {Prochaska}, {Cantalupo}, {Fox}, {Oppenheimer}, {Tumlinson}, {Tripp}, {Lehner}, \& {McQuinn}}]{Werk_2016}
{Werk}, J.~K., {Prochaska}, J.~X., {Cantalupo}, S., {et~al.} 2016, \apj, 833, 54, \dodoi{10.3847/1538-4357/833/1/54}

\bibitem[{Wilde {et~al.}(2021)Wilde, Werk, Burchett, Prochaska, Tchernyshyov, Tripp, Tejos, Lehner, Bordoloi, O'Meara, \& Tumlinson}]{Wilde_2021}
Wilde, M.~C., Werk, J.~K., Burchett, J.~N., {et~al.} 2021, The Astrophysical Journal, 912, 9, \dodoi{10.3847/1538-4357/abea14}

\bibitem[{{Zabl} {et~al.}(2021){Zabl}, {Bouch{\'e}}, {Wisotzki}, {Schaye}, {Leclercq}, {Garel}, {Wendt}, {Schroetter}, {Muzahid}, {Cantalupo}, {Contini}, {Bacon}, {Brinchmann}, \& {Richard}}]{Zabl_2021}
{Zabl}, J., {Bouch{\'e}}, N.~F., {Wisotzki}, L., {et~al.} 2021, \mnras, 507, 4294, \dodoi{10.1093/mnras/stab2165}

\bibitem[{{Zahedy} {et~al.}(2019){Zahedy}, {Chen}, {Johnson}, {Pierce}, {Rauch}, {Huang}, {Weiner}, \& {Gauthier}}]{Zahedy_2019}
{Zahedy}, F.~S., {Chen}, H.-W., {Johnson}, S.~D., {et~al.} 2019, \mnras, 484, 2257, \dodoi{10.1093/mnras/sty3482}

\end{thebibliography}
\bibliographystyle{aasjournal}

\vspace{5mm}





 \appendix
 \section{Model comparison} 
 \label{sec:model_comp}

In section \ref{N_prof}, we have computed the model averaged $\left<N(\OVI)\right>$ following a modified exponential profile. In literature, several other functional forms are adopted to determine the $N(\OVI)-$profiles. Here, we present a brief overview of two such models and compare the results.

 \subsection{\citet[]{Qu_2024}}

\begin{figure*}
 \centering
     \includegraphics[width=0.33\linewidth]{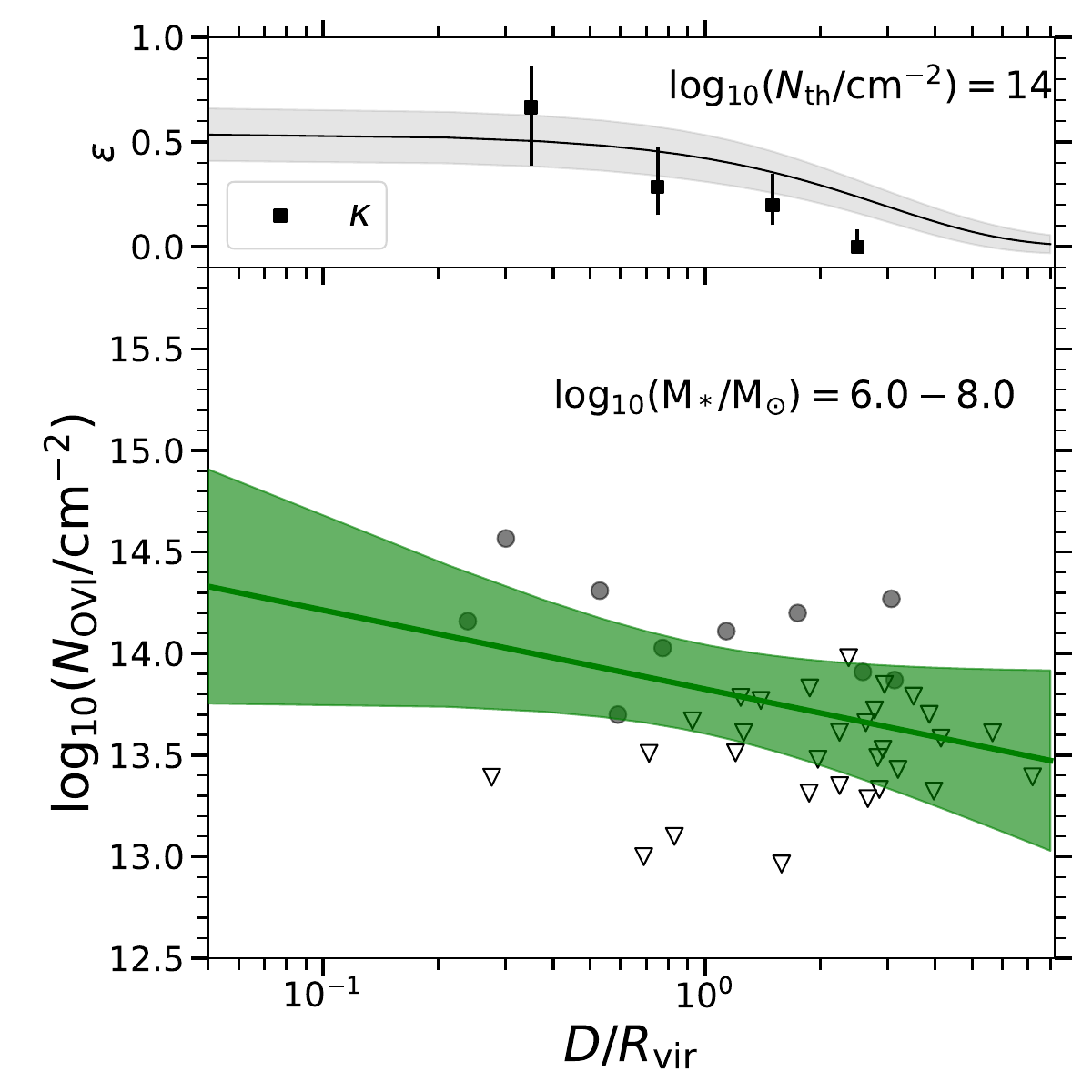}%
     \includegraphics[width=0.33\linewidth]{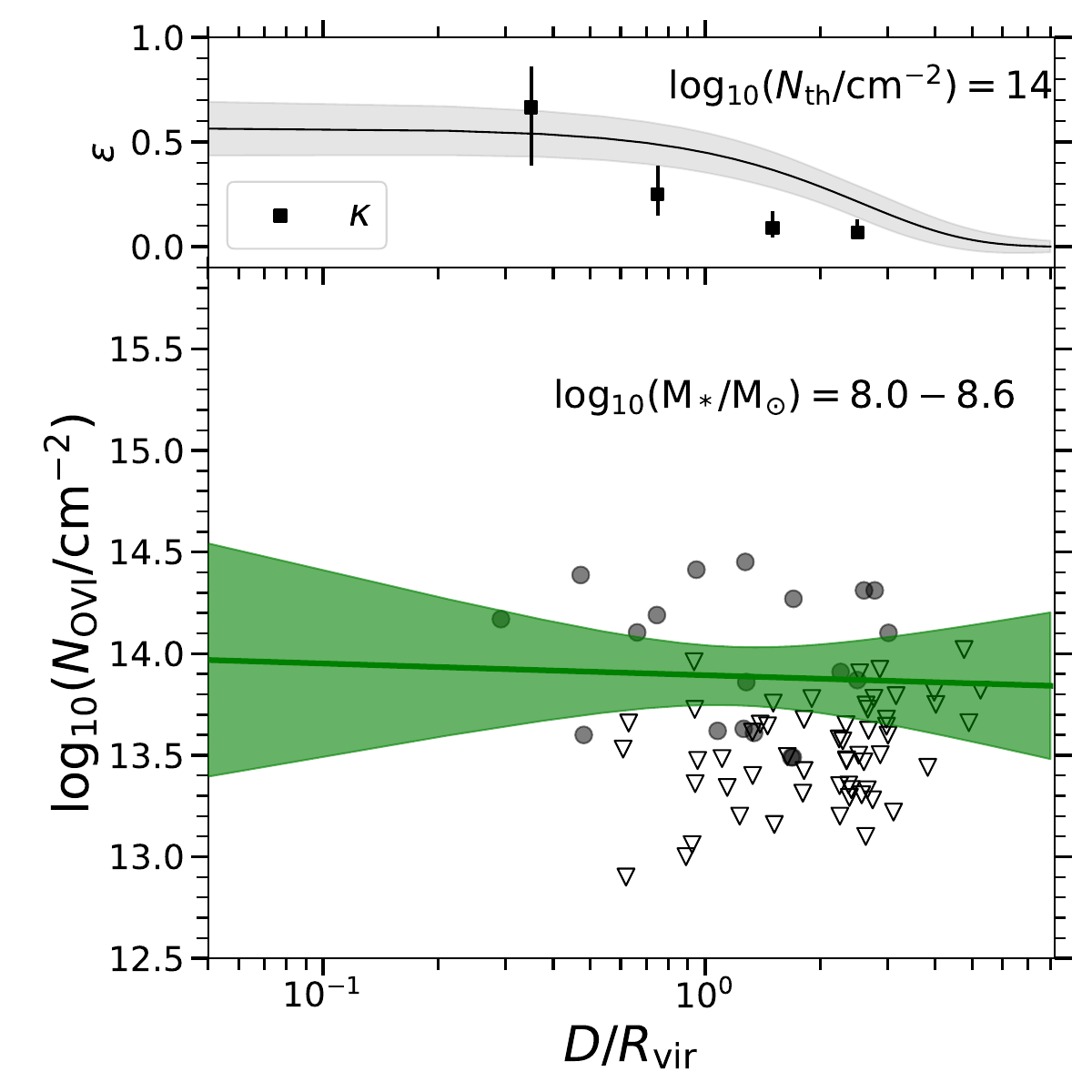}%
     \includegraphics[width=0.33\linewidth]{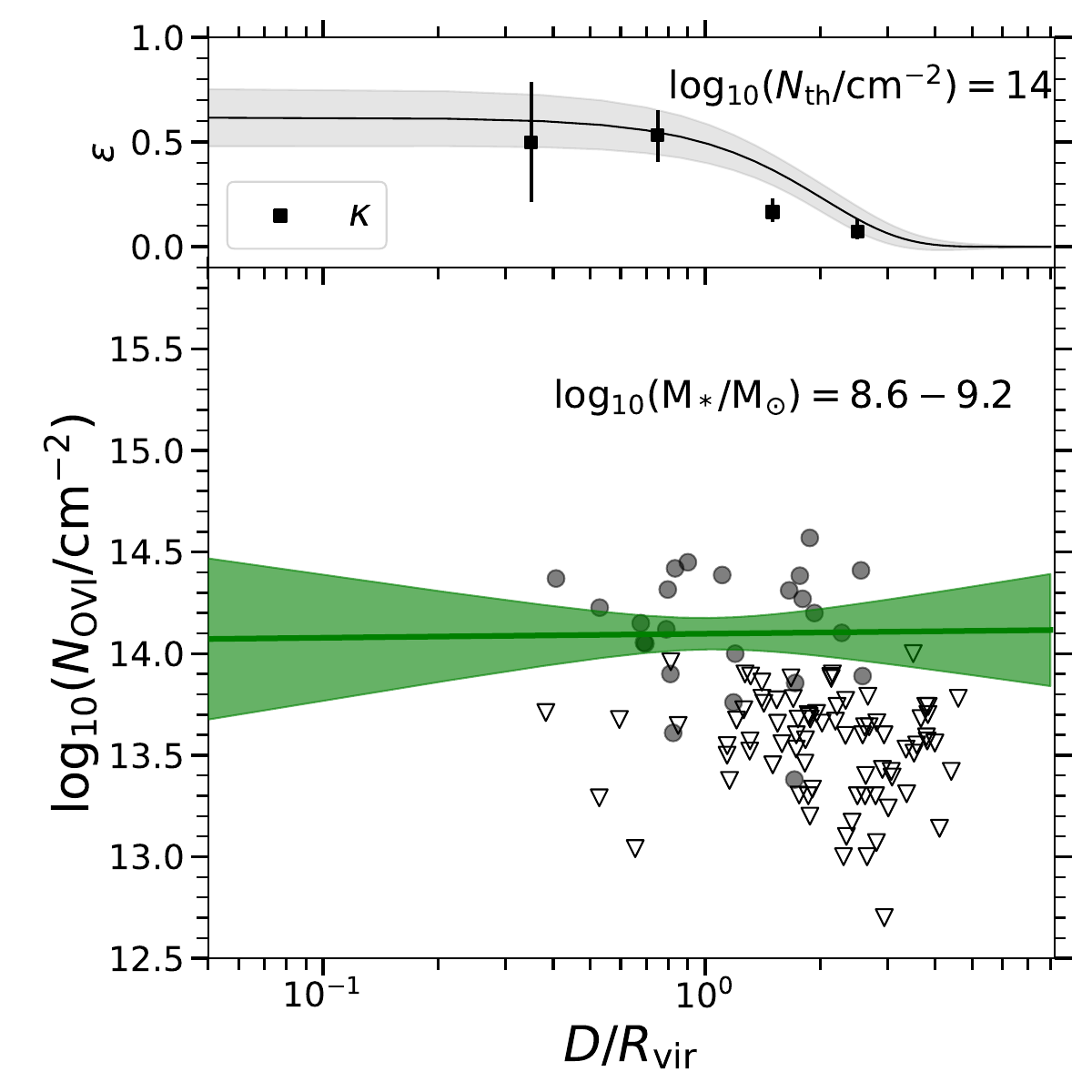}
     \includegraphics[width=0.33\linewidth]{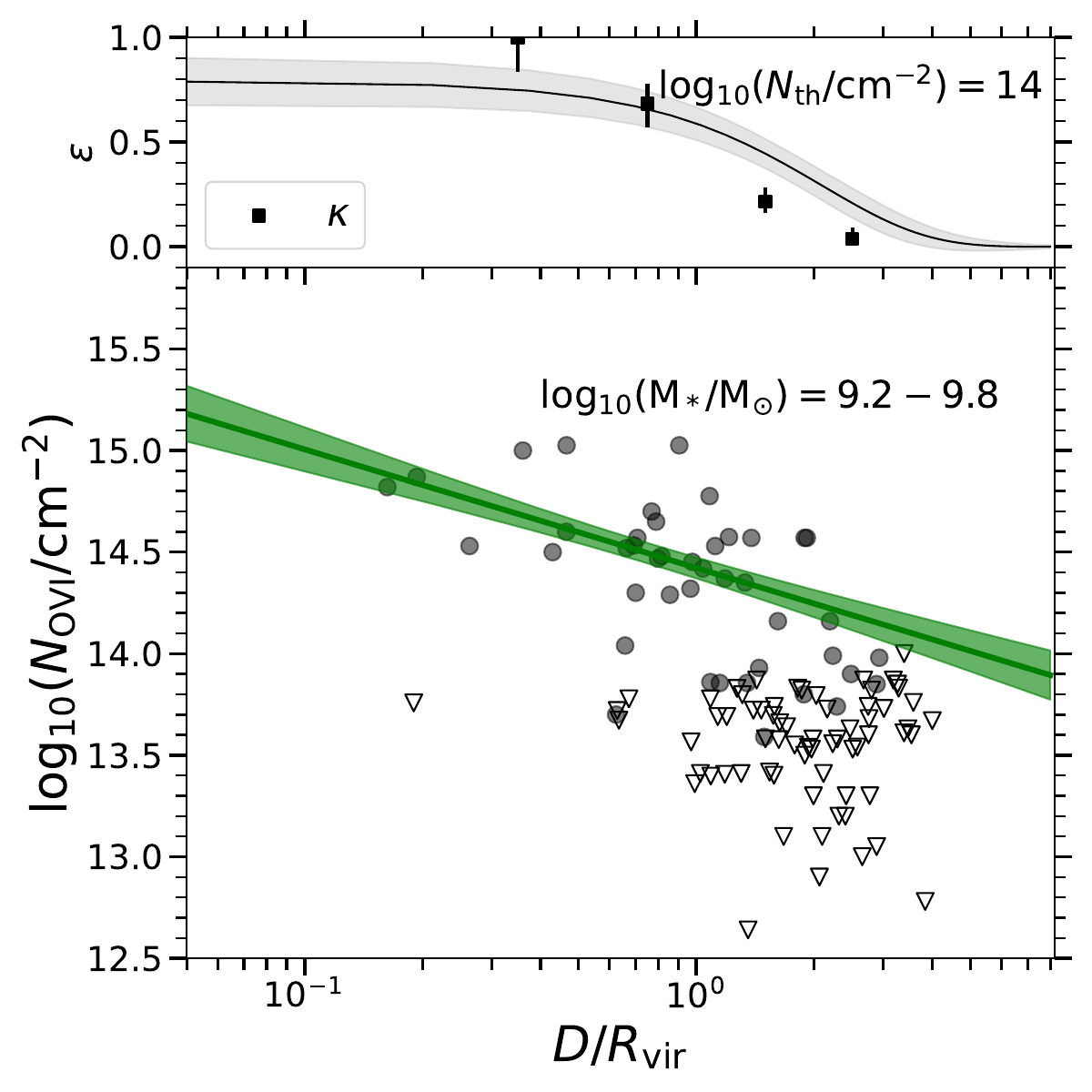}%
     \includegraphics[width=0.33\linewidth]{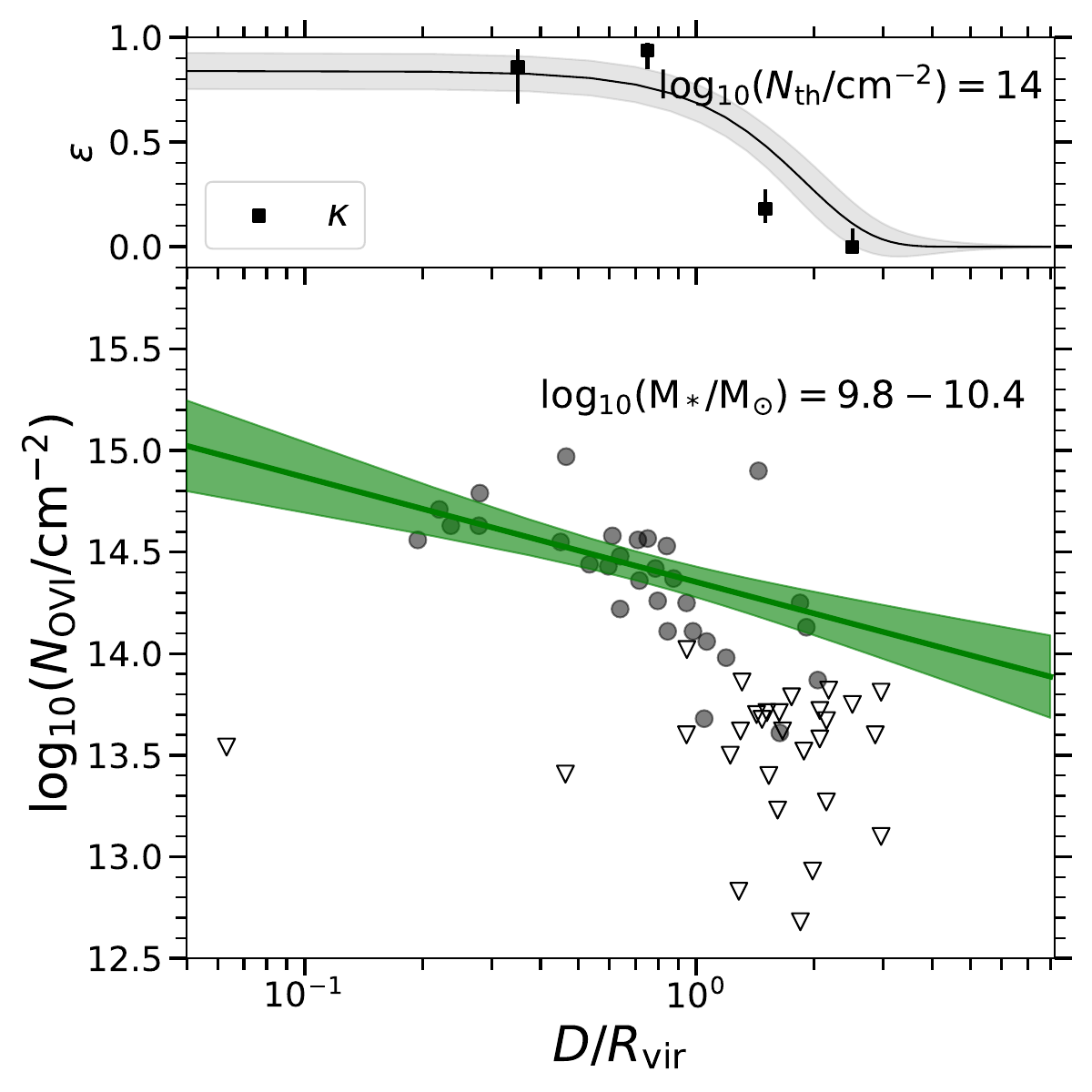}%
     \includegraphics[width=0.33\linewidth]{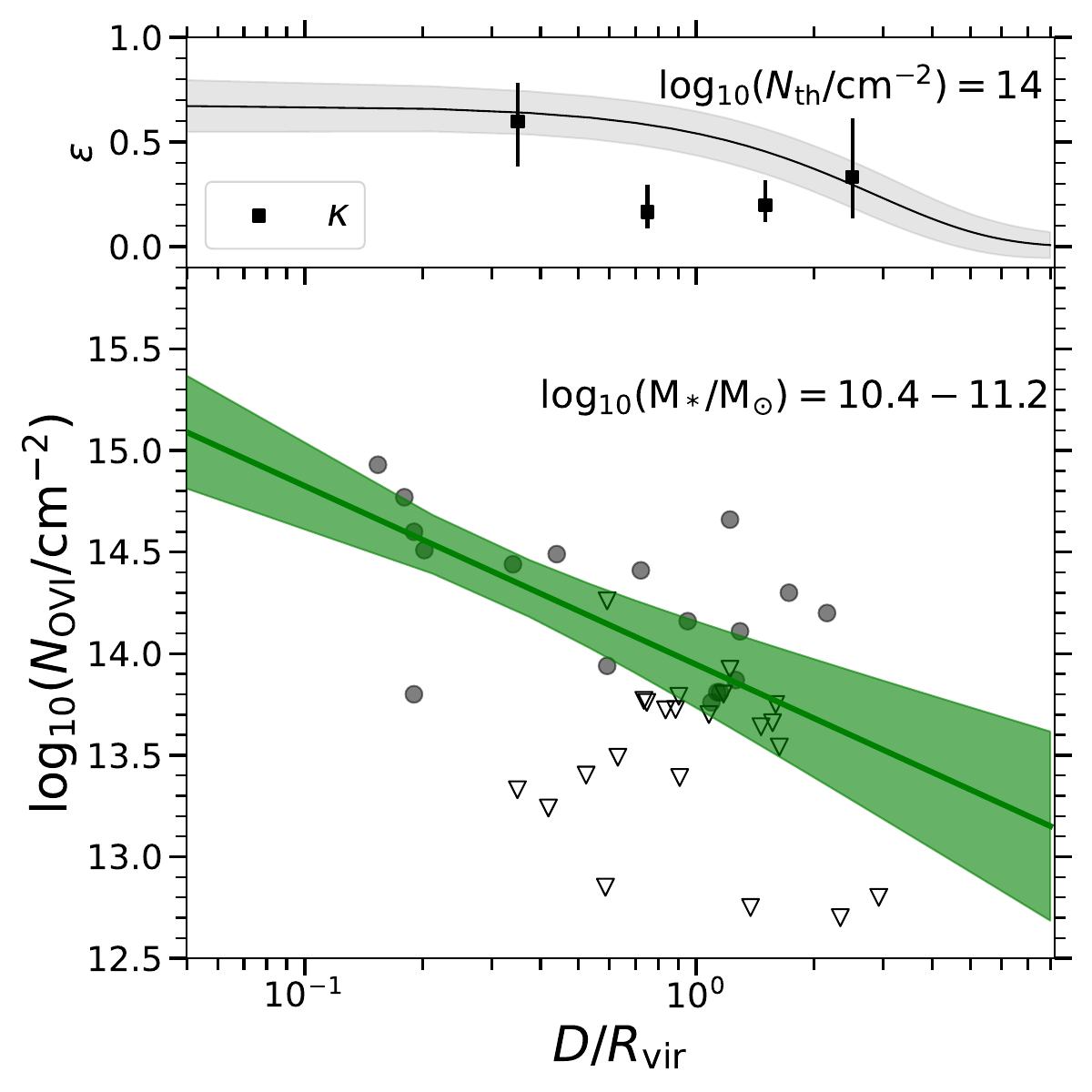}  

     \caption{Model $N(\OVI)-$profiles for different mass bins following the procedure of \citet[]{Qu_2024}. The smaller upper panels show the variation of the parameter $\epsilon$ (see Eq.~\ref{eq:epsilon}) with normalized impact parameter.
     The square symbols represent binned covering fraction with a threshold column density of \lognovi=14.}    
     \label{fig:qu_novi}
 \end{figure*}

 \citet[]{Qu_2024} adopted a simple power-law (i.e., $N \propto (D/R_{\rm vir})^{-\alpha}$) to model the $N(\OVI)-$profile, but added an intrinsic covering fraction to account for the non-detections at smaller impact parameters. The two-component model, similar to the one used by \citet[]{Huang_2021} has a modified exponential covering fraction given by 
 \begin{equation}
     \epsilon(D/R_{\rm vir}) = {\rm Min}\left[1, \epsilon_0~{\rm exp}\left( -\left(\frac{D/R_{\rm vir}}{x_0} \right)^{\beta} \right)\right]~. 
     \label{eq:epsilon} 
 \end{equation}
The best-fit $N(\OVI)-$profiles for this model are shown in Fig.~\ref{fig:qu_novi}. The $\left<N(\OVI)\right>$ and $M(\OVI)$ obtained with this model are shown with black pentagons in the left and right panels of Fig.~\ref{fig:model_comp} respectively.

 \subsection{\citet[]{Tchernyshyov_2022}}
 \label{sec:appendix_tch23} 
 
\begin{figure*}
 \centering
     \includegraphics[width=0.33\linewidth]{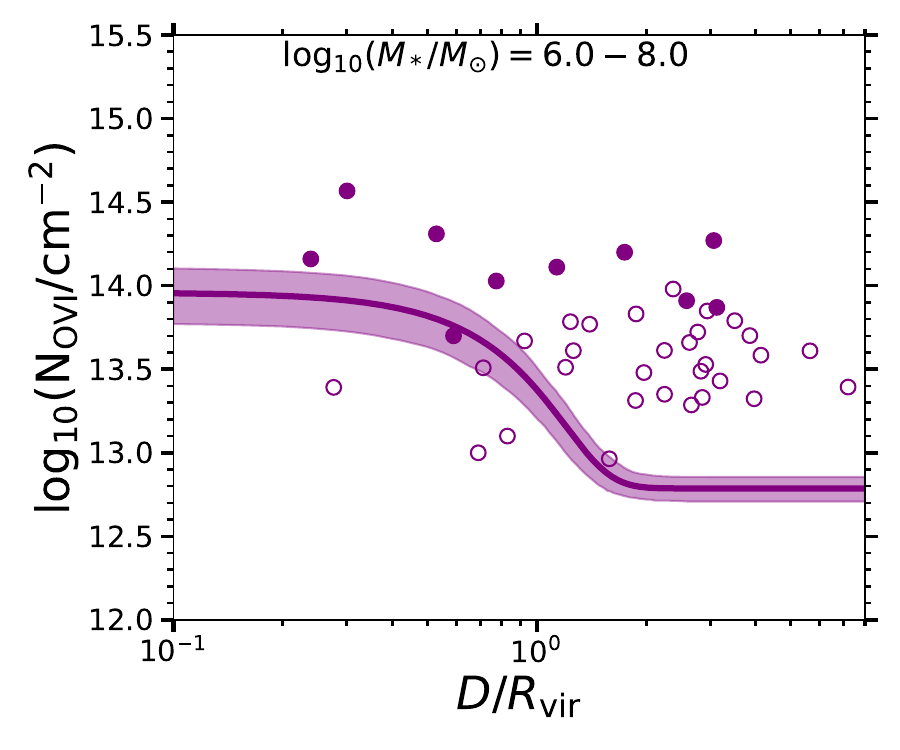}%
     \includegraphics[width=0.33\linewidth]{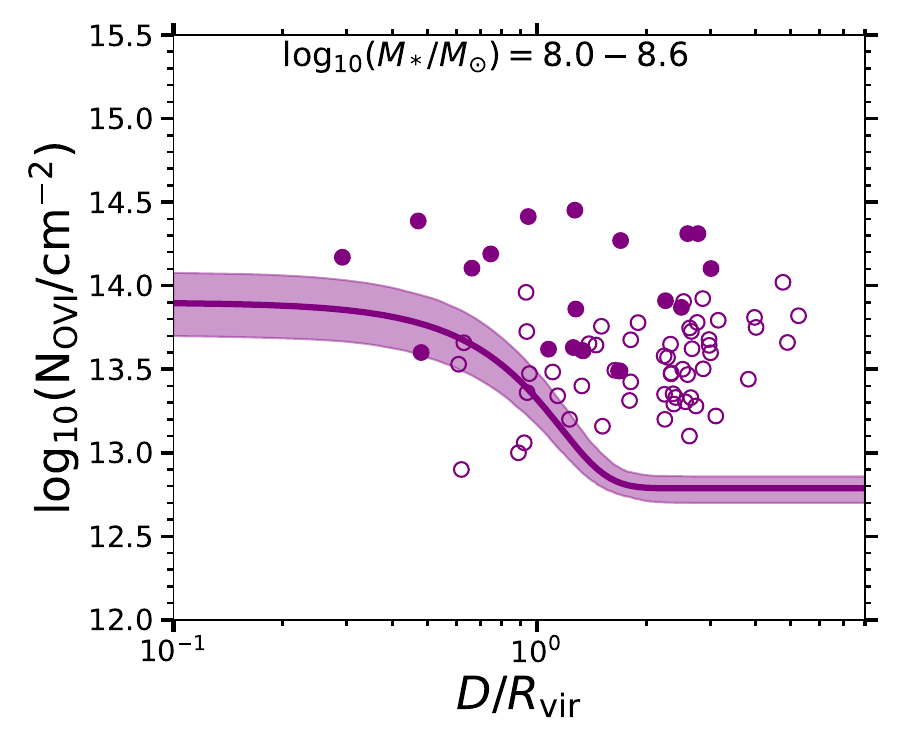}%
     \includegraphics[width=0.33\linewidth]{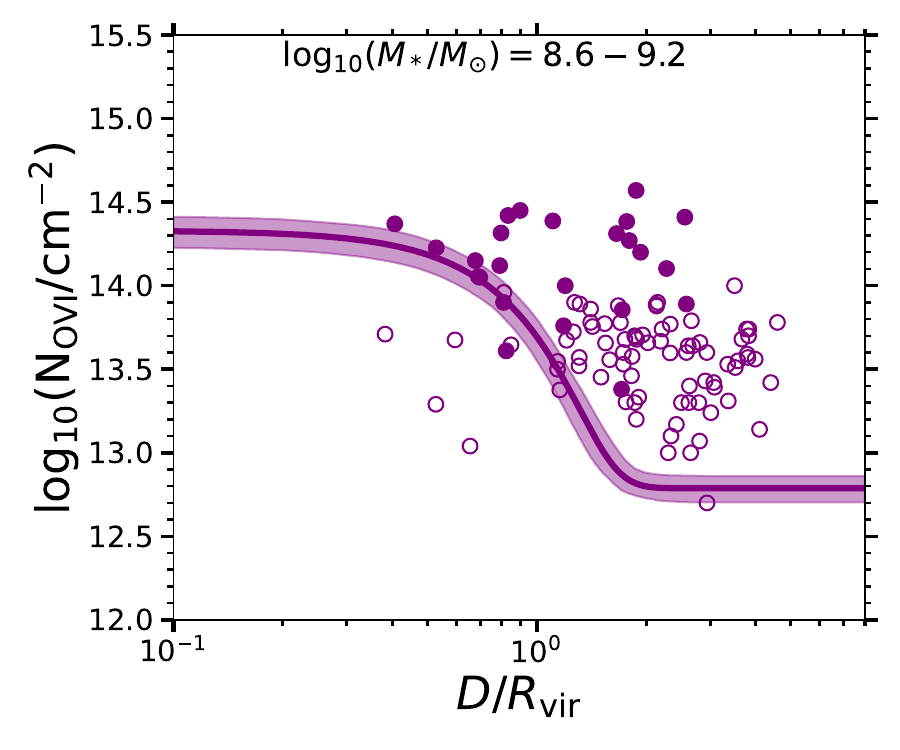}
     \includegraphics[width=0.33\linewidth]{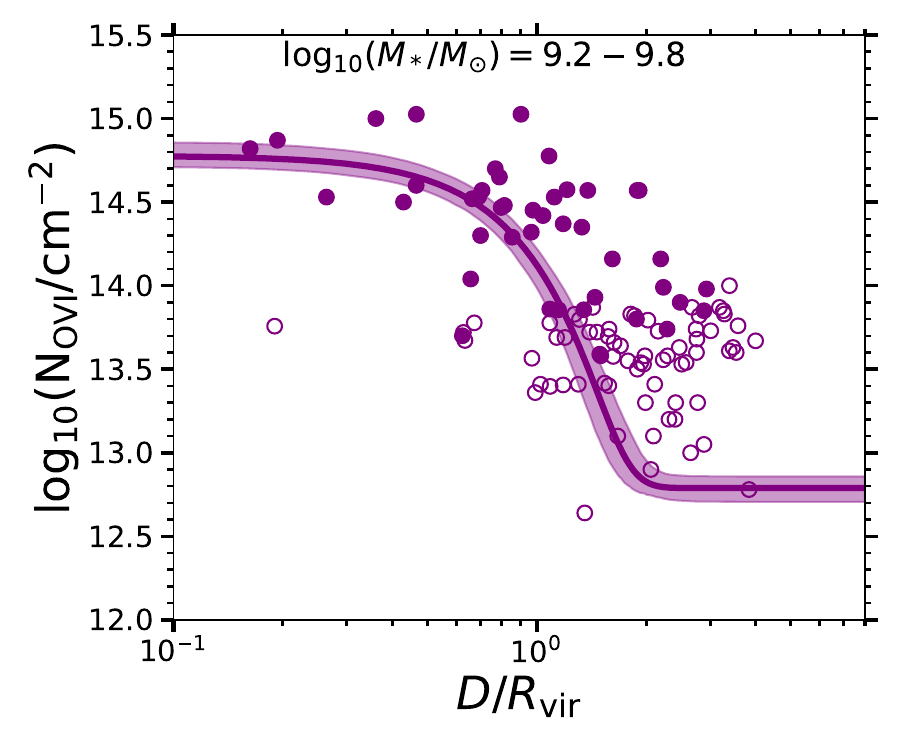}%
     \includegraphics[width=0.33\linewidth]{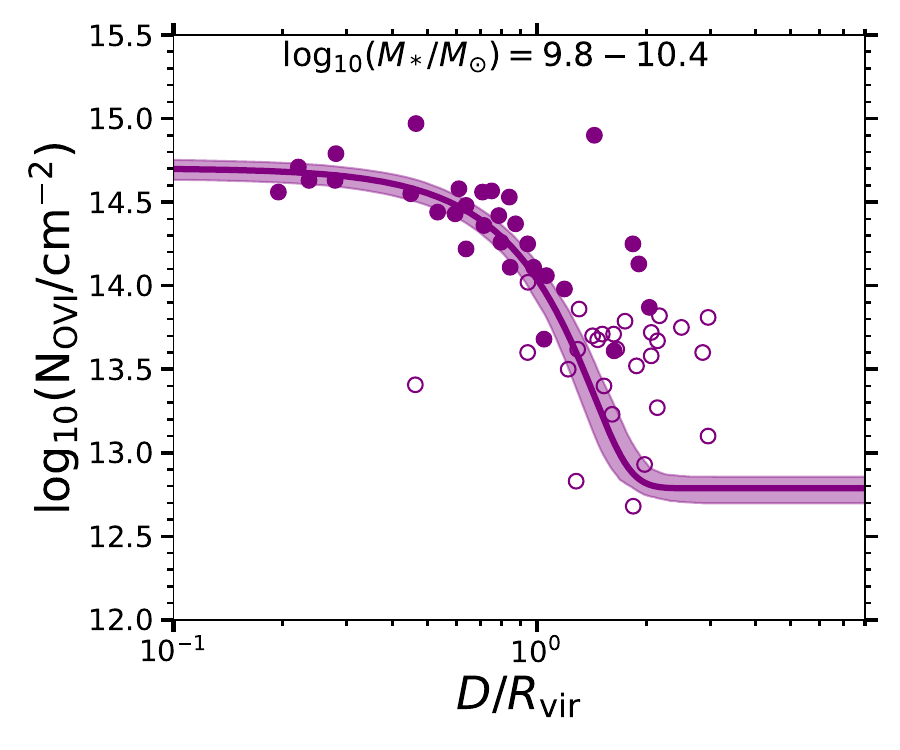}%
     \includegraphics[width=0.33\linewidth]{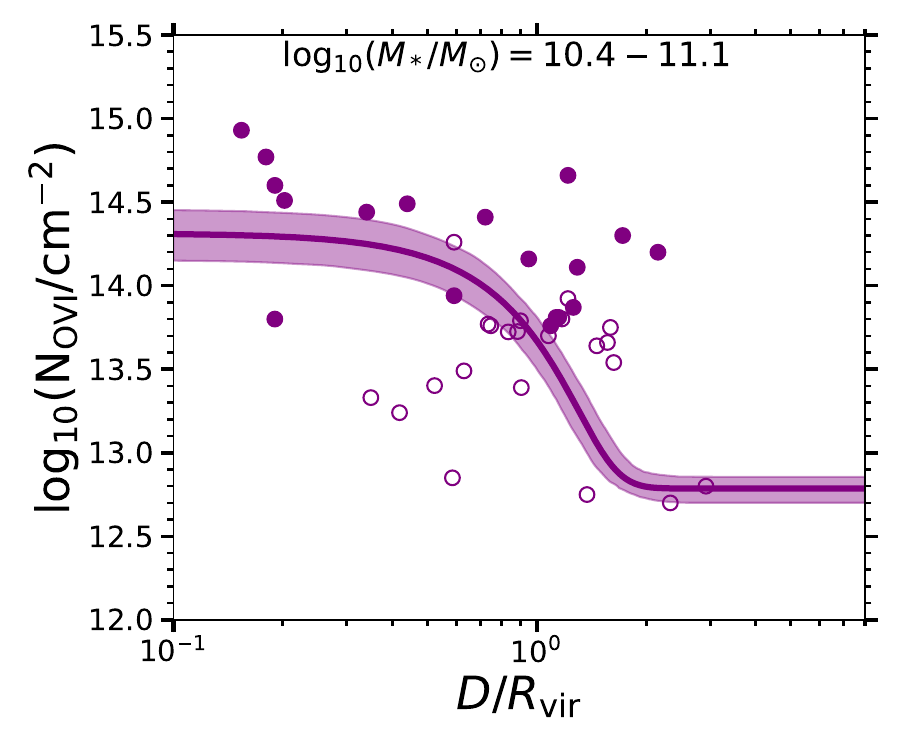}  
     \caption{Model $N(\OVI)-$profiles for different mass bins following the procedure of \citet[]{Tchernyshyov_2022} briefly outlined in Section~\ref{sec:appendix_tch23}.}
     \label{fig:tch_novi}
 \end{figure*}

\begin{figure*}
    \centering
    \includegraphics[width=0.55\linewidth]{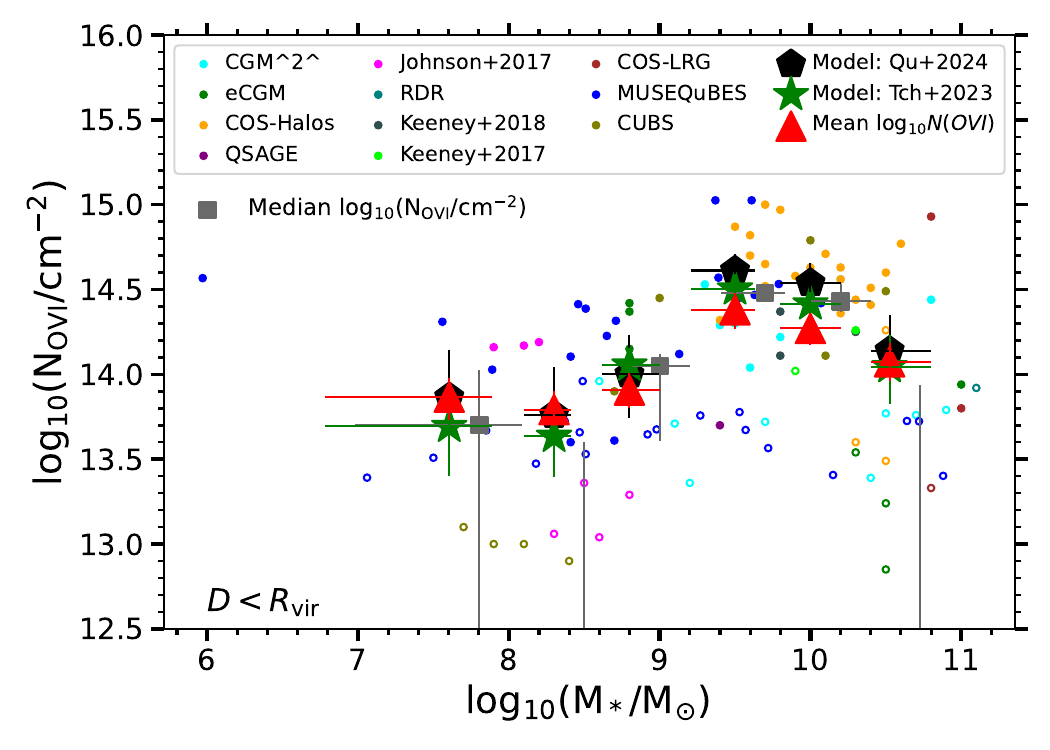}%
    \includegraphics[width=0.48\linewidth]{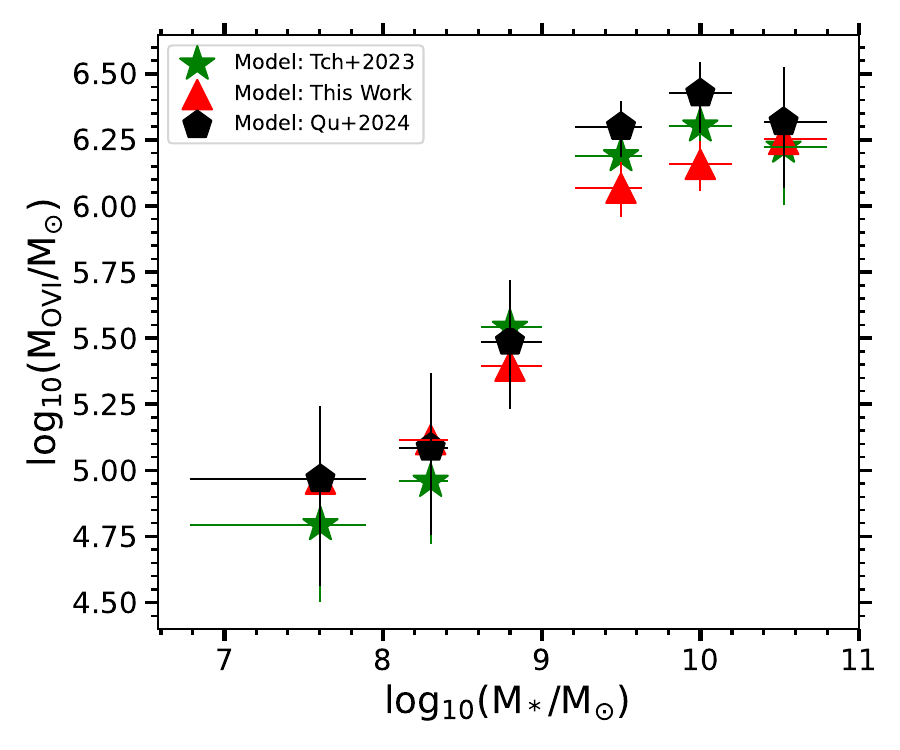}
    \caption{{\tt Left:} Same as Fig.~\ref{fig:avgOVI_N} but with added measurements following the prescriptions of \citet[][green star symbols]{Tchernyshyov_2022} and \citet[][black pentagons]{Qu_2024}. Our measurements (red triangles) are in good agreement with the measurements obtained from their prescriptions. {\tt Right:} $M(\OVI)$ within $R_{\rm vir}$ for the star-forming galaxies obtained from the model-averaged $N(\OVI)$ shown on the right panel plotted against the \logm\ of host galaxies. 
    }
    \label{fig:model_comp}
\end{figure*}

\citet[]{Tchernyshyov_2022} adopted a modified exponential profile similar to our work with a baseline column density. Their $N(\OVI)-$profile is given by: 
\begin{equation}
\begin{split}
\label{eq:Novi}
    {\rm log_{10}} \novie (D/R_{\rm vir})= 
    {\rm log_{10}}\left(10^{\mathcal{N}_0} e^{-(\frac{D/R_{\rm vir}}{L_s})^{\gamma}} + 10^{\mathcal{N}_b}  \right)~.  
\end{split}    
\end{equation}
Following their methodology, we have generated the $N(\OVI)-$ profiles keeping the baseline $\mathcal{N}_b$, slope $\gamma$ and length-scale $L_s$ tied in the six mass bins, only allowing the normalization $\mathcal{N}_0$ to vary in different \logm\ bins. The best-fit profiles are shown in Fig.~\ref{fig:tch_novi}. The $\left<N(\OVI)\right>$ and $M(\OVI)$ are shown with green stars in the left and right panels of Fig.~\ref{fig:model_comp} respectively.

 \begin{figure*}
  \centering

     \includegraphics[width=0.33\linewidth]{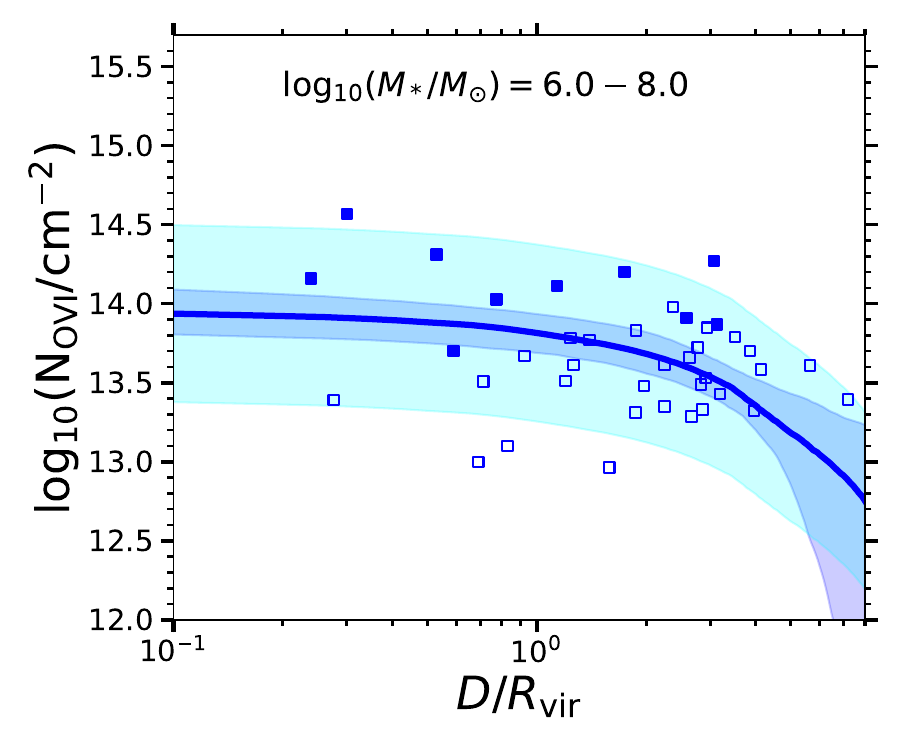}%
     \includegraphics[width=0.33\linewidth]{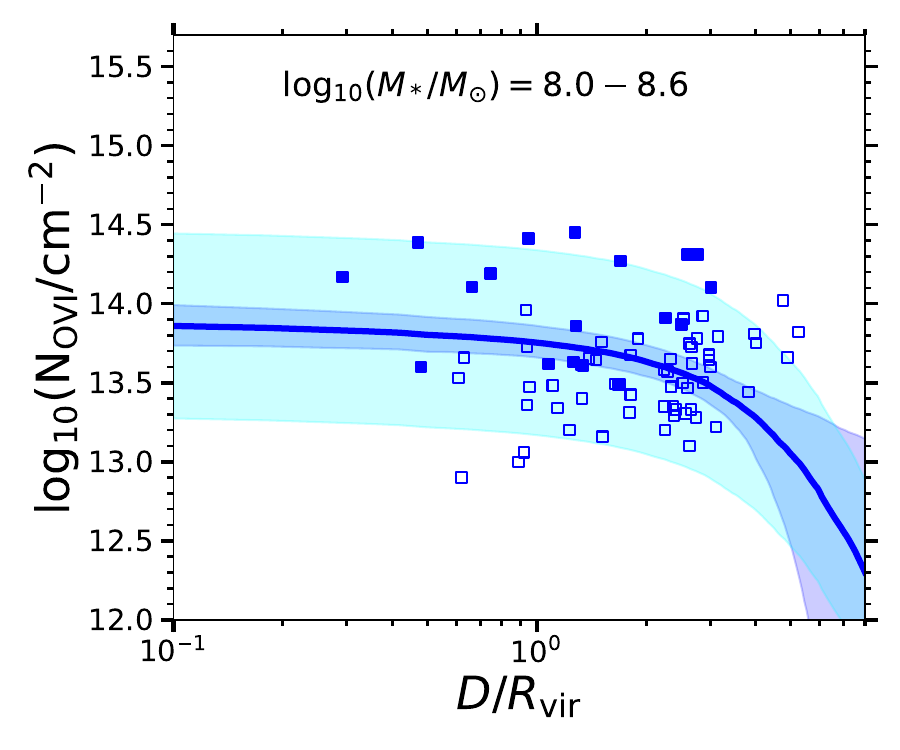}%
     \includegraphics[width=0.33\linewidth]{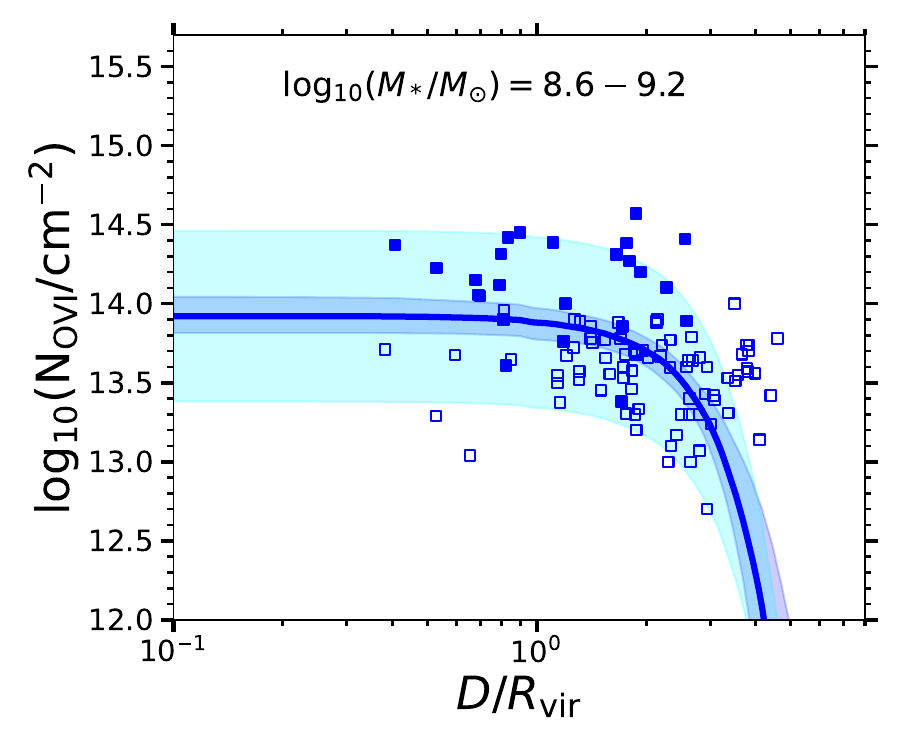}
     \includegraphics[width=0.33\linewidth]{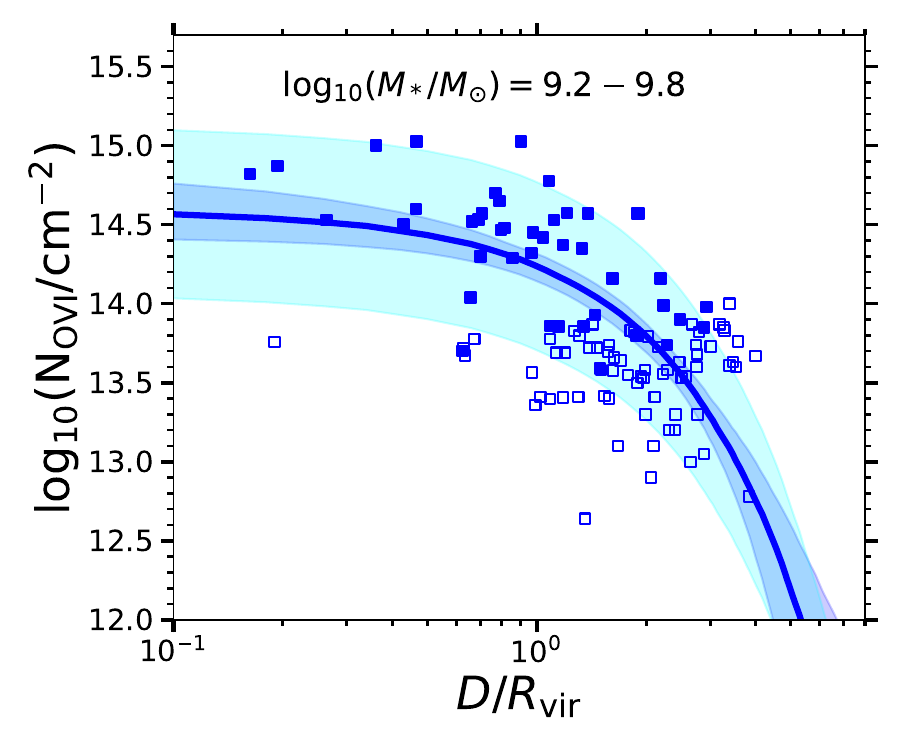}%
     \includegraphics[width=0.33\linewidth]{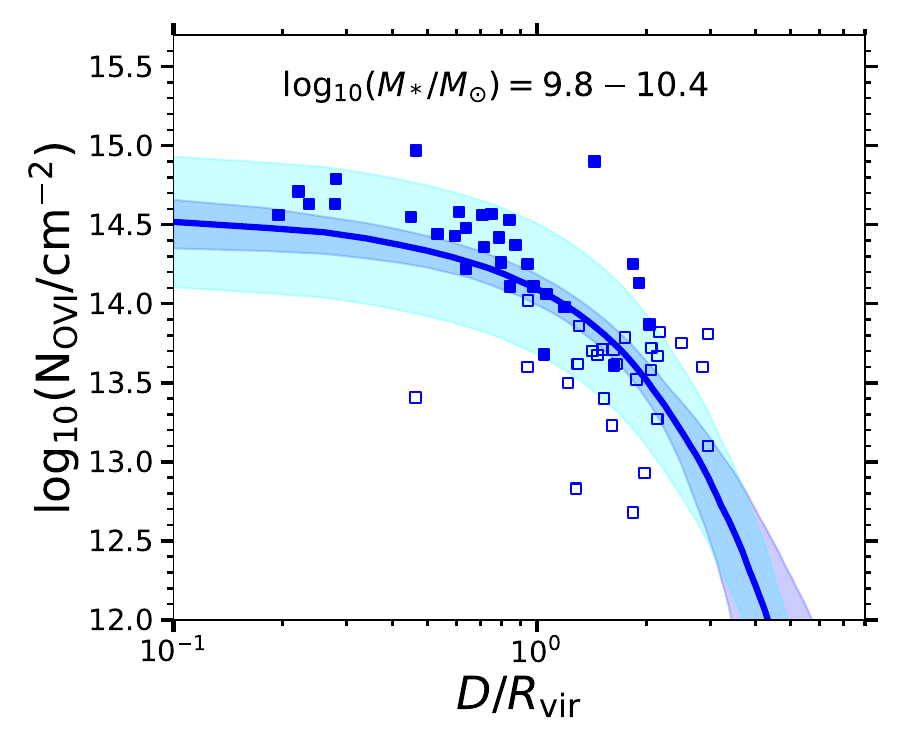}%
     \includegraphics[width=0.33\linewidth]{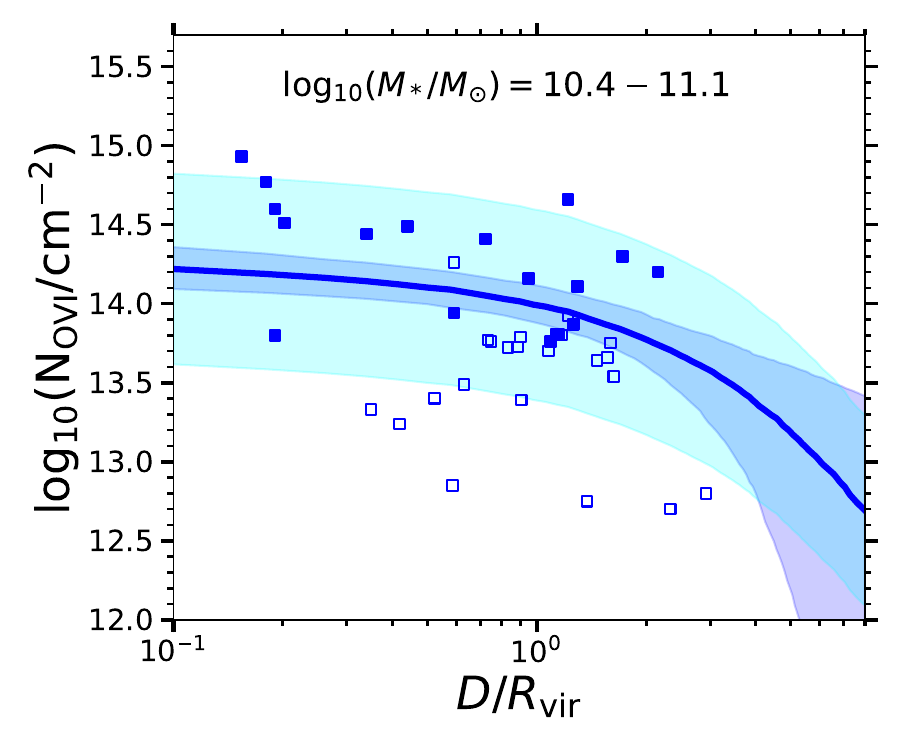}  
     \caption{Same as the right panel of Fig.~\ref{fig:Novi-prof}, but for different mass bins as indicated in the panels. We adopted a modified exponential function given in Eq.~\ref{eq:novi_prof} to model the profiles.}   
 \end{figure*}

\section{$\left<N(\OVI)\right>$ comparison for star-forming and passive galaxies} 
\label{sec:novi_sf_e}

\begin{figure}
    \centering
    \includegraphics[width=0.7\linewidth]{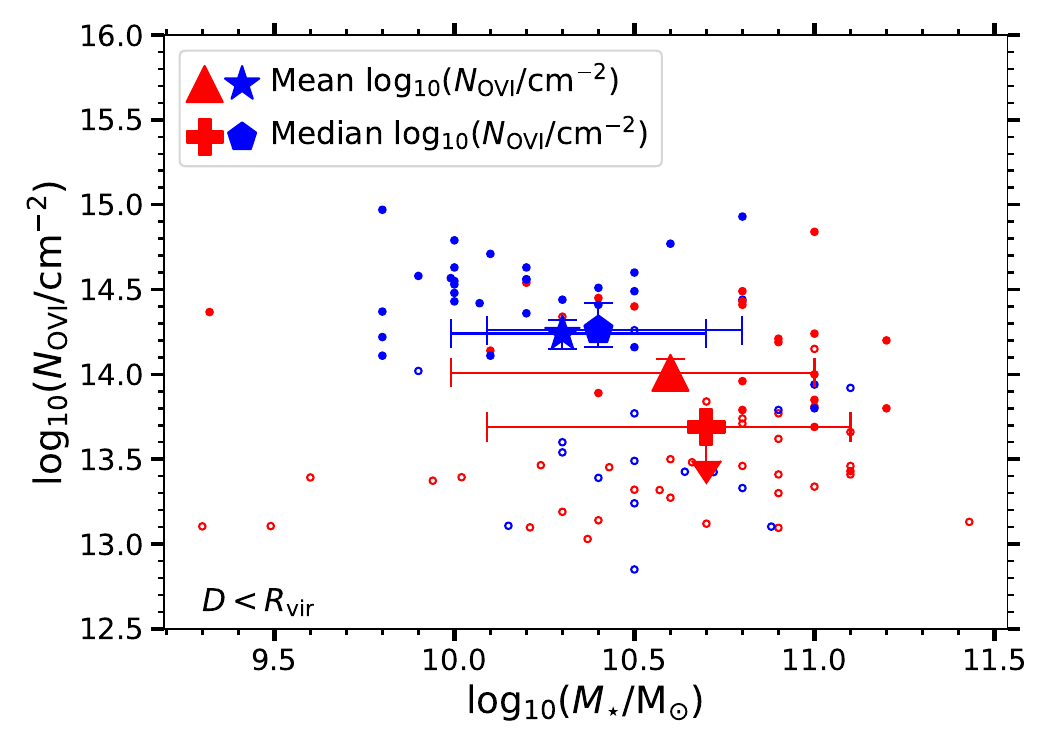}
    \caption{\OVI\ column density against stellar mass for the combined sample of MUSEQuBES and literature. The blue star symbol and red triangle indicate the model-averaged $N(\OVI)$ within $R_{\rm vir}$ for the star-forming and passive galaxies, respectively. The blue pentagon and red plus with a downward arrow represent the non-parametric median $N(\OVI)$ within $R_{\rm vir}$ for star-forming galaxies and the 95\% upper limit on the median for the passive galaxies, respectively, obtained with the {\sc Lifelines} package of {\sc Python}.   
    The blue and red small circles represent individual measurements for star-forming and passive galaxies within $R_{\rm vir}$. The open circles represent $3\sigma$ upper limits. Star-forming galaxies with \logm$>9.8$ are only considered here to control the stellar mass.}
    \label{fig:sf-q_comp}
\end{figure}

The \OVI\ column densities for passive galaxies in MUSEQuBES sample do not show any significant anticorrelation with $D/R_{\rm vir}$ (see Fig.~\ref{fig:Novi-prof}). This is also true for the combined sample of passive galaxies from  MUSEQuBES and the literature ($\tau=-0.1,~p=0.08$). In order to quantify the difference of $\left<N(\OVI)\right>$ inside $R_{\rm vir}$ for the star-forming and passive galaxies, we adopted the modified exponential function used in this work to model the $N(\OVI)$--profile of passive galaxies.
The model-averaged $\left<N(\OVI)\right>$ for  passive galaxies is $\approx10^{14}~{\rm cm}^{-2}$ within $R_{\rm vir}$. In order to compare with the star-forming galaxies of similar mass, we combined the last two \logm\ bin of star-forming galaxies from section \ref{sec:4.1.1}. The $\left<N(\OVI)\right>\approx10^{14.3}~{\rm cm}^{-2}$ within $R_{\rm vir}$ of star-forming galaxies is significantly enhanced compared to the passive galaxies (see Fig.~\ref{fig:sf-q_comp}).

\end{document}